\documentclass[twocolumn]{article}

\usepackage{arxiv}

\usepackage[utf8]{inputenc} 
\usepackage[T1]{fontenc}    
\usepackage{hyperref}       
\usepackage{url}            
\usepackage{booktabs}       
\usepackage{amsfonts}       
\usepackage{nicefrac}       
\usepackage{microtype}      
\usepackage{lipsum}		
\usepackage{graphicx}
\usepackage[sort, numbers]{natbib}
\usepackage{doi}

\usepackage{hyperref} 

\usepackage[table,xcdraw]{xcolor}
\usepackage[edges]{forest} 
\usepackage{float}
\usepackage{multirow}
\usepackage{adjustbox}
\usepackage{tabularx}
\usepackage{hyperref}
\usepackage{lscape}
\usepackage{multirow}
\usepackage{graphicx}
\usepackage{longtable}

\usepackage[normalem]{ulem}
\useunder{\uline}{\ul}{}
\usepackage{lscape}
\usepackage{float}
\floatstyle{plaintop}
\restylefloat{table}
\usepackage{caption}
\usepackage{booktabs}
\usepackage{float}

\usepackage{titlesec}

\titlespacing{\section}{0pt}{\parskip}{-\parskip}
\titlespacing{\subsection}{0pt}{\parskip}{-\parskip}
\titlespacing{\subsubsection}{0pt}{\parskip}{-\parskip}

\makeatletter
\newcommand\subsubsubsection{\@startsection{paragraph}{4}{\z@}{-0.75ex\@plus -1ex \@minus -.25ex}{0.001ex \@plus .00ex}{\normalfont\normalsize\bfseries}}
\makeatother

\title{Domain Adaptation and Generalization on Functional Medical Images: A Systematic Survey}

\author{%
\textbf{Gita Sarafraz}\textsuperscript{*},
\textbf{Armin Behnamnia}\textsuperscript{*},
\textbf{Mehran Hosseinzadeh}\textsuperscript{*},
\textbf{Ali Balapour}\textsuperscript{*},
\textbf{Amin Meghrazi}\textsuperscript{*},\\
and \textbf{Hamid R. Rabiee}\textsuperscript{**}
\\[1ex]
Sharif University of Technology
\\[1ex]
Department of Computer Engineering
\\[1ex]
\{gita.sarafraz, armin.behnamnia, mehran.hosseinzadeh1, 
\\  ali.balapour, amin.meghrazi, rabiee\}@sharif.edu
}

\date{}

\renewcommand{\shorttitle}{\textit{arXiv} Template}

\begin{document}
\maketitle
\def\thefootnote{*}\footnotetext{Equal Contribution}\def\thefootnote{\arabic{footnote}}
\def\thefootnote{**}\footnotetext{Corresponding Author}\def\thefootnote{\arabic{footnote}}

	\hspace{7pt}  \textbf{\textit{Abstract}--
 Machine learning algorithms have revolutionized different fields, including natural language processing, computer vision, signal processing, and medical data processing. Despite the excellent capabilities of machine learning algorithms in various tasks and areas, the performance of these models mainly deteriorates when there is a shift in the test and training data distributions. This gap occurs due to the violation of the fundamental assumption that the training and test data are independent and identically distributed (i.i.d). In real-world scenarios where collecting data from all possible domains for training is costly and even impossible, the i.i.d assumption can hardly be satisfied. The problem is even more severe in the case of medical images and signals because it requires either expensive equipment or a meticulous experimentation setup to collect data, even for a single domain. Additionally, the decrease in performance may have severe consequences in the analysis of medical records. As a result of such problems, the ability to generalize and adapt under distribution shifts (domain generalization (DG) and domain adaptation (DA)) is essential for the analysis of medical data. This paper provides the first systematic review of DG and DA on functional brain signals to fill the gap of the absence of a comprehensive study in this era. We provide detailed explanations and categorizations of datasets, approaches, and architectures used in DG and DA on functional brain images. We further address the attention-worthy future tracks in this field.
}


\section{Introduction}
 Machine Learning (ML) is the process of guiding a computer system on how to make accurate predictions for a specific task when fed with data. Given the popularity of previous machine learning approaches, the main challenge in using them is how to choose features that fit more information and overlap less before learning. Deep Learning (DL) is a subset of machine learning techniques that achieve precise performance and flexibility in several learning tasks, such as medical image analysis, without the need to specify features before the learning process; these models are trained based on the assumption that the training and testing data are identically and independently distributed (i.i.d.) \cite{domaingeneralizationsurvey2022_second}. Due to the i.i.d assumption and the emergence of a variety of datasets in every machine learning task, models trained on a particular domain data work poorly on the data from other domains and cannot function well on new data samples. The lack of domain generalization, which is the ability of the model to work well on new data samples from different domains, makes many deep neural networks and traditional ML models impractical and unusable for real-world applications.
 
 The generalization issue is even more apparent in medical image analysis. On the one hand, because of the vast range of conditions, priors, and affecting factors for each data sample, expecting the model to work on data measured in a different situation or from a different subject is often impracticable. On the other hand, considering that the study of medical images directly concerns people's health, even small mistakes are unacceptable and can lead to severe consequences. Hence, in these tasks, the ability to adapt the model trained on single or multiple source domains into a new target domain, known as domain adaptation (DA), and train generalizable models, known as domain generalization (DG), is crucial.

In this work, we present the first comprehensive review of methods for establishing domain generalization/adaptation for medical images, focusing on functional brain data.

In this survey, each model is categorized by:

\begin{enumerate}

    \item \textbf{Approach}: main idea for DG/DA,

    \item \textbf{Architecture}: the building block using which the model is trained to make a DG/DA system,

    \item \textbf{Domain}: the type of domain defined in the generalization/adaptation task,

    \item \textbf{Task}: the main task that the model is required to solve in a DG/DA fashion,

    \item \textbf{Multi/single source}: whether the work tackles the situation in which we have multiple sources (multi-source) or not (single-source).

\end{enumerate}

We also collect the popular and mainly used datasets in the literature and provide a brief explanation of each as well as a comparison by different properties, such as the number of subjects and size of the dataset.

There are several review papers on domain generalization or adaptation methods in general concept \cite{domaingeneralizationsurvey2022}, \cite{domaingeneralizationsurvey2022_second}, \cite{domainadatationsurvey}, \cite{domainadaptationsurvey2}, and one survey paper specialty on domain adaptation in medical images \cite{domainadaptationmedical} which mainly focuses on models built for structural brain data such as MRI. Nevertheless, this work focuses on functional brain data, which is more inclusive and vital, and it also investigates recent models more thoroughly and systematically.

This paper is organized as follows: In section \ref{sec:background}, we briefly review the concepts, notations, and fields related to DG/DA and medical images and signals analysis. Section \ref{sec:problem_Definition} describes the applications and studied tasks of domain adaptation and generalization in medical image analysis. In the following, the most widely adopted architectures in the methods reviewed in this study are presented in Section \ref{sec:architectures}. Next, Section \ref{sec:methods} reviews remarkable recent DA and DG methods used to process medical image data. In this section, we provide a comprehensive hierarchy of the approaches followed in the literature that semantically categorizes recent studies in this field. In Section \ref{sec:datasets}, we go through the popular public datasets used as benchmarks for domain adaptation or generalization of medical image data. Lastly, in Section \ref{sec:futuredirection}, we propose potential future works that are suggested to be followed according to our studies, and in Section \ref{sec:conclusion} we conclude the article.


\section{Background}
\label{sec:background}
In this section, we briefly describe issues, notations, and categories in domain adaptation and domain generalization. In addition, we also describe tasks and problems associated with medical image analysis and signal analysis.

\subsection{Domain Adaptation}
\begin{figure*}[]
\centerline{\includegraphics[scale=0.8]{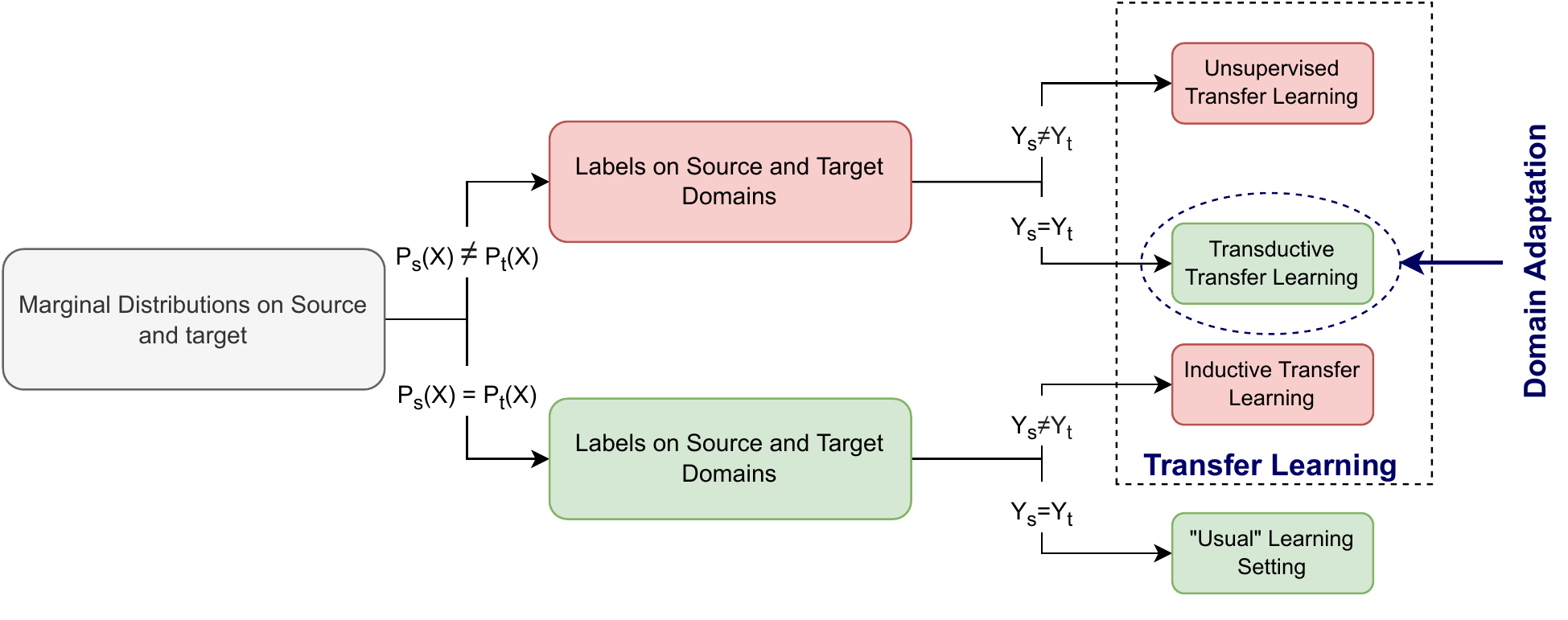}}
\caption{Different types of machine learning methods based on marginal distributions and tasks of source and target domains. Three out of four types are considered transfer learning methods. If the marginal distribution of source and target domains are not equal, but their tasks (label domain) are equal, it can be considered domain adaptation. \cite{bg_da_Pan_2010}}
\label{fig:bg_da_TL_categories}
\end{figure*}

With an increasing amount of massive data, deep learning models are being pushed to get the most accurate results in a wide range of fields and applications. However, a significant portion of the available data is unlabeled, and preparing and labeling proper data for deep neural networks is challenging and time-consuming \cite{bg_da_wu_2019}. Furthermore, in some fields, like medical image analysis, acquiring data is challenging, and tagging the data requires the collaboration of several experts. \\
In order to address this problem, deep models may be trained using labeled datasets and used directly on the target dataset for inference (called direct transfer). In spite of this, direct transfer cannot effectively transfer knowledge between datasets. It has been demonstrated by the \cite{bg_da_Zhao_2020} that direct transfer in digit recognition and semantic segmentation does not perform as well as traditional supervised learning methods. In direct transfer, the model's performance deteriorates due to the domain shift between the source and target datasets. \\
Alternatively, transfer learning can be used to resolve the problem. Transfer learning involves transferring a well-trained model from a dataset with large numbers of labeled samples to a target dataset with fewer or sparse labels. 
Figure \ref{fig:bg_da_TL_categories} illustrates the different subcategories of transfer learning.
Domain adaptation is referred to as a group of machine learning methods that can transform learned information from one or several fully labeled source datasets to a target dataset, defined on the same task, considering the existence of domain shift. As a result, DA methods are effective tools for learning general knowledge from the source domain(s) and transferring it to target domains.


Based on \cite{bg_da_Pan_2010}, DA is, in fact, a transfer learning method. In DA, the label space, or tasks, are mostly the same between source and target domains, but the marginal distributions of domains are different. Domain differences may exist between the source and target domains as well as between the source domains themselves.
\subsubsection{Notation}

To define the DA problem, we should explain the source and target domains. Suppose a shared space $\mathcal{X} \times \mathcal{Y}$ where $\mathcal{X}$ is the space of feature values and $\mathcal{Y}$ is the space of label values. A domain is a collection of paired data and labels sampled from a distribution. The samples of training or test data come from their corresponding domains. Assume that $D_s=\{(\mathbf{x}^{(i)}_s, y^{(i)}_s)\}^{N_s}_{i=1}$ is a domain where $(\mathbf{x}^{(i)}_s, y^{(i)}_s)$ is sampled from joint distribution $P_{X_sY_s}$ defined on $\mathcal{X} \times \mathcal{Y}$. \\
Consider there are $N$ source domains $S_i$, where $N \geq 1$, and one target domain $T$- note that in some scenarios, we can have more than one target domain, but for simplicity, we consider a single-target domain problem. Each source domain is denoted by $D_{s_i}=\{(\mathbf{x}^{(j)}_{s_i}, y^{(j)}_{s_i})\}^{N_{s_i}}_{j=1}$ where $(\mathbf{x}^{(j)}_{s_i}, y^{(j)}_{s_i})$ is drawn from the joint distribution $P_{X_{s_i} Y_{s_i}}$ on $\mathcal{X} \times \mathcal{Y}$. We consider the target domain denoted as $D_t = \{(\mathbf{x}^{(i)}_t, y^{(i)}_t)\}^{N_t}_{i=1}$, drawn from distribution $P_{X_tY_t}$ on $\mathcal{X} \times \mathcal{Y}$. For the task of unsupervised domain adaptation which is the focus of this study, we only have the unlabeled target domain $D^u_t = \{\mathbf{x}^{(i)}_t\}^{N_t}_{i=1}$.\\
In the DA problem, unlike other categories of transfer learning, the conditional distribution of source and target domain is the same, that is $P(Y_{s_i}|X_{s_i}) = P(Y_t|X_t)$, but the marginal distribution, $P_{X_{s_i}}$ for $i \in \{1, 2, ..., N\}$  and $P_{X_t}$ are different; in other words, $P(X_{s_i}) \neq P(X_t)$. This discrepancy is known as domain shift. Note that this discrepancy also exists among each pair of source domains: $P(X_{s_i}) \neq P(X_{s_j})$ for $i, j \in \{1, 2, ..., N\}$ and $i \neq j$. \\
The goal of domain adaptation is to reduce the negative effects caused by domain shifts between source and target domains. In other words, given $D_{DA} = \{D^u_t, D_{s_1}, D_{s_2}, ..., D_{s_N}\}$ a domain adaption algorithm $\mathcal{L}_{DA} : D_{DA} \rightarrow \mathcal{Y}^\mathcal{X}$ proposes a generalizable and robust function $f: \mathcal{X} \rightarrow \mathcal{Y}$ that gets the minimum prediction error on unseen samples which are drawn from the target domain:

$$ \min_{f} \mathbb{E}_{(x, y) \in D_t} \left [l\left( f (x), y \right) \right] $$

\subsection{Domain Generalization}
In the previous subsection, the issue of domain adaptation was discussed. As with domain adaptation techniques, domain generalization approaches attempt to address the challenging problem where test data distributions differ from training ones. Differentiating these two categories, DG cases have an unknown distribution of the test data while DA cases have a known distribution of the target data.
\subsubsection{Notation}
The same notation as the previous section applies to domain generalization. The only difference is that the target domain is unknown and can be drawn from an arbitrary distribution on $\mathcal{X} \times \mathcal{Y}$. So here the domain generalization algorithm $\mathcal{L}_{DG} : D_{DG} \rightarrow \mathcal{Y}^\mathcal{X}$ uses only the set of source domains $D_{DG} = \{D_{s_1}, D_{s_2}, ..., D_{s_N}\} = D_{DA} - \{D^u_t\}$ to estimate a generalizable robust function $f: \mathcal{X} \rightarrow \mathcal{Y}$ which minimizes the prediction error on any arbitrary target domain:
$$
\min_{f} \max_{D_t} \mathbb{E}_{(x, y) \in D_t} \left [l\left(f(x), y \right)\right]
$$

\subsection{Domain Adaptation and Domain Generalization Categories}
Based on several factors, scenarios, limitations, and algorithms, DG/DA methods can be categorized into independent groups. The following section discusses the most significant categories related to our issue, based on three different settings: the availability of labeled data, the number and distribution of source domains, and the distribution of label space:
\subsubsection{Labeled Data Availability}
According to the availability of labeled target data, we can have three classes of DG/DA methods:
\begin{itemize}
\item \textbf{Supervised DG/DA}: All data from the target domain have labels.
\item \textbf{Semi-supervised DG/DA}: A small number of target domain samples have labels.
\item \textbf{Unsupervised DG/DA}: There is no labeled data in the target domain.
\end{itemize}
This paper refers to DG/DA as \textbf{unsupervised} DG/DA.
\subsubsection{Number of Source Domains}
DG/DA methods can be divided into two varieties based on the number of source domains: single-source and multi-source.
\begin{itemize}

\item\textbf{Single-Source DG/DA}

The single-source DG/DA setting assumes that the training data are selected from a single distribution. This category includes data collected from one or more domains but does not require domain labels, so it can be applied to multi-source scenarios. Despite the fact that single-source techniques are generally less complicated than multi-source ideas, they may not be efficient when there are multiple sources from different domains available \cite{multi2020}.

\item\textbf{Multi-Source DG/DA}
Multi-source setting assumes that multiple distinct relevant domains are available. The corresponding methods are based on the assumption that labeled training data collected from multiple sources will have different distributions. As a trivial matter, it is possible to combine the sources into a single source and discard the differences between them. However, source-combined DG/DA often results in lower performance than simply using one or a collection of suitable sources and discarding the others, as this approach neglects the fundamental variations among multiple sources that affect the DG/DA algorithm.
In addition, by selecting a small subset of domains and learning the relationship between them, a model can learn patterns that are more stable across source domains. This helps it generalize more effectively to unknown domains. Domain shifts among different sources can be addressed in a variety of ways; for example, latent space transformation can be used to align and transform source domains and target domains into latent spaces. An alternative approach to handling inter-source domain shifts is to align each source domain separately with the target domain, and then aggregate the adapted sources.\cite{multi2020}
\end{itemize}
\subsubsection{Label Distribution}
\begin{figure}[]
\centerline{\includegraphics[scale=0.45]{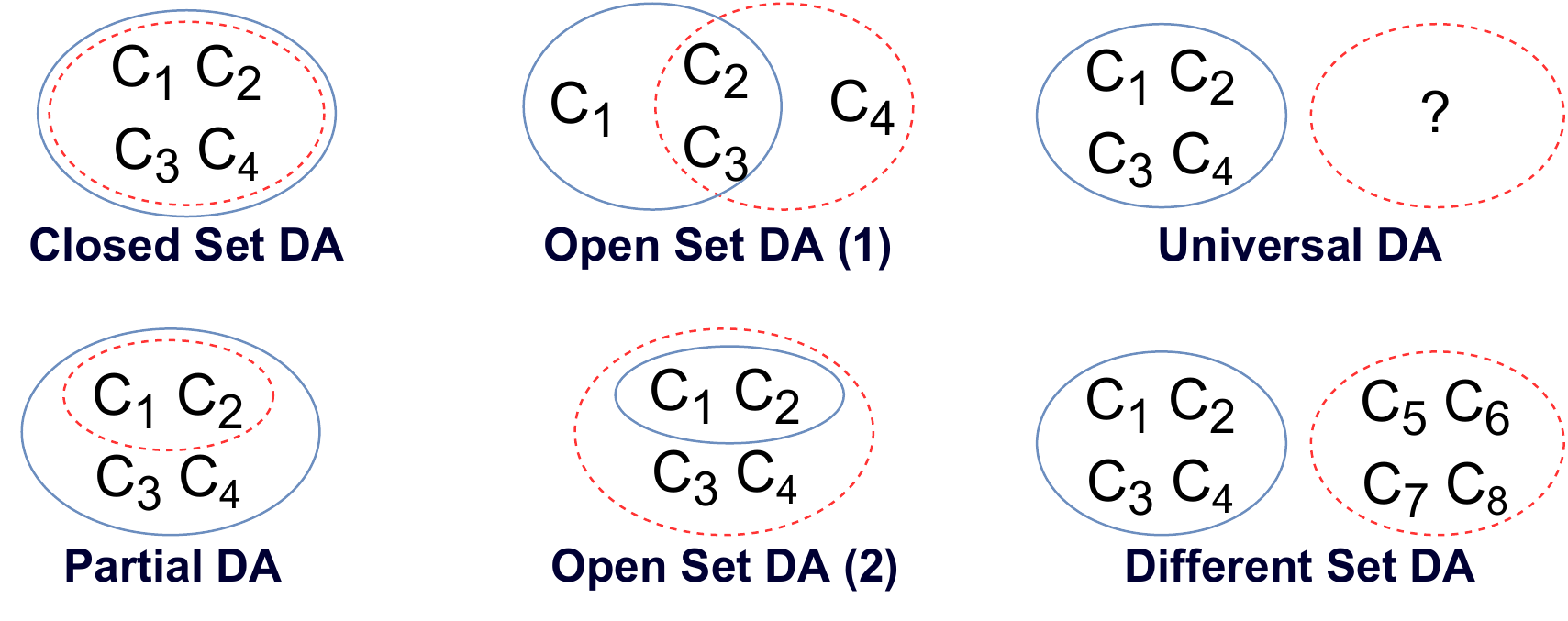}}
\caption{Different DA scenarios based on label distributions. The straight blue line represents the source domain label set, and the dotted red line represents the target domain label set. \cite{bg_da_He_2020}}
\label{fig:bg_da_DA_label_dists}
\end{figure}

Based on \cite{bg_da_He_2020}, labels of source and target domains can consist of the same or different classes. As illustrated in Figure \ref{fig:bg_da_DA_label_dists}, the variability of this difference creates several scenarios for DG/DA methods:
\begin{itemize}
\item \textbf{Closed set DG/DA}: Labels of source and target domains come from the same classes,
\item \textbf{Partial DG/DA}: Target domain's classes are a subset of source domains' classes,
\item \textbf{Open set DG/DA}: defined in two different ways in the literature:
\begin{enumerate}
\item Source and target domains have common classes. However, each of them can have other classes \cite{bg_da_Busto_2017}.
\item Source domains' label set is a subset of the target domain's label set \cite{bg_da_Saito_2018}.
\end{enumerate}
\item \textbf{Universal DG/DA}: Target domain's label set is unknown and might have a number of common classes with the source domains' label set \cite{bg_da_You_2019}.
\item \textbf{Different set DG/DA}: Source and target domains have completely different classes.
\end{itemize}

\subsection{Medical Images and Signals Analysis}
The automatic analysis of medical images and signals involves the study of medical measurements of different body parts, whether in the form of images or electromagnetic signals. This is done in order to provide an intelligent diagnosis of various medical conditions.
Due to the development of systems based on Artificial Intelligence (AI) and the rapid increase in computational power, it has become more and more common to process high-resolution medical images intelligently. By using ML systems trained on large datasets of medical recordings, the analysis of medical images can be made faster and more accurate. These models can assist physicians when they have doubts about their diagnosis or miss a critical clue in the recordings. 

There are two types of medical images and signals: structural and functional. Structural medical images are the ones that only record the state of the body in a single unit of time. They focus on the spatial structure of the body part in the form of an image. They include Computed Tomography (CT), Magnetic Resonance Imaging (MRI), Pathology, Endoscopy, Colonoscopy, Automated Breast Volume Scan (ABVS), Gastroscopy, Cytology, and X-Ray images. Functional medical images and signals contain spatial and temporal information, capturing measurements of body processes rather than body states. The majority of these measures are captured from the brain, such as EEG, fMRI, MEG, and fNIRS. Other functional medical signals are not measured from the brain, such as ECG (from the heart) and EOG (from the eyes). Structural medical images have been used to diagnose cancers and abnormalities in body organs, including brain tumors.
(\cite{bg_med_havaei_brain_2017}, \cite{bg_med_ottom_znet:_2022}, \cite{bg_med_jiang_novel_2021}, \cite{bg_med_khan_brain_2021}, \cite{bg_med_ullah_cascade_2022}, \cite{bg_med_jiang_swinbts:_2022}),   breast tumors (\cite{bg_med_breast_han_deep_2017}, \cite{bg_med_breast_laar_brainbrush_2013}, \cite{bg_med_breast_nahid_histopathological_2018}, \cite{bg_med_breast_rasti_breast_2017}, \cite{bg_med_breast_samala_evolutionary_2018}), lung, liver and kidney diseases (\cite{bg_med_abd_ben-cohen_fully_2018},
\cite{bg_med_abd_causey_ensemble_2021}, \cite{bg_med_abd_hou_triple-stage_2020},
\cite{bg_med_abd_lung_2016},
\cite{bg_med_abd_segmentation_2017},
\cite{bg_med_abd_wei_two-phase_2019},
\cite{bg_med_abd_wu_segmentation_2022}). \\
This review focuses on \textbf{functional} medical images and signals. \\
Although not as popular, recordings from non-brain areas of the body have also been studied in the literature. For example, ECG signals are analyzed to detect heart problems (\cite{bg_med_heart_kiranyaz_real-time_2016}, \cite{bg_med_heart_kutlu_multistage_2016}, \cite{bg_med_heart_lih_comprehensive_2020}, \cite{bg_med_heart_liu_multiple-feature-branch_2018}, \cite{bg_med_heart_tan_application_2018}), or EOG signals. \\
Brain-Computer-Interface (BCI) models are the systems that analyze and use brain-related medical images, which constitute the majority of models based on functional medical images. We have categorized different tasks in brain signal analysis as follows:
\subsubsection{Motor Imagery (MI)}
One of the most important applications of BCI systems is the rehabilitation or diagnosis of patients through Motor Imagery (MI), a mental process in which the patient imagines moving a part of their body(\cite{bg_med_mi_amin_deep_2019},
\cite{bg_med_mi_hassanpour_novel_2019},
\cite{bg_med_mi_li_channel-projection_2019},
\cite{bg_med_mi_ortiz-echeverri_new_2019},
\cite{bg_med_mi_zhang_making_2020},
\cite{bg_med_mi_zhang_novel_2019}).
\subsubsection{Bain-related Disease Diagnosis (BDD)}
Another application of BCI systems is the detection of
brain-related diseases and conditions such as Parkinson's Disease (\cite{bg_med_parkinson_arasteh_deep_2021},
\cite{bg_med_parkinson_identification_2022},
\cite{bg_med_parkinson_lee_convolutional-recurrent_2021},
\cite{bg_med_parkinson_oh_deep_2020}), Alzheimer (\cite{bg_med_alzheimer_bi_random_2018},
\cite{bg_med_alzheimer_ieracitano_convolutional_2019},
\cite{bg_med_alzheimer_nakamura_electromagnetic_2018},
\cite{bg_med_alzheimer_sheng_novel_2019}),  Schizophrenia (\cite{bg_med_schizophrenia_goshvarpour_schizophrenia_2020},
\cite{bg_med_schizophrenia_phang_multi-domain_2020},
\cite{bg_med_schizophrenia_santos-mayo_computer-aided_2017},
\cite{bg_med_schizophrenia_shim_machine-learning-based_2016}), and Autism Spectrum Disorder (ASD) (\cite{bg_med_asd_baygin_automated_2021},
\cite{bg_med_asd_bi_classification_2018},
\cite{bg_med_asd_ranjani_classifying_2021},
\cite{bg_med_asd_tawhid_spectrogram_2021}). Effective and precise detection of such diseases can be very helpful for early detection and diagnosis of them, leading to more effective treatments.
\subsubsection{Emotion Recognition (ER)}
Affective computing, mostly represented by emotion recognition, that is, predicting a person's emotional state and utilizing it as a part of an intelligent system, is also one of the important applications of BCI systems (\cite{bg_med_emotion_islam_eeg_2021},
\cite{bg_med_abd_causey_ensemble_2021},
\cite{bg_med_emotion_liu_multi-channel_2020},
\cite{bg_med_emotion_luo_data_2020},
\cite{bg_med_emotion_xiao_4d_2022},
\cite{bg_med_emotion_yin_eeg_2021},
\cite{bg_med_emotion_zhong_eeg-based_2022}).
\subsubsection{Seizure Analysis (SA) and Mental State Diagnosis (MSD)}
Monitoring people's mental state also has many applications such as seizure and epilepsy prediction and detection (\cite{bg_med_seizure_dissanayake_geometric_2022},
\cite{bg_med_seizure_gao_pediatric_2022},
\cite{bg_med_seizure_jana_deep_2021},
\cite{bg_med_seizure_li_eeg-based_2022},
\cite{bg_med_seizure_priya_prathaban_dynamic_2021},
\cite{bg_med_seizure_ra_novel_2021},
\cite{bg_med_seizure_yang_effective_2021},
\cite{bg_med_seizure_zhang_lightweight_2021}). Other popular applications
include mental workload classification and assessment, mental state prediction, and diagnosis of some mental diseases such as Tinnitus.
\subsubsection{Awareness Monitoring (AM)}
One aspect of awareness monitoring is driver awareness validation, fatigue, and drowsiness detection (\cite{bg_med_fatigue_barua_automatic_2019},
\cite{bg_med_fatigue_wang_analysis_2018},
\cite{bg_med_fatigue_xiong_classifying_2016},
\cite{bg_med_fatigue_zhang_design_2017}). Awareness estimation is also referred to as vigilance estimation in the literature (\cite{bg_med_vigilance_cheng_vigilancenet:_2022}, \cite{bg_med_vigilance_luo_wasserstein-distance-based_2021}, \cite{bg_med_vigilance_song_deep_2021},
\cite{bg_med_vigilance_wang_vigilance_2021}).
\subsubsection{Sleep Diagnosis (SD)}
Sleep diagnosis based on brain signals is a popular study that utilizes measures signals such as EEG from the brain and tries to recognize different sleep stages or events, or detect sleep disorders (\cite{bg_med_sleep_fu_deep_2022},
\cite{bg_med_sleep_phan_sleeptransformer:_2022},
\cite{bg_med_sleep_sudhakar_sleep_2021},
\cite{bg_med_sleep_wang_bi_2022}).
\subsubsection{Visual Perception Analysis (VPA)}
The analysis of human visual imagination and understanding of the surroundings using brain signals is an interesting field of research in brain signal analysis. It includes the study of Steady-State Visual-Evoked Potentials (SSVEP) (\cite{bg_med_ssvep_bhuvaneshwari_bio-inspired_2023}, \cite{bg_med_ssvep_karunasena_single-channel_2021}, \cite{bg_med_ssvep_liu_align_2022}, \cite{bg_med_ssvep_sun_cross-subject_2022}, \cite{bg_med_ssvep_tabanfar_subject-independent_2023}, \cite{bg_med_ssvep_yan_cross-subject_2022}, \cite{bg_med_ssvep_zhu_eegnet_2021}) and visual recognition (\cite{bg_med_vpa_bagchi_adequately_2021}, \cite{bg_med_vpa_bagchi_eeg-convtransformer_2022}, \cite{bg_med_vpa_kumari_automated_2022}, \cite{bg_med_vpa_lee_inter-subject_2022}, \cite{bg_med_vpa_zheng_attention-based_2021}).
\subsubsection{Human Thought Analysis (HTA)}
There are also interesting ongoing studies based on human thought prediction, such as creative drawing (\cite{bg_med_thought_laar_brainbrush_2013}) and imagined speech recognition (\cite{bg_med_imagined_speech_agarwal_electroencephalographybased_2022}, \cite{bg_med_imagined_speech_kumar_deep_2021}, \cite{bg_med_imagined_speech_li_decoding_2021}). However, these areas of study are still at a very primitive stage.

\begin{figure}[]
\centerline{\includegraphics[width=0.5\textwidth]{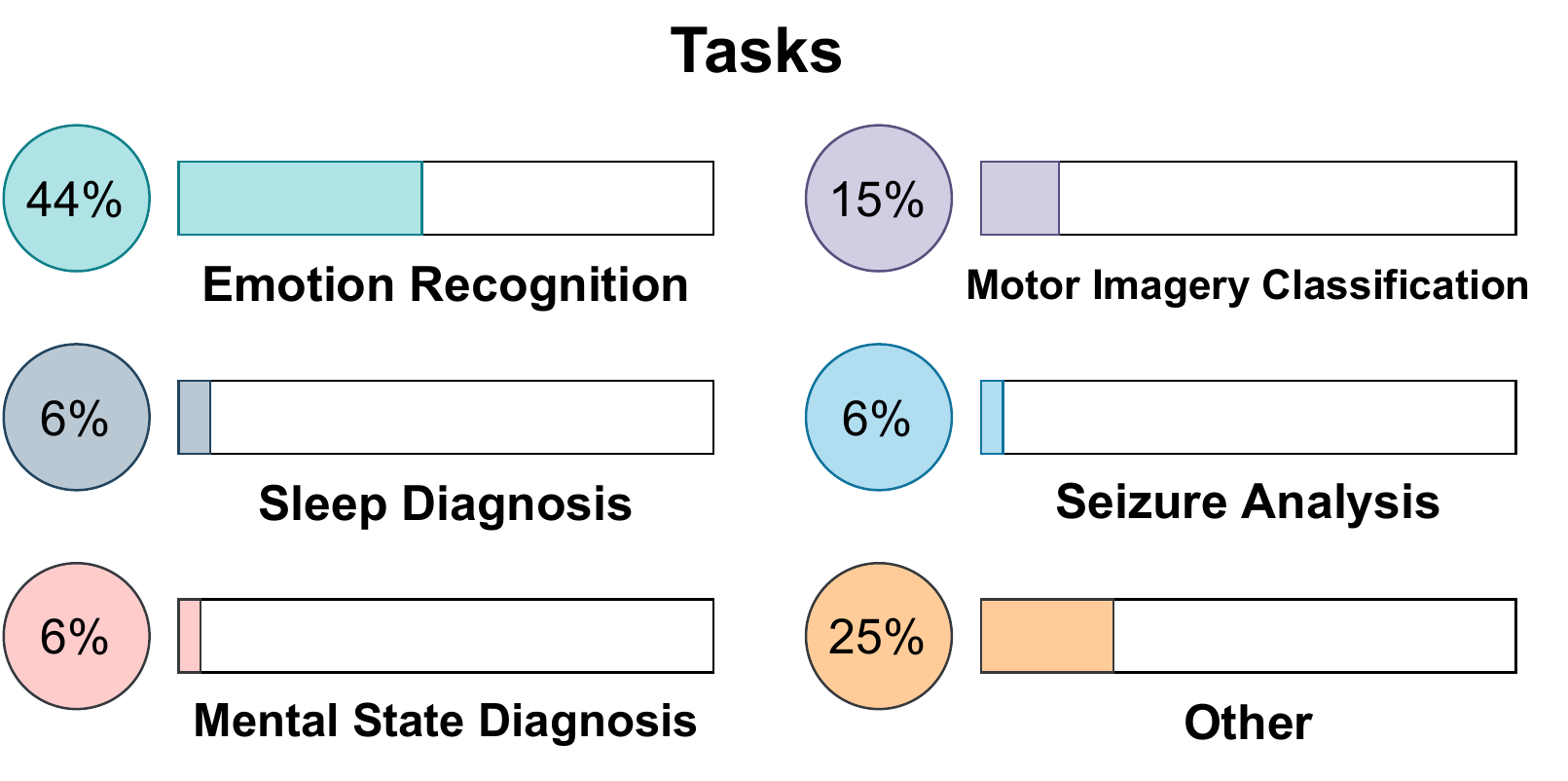}}
\caption{Distributions of recent DG and DA studies on EEG modality based on tasks}
\label{fig:EEG_tasks_dist}
\end{figure}

\begin{figure}[]
\centerline{\includegraphics[width=0.5\textwidth]{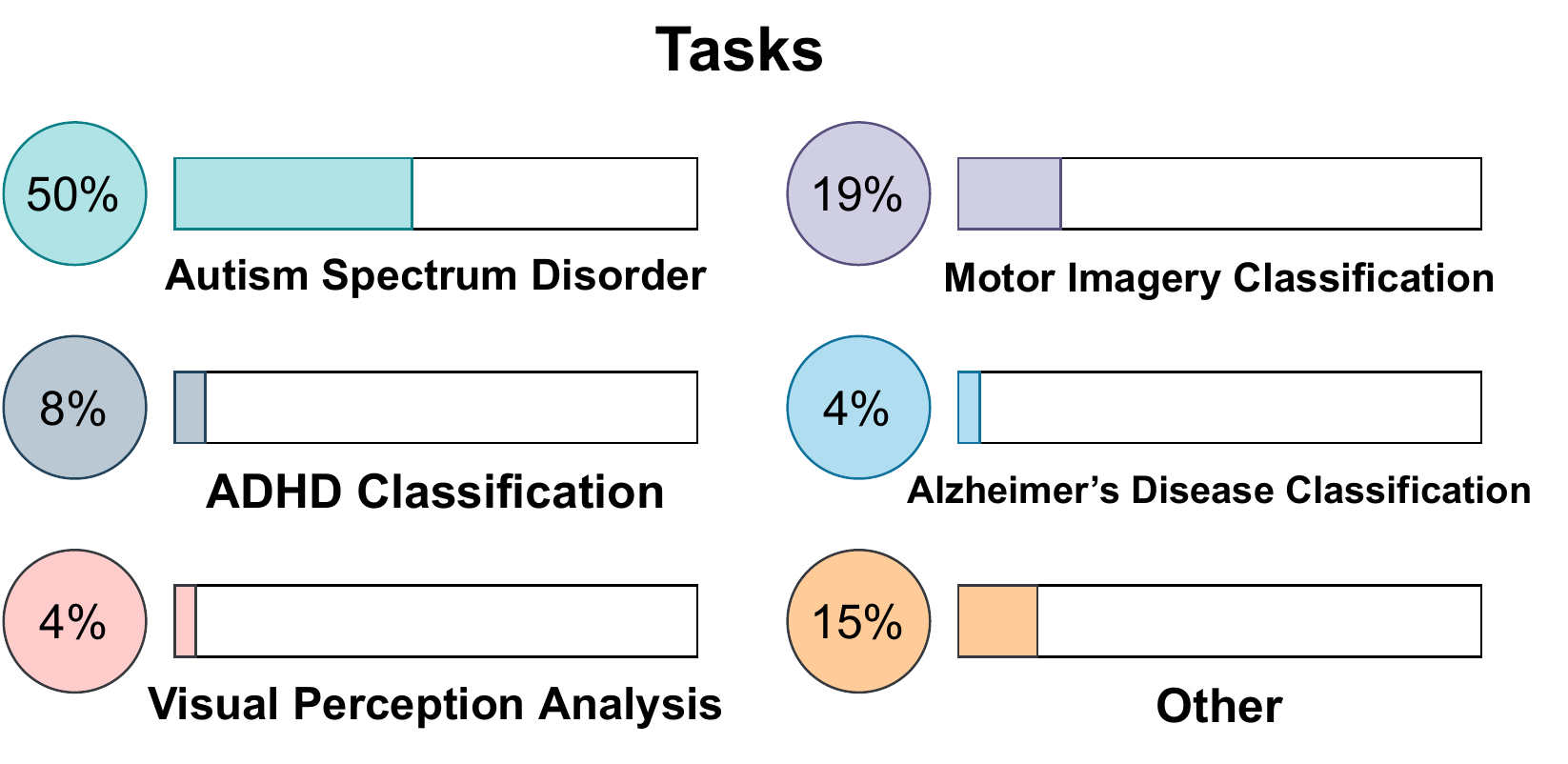}}
\caption{Distributions of recent DG and DA studies on fMRI modality based on tasks}
\label{fig:fMRI_tasks_dist}
\end{figure}

It is worth mentioning that the distribution of the most frequently used tasks in recent DG/DA-related EEG and fMRI papers is shown in Figure \ref{fig:EEG_tasks_dist} and Figure \ref{fig:fMRI_tasks_dist} respectively.


\section{Domain Adaptation and Generalization for Medical Image Analysis}
\label{sec:problem_Definition}

Machine Learning techniques in medical field analysis usually suffer from domain shift. This problem can arise from various factors, such as different centers, different devices in the same center, different subject populations, or different experimental conditions. Moreover, gathering substantial medical data can be very time-consuming and expensive. Some medical signals require costly measurement devices (e.g., MEG), while others need a meticulous and stable experimentation setup (e.g., EEG). Hence, gathering a reasonable amount of data on every new site or for every new subject is not affordable in many cases. Thus, the challenge of domain shift is unavoidable in the case of medical image analysis.

 Due to the direct impact of this problem on some crucial medical diagnoses, the use of domain adaptation in the medical field is undisputed. Additionally, there is a common issue in most of the ML models that they should have high performance on newly gathered data. This will be more relevant in the medical field because of the higher frequency of facing data from a new domain, such as a new patient. Therefore, domain generalization is also of great importance in this field.

 \subsection{DG and DA Tasks in Medical Image Analysis}
 Different adaptation or generalization tasks can be defined between different types of domains, such as subjects, datasets, sessions, etc. The cross-subject task is the most common in DA or DG on medical data, which considers the variability of data across subjects and tries to eliminate shifts between different subjects. Cross-dataset is another common DG/DA task on the medical data related to existing domain shifts between datasets. This task aims to learn various aspects of these differences across medical image datasets. Cross-session task is also frequent in medical image analysis and is defined when the goal of DA or DG is to consider intra-subject data variabilities that emerged during different experimental circumstances. Some other DA or DG tasks are less common than those mentioned above; for instance, the cross-day task is analogous to the cross-session task. Also, it is worth noting that the cross-device task may also be studied, which considers the data variability caused by different devices used to measure the subject’s signals.

\begin{figure}[]
\centerline{\includegraphics[width=0.5\textwidth]{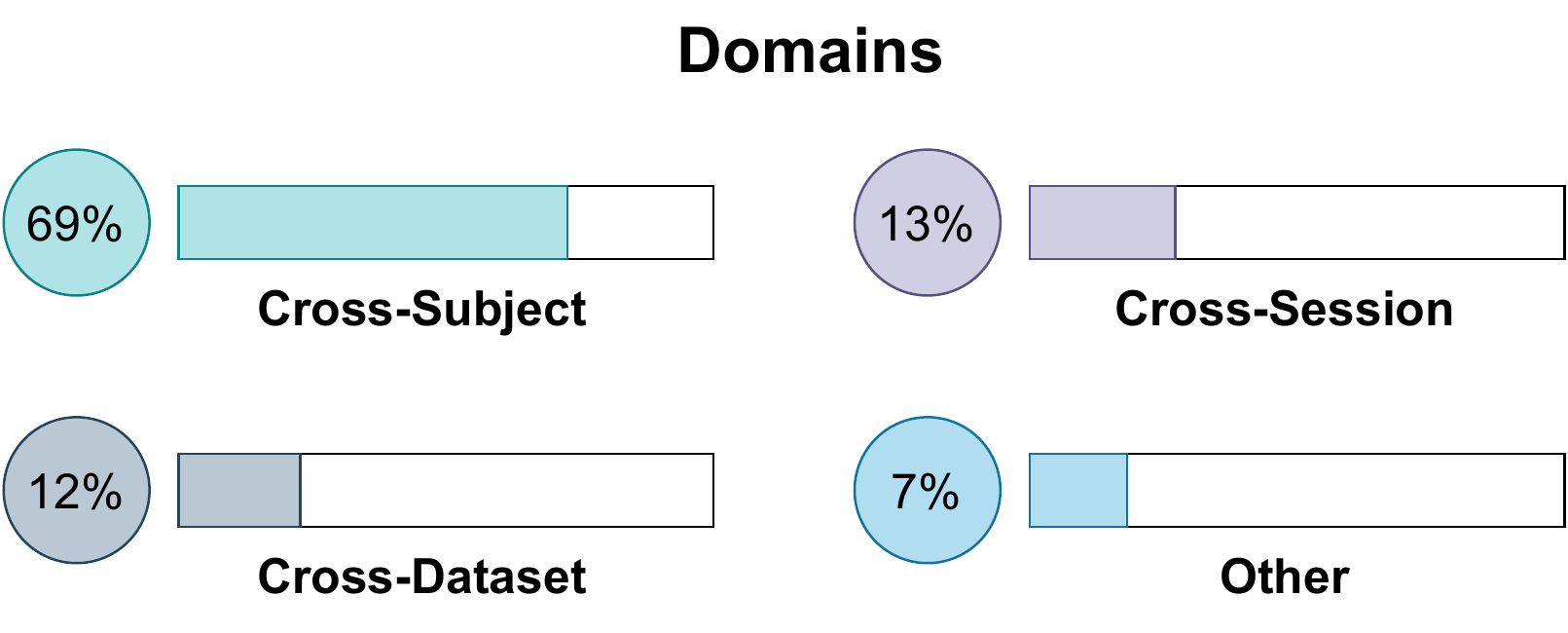}}
\caption{Distributions of recent DG and DA studies on EEG modality based on domains}
\label{fig:EEG_domains_dist}
\end{figure}

\begin{figure}[]
\centerline{\includegraphics[width=0.5\textwidth]{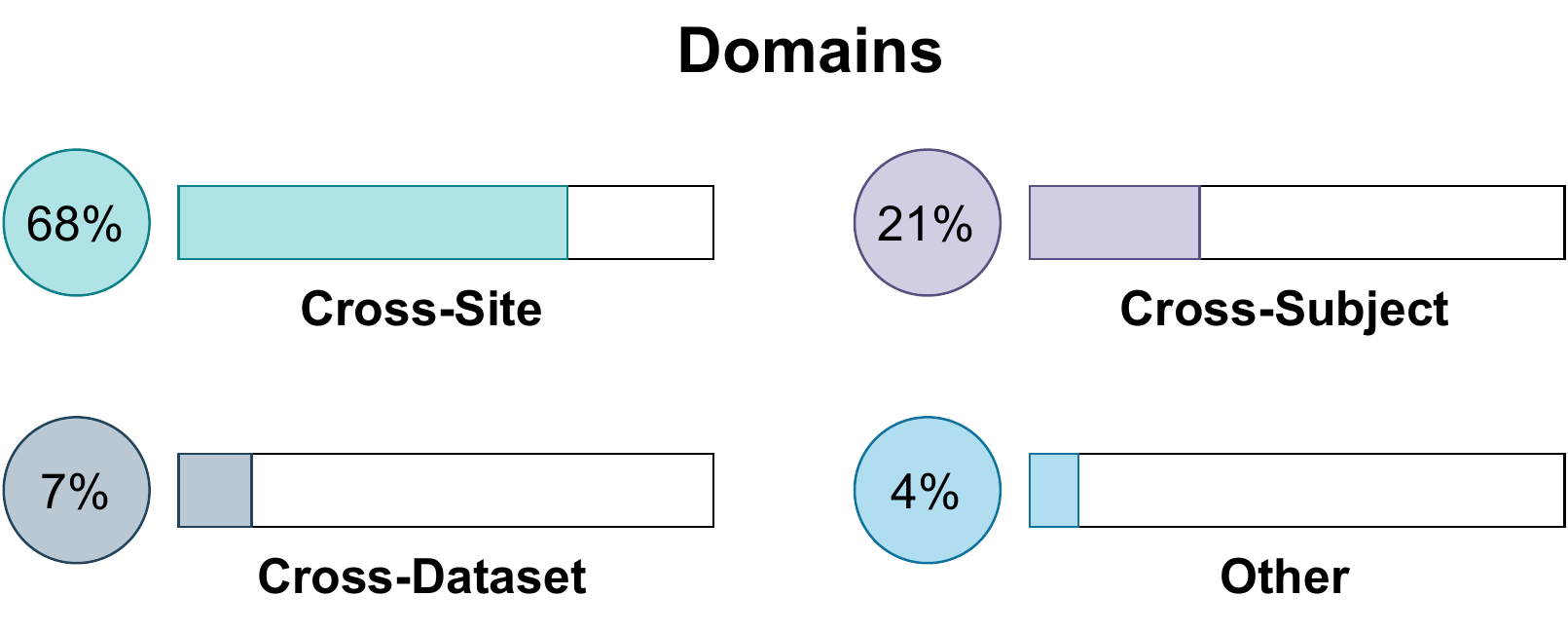}}
\caption{Distributions of recent DG and DA studies on fMRI modality based on domains}
\label{fig:fMRI_domains_dist}
\end{figure}

 Furthermore, in Figures \ref{fig:EEG_domains_dist} and \ref{fig:fMRI_domains_dist}, we show the distribution of the most frequently used medical domains in recent DG/DA-related EEG and fMRI studies respectively.

 \subsection{DG vs. DA in Medical Image Analysis}
 There are fundamental differences between domain adaptation and domain generalization, which cause different applications. As mentioned before, good performance for unseen medical data is almost vital; it is very time-consuming to learn a different model for a new subject or patient. Nevertheless, generalization is not always the desired function in the medical field; sometimes, we face specific domains, such as data from the same organ acquired by different devices or from different subjects. In these situations, what is most needed is the minimization of domain shift between these related but different domains, so domain adaptation is the key. Sometimes in adaptation, there are a few seen target data; sometimes, all the target data is unseen. The goal is to use this knowledge to learn about the target domain. To conclude, the main difference between adaptation and generalization is the access to target data during the training process; in other words, in adaptation, we aim to complete our diagnosis for a related domain, and to that end, we take advantage of our current knowledge on source data, and the structure of target data. In contrast, in generalization, we are only allowed to use our knowledge of accessed source data and expand it for an unseen domain of data whose structure is even unknown.
 
 Generalization of the model is usually an essential desire of researchers, especially in medical image analysis. As mentioned, the better our model performs on new and unseen data, the stronger it is and will become more valuable and practical. So the importance of generalization is inevitable even when the main task is adaptation. To handle this, some papers with the main task of adaptation exploit generalization ideas as well. Therefore, we will explain both their adaptation task and the different ideas used for generalization.

\section{Architectures}
\label{sec:architectures}
\begin{figure*}[t]
\centerline{\includegraphics[width=\textwidth]{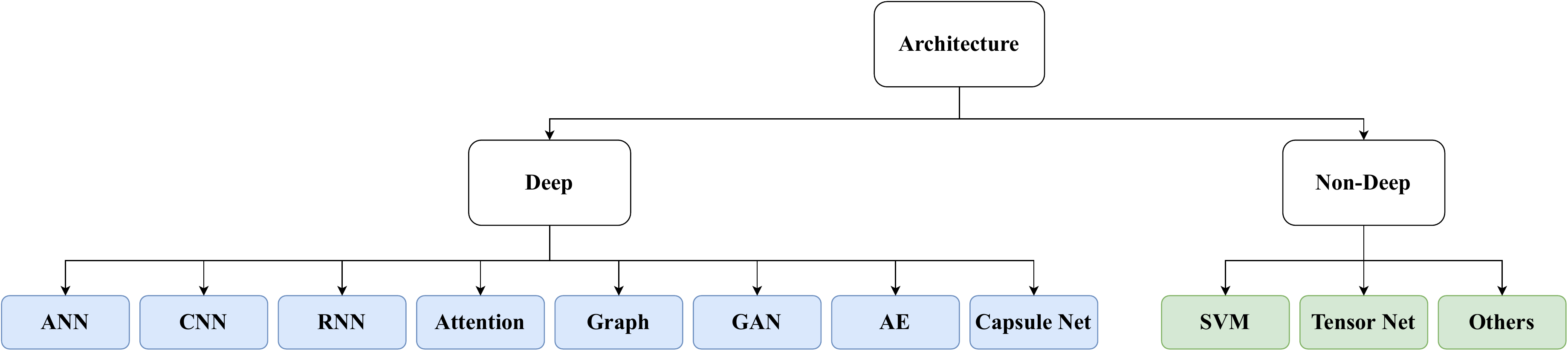}}
\caption{Hierarchy of common machine learning and deep learning architectures in functional medical imaging}
\label{fig:architectures_hierarchy}
\end{figure*}
In this study, we have thoroughly investigated the most commonly used architectures in the works on DG/DA of functional medical data. These architectures are categorized and illustrated in Figure \ref{fig:architectures_hierarchy} as follows:

\textbf{Deep Architectures}: This category includes most of the recent works in the literature.
   \begin{itemize}
       \item \textbf{Artificial Neural Network (ANN)}: Some methods are implemented as simple (feed-forward) neural networks.
       \item \textbf{Convolutional Neural Network (CNN)}: The majority of the methods exploit CNNs for processing multi-dimensional functional medical data.
       \item \textbf{Recurrent Neural Network (RNN)}: These architectures enable the use of temporal dependencies in the data sequences.
       \item \textbf{Attention-based Models}: Attention ideas are sometimes performed to introduce weights for different channels or dimensions (and sometimes timesteps) in the data
       \item \textbf{Graph-based Architectues}: Using graphs for modeling data helps model interactions among different devices or electrodes, especially in EEG data.
       \item \textbf{Generative Adversarial Network (GAN)}: GANs may also be used for augmenting data in medical use cases.
       \item \textbf{Auto-encoder (AE)}: Some methods use the latent representations of the medical data extracted by an auto-encoder in an unsupervised manner.
       \item \textbf{Capsule Networks \cite{sabour2017dynamic}}: The recently-introduced architectures of capsule networks are implemented in recent models.
   \end{itemize}
\textbf{Non-Deep Architectures}: Some non-deep models are still widely used in the proposed models for DG/DA on functional medical data.
   \begin{itemize}
       \item \textbf{Support Vector Machine (SVM)}: SVM is one of the most famous and broadly adopted non-deep models in this era. 
       \item \textbf{Tensor Network}: Some works adopt tensor networks that have the advantage of graph-based data modeling.
       \item \textbf{Others}: There are other non-deep models, such as KNNs, random forests, etc., also used in the literature.
   \end{itemize}
The architectures used in each studied model are depicted in Table \ref{eeg-papers-table} and Table \ref{fmri-papers-table} for EEG and fMRI data, respectively.


\section{Methods}
\label{sec:methods}

The most recent methods used for the adaptation and generalization of tasks on functional medical images are categorized and introduced in this section. These methods are explained based on two perspectives, DA and DG, in the following.
\subsection{DA Approaches}

\begin{figure*}[t]
\centerline{\includegraphics[width=\textwidth]{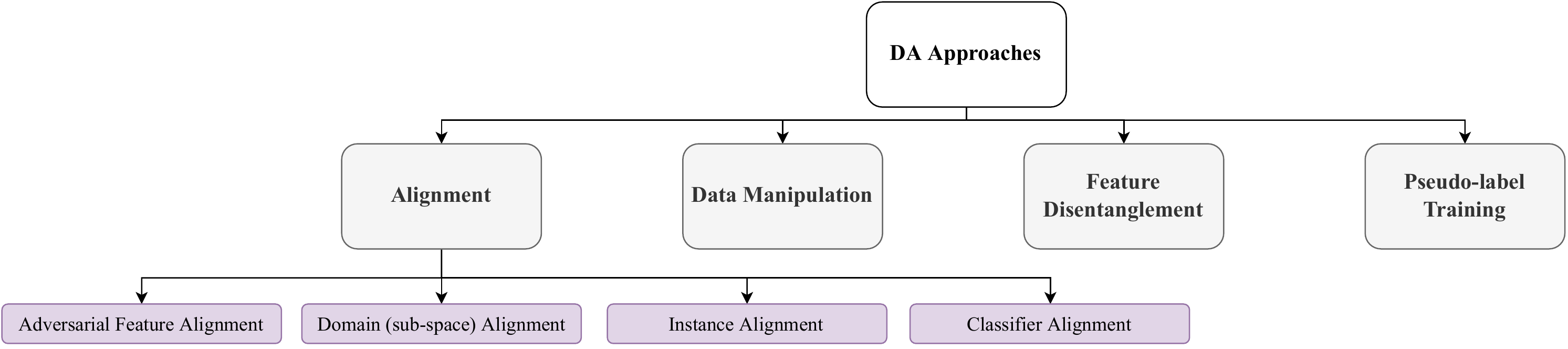}}
\caption{Hierarchy of Domain Adaptation approaches in functional medical imaging}
\label{fig:da_approaches_hierarchy}
\end{figure*}

We have studied the latest research seeking domain adaptation in the context of functional medical images and based on their design ideas, these methods are classified as in the hierarchy depicted in Figure \ref{fig:da_approaches_hierarchy}, including alignment, data manipulation, feature disentanglement, and pseudo-label training. Summarized information about the methods discussed in this section can be found in Table \ref{eeg-papers-table}. In the following parts of this section, these approaches and works following their ideas are described.
\subsubsection{Alignment}
One of the most common domain adaptation strategies arises from aligning the model's input at test time with previously seen data or features. A majority of approaches rely on these techniques so that the inputs (or secondary features) to the model are kept aligned with a fixed network architecture. Consequently, the same architecture can yield relatively similar performance for source and target data. Alignment-based methods consist of adversarial alignment (alignment using an adversarial objective), domain alignment (aligning the distribution of target and source data), instance alignment (aligning source and target sample by sample), and classifier alignment (adapting the classifier model to the target domain).
\subsubsubsection{Adversarial Feature Alignment}
This approach is implemented in a substantial number of papers focusing on aligning features between source and target domains. The objective of these methods is to extract features that are similar between target and source data using an adversarial training setup. Inspired by the Domain Adversarial Neural Network (DANN) \cite{ganin2016domain}, in most of them, a common feature encoder is trained in a min-max game with a domain classifier. Essentially, the feature encoder learns to extract features such that the domain classifier is unable to distinguish between source and target data. This procedure results in achieving a common feature space between data from different domains.

In \cite{Paper_Zhao_26} and \cite{Paper_Lebedeva_RP44}, the idea of DANN is applied by training a domain classifier whose loss is inverted by a Gradient Reversal Layer (GRL) \cite{ganin2015unsupervised} and forcing the feature extractor to remove domain-variant features while improving the classification accuracy of the main task. In \cite{Paper_Su_fMRI_2}, Su et al. employ an adversarial discriminator that is trained to be challenged by a pre-trained feature extraction for brain anomaly detection on fMRI data. Heremans et al. adopt an akin approach in \cite{Paper_Heremans_RP43} to enhance the performance of common neural networks used for sleep stage classification by using an adversarial domain classifier on the feature extraction backbone. In \cite{Paper_Zhao_92}, multi-view features are extracted in the time and frequency domain and then, combined with the original data, are used in an adversarial learning module with two generators for separating patient and seizure features alongside discriminators ensuring this separation. Also, in \cite{Paper_He_20}, an Adversarial Discriminative Temporal Convolutional Network (AD-TCN) is proposed, where initially, an encoder and classification layer are trained on the source data. Secondly, the adversarial loss is employed via a domain classifier applied to the source encoded features and a distinct target encoder, making the target encoder able to be combined with this new classifier for target inference. In \cite{Paper_Liu_48}, in one branch, features obtained from a pure-info encoder are fed into a classifier and an adversarial side discriminator so that data from the two ears are aligned and processed efficiently together in the classifier. In another branch, after applying a domain-variance encoder, the resulting features plus the ones from the first branch are combined to reconstruct the data, where a domain discriminator is further adversarially trained. \cite{Paper_Wang_49} proposes adversarial adaptation in multi-source setup by first selecting the source samples most correlated with the target sample, and then mapping their corresponding features in a common space, with the aid of a discriminator intended to not be able to differentiate domains. In \cite{Paper_Pominova_fMRI_15}, the Fader network method \cite{pominova2021fader} is used for domain adaptation and removing task-irrelevant features in fMRI data. In this method, an auto-encoder is utilized whose output encoding is used for the final classification task, as well as the domain classification in an adversarial manner. Furthermore, Li et al. use an adversarial subject classifier in \cite{Paper_Li_55} to ensure the subject independence of the extracted features for emotion recognition. Moreover, two different RNNs are also employed, for the right and left brain hemispheres, in each of the two vertical and horizontal streams over the electrodes to maintain structural information. Eldele et al. \cite{Paper_Eldele_78} utilize adversarial training along with self-attention and self-training in their method where the extracted features are passed through unshared attention-based modules to retain domain-specific features as well as task-related ones, as domain-specific features may also be helpful for label prediction. The adversarial domain classifier further encourages data alignment while preserving domain-related features. In \cite{Paper_Bao_91}, Bao et al., in addition to Maximum Mean Discrepancy (MMD) \cite{gretton2012kernel} minimization, further use a domain classifier so that it fails to separate source and target domains, as they mention that merely MMD will not guarantee multi-source DA. In \cite{Paper_Avramidis_46}, separate branches are designed for extracting features from EEG data and the music used for data collection. The representation from these two branches is further aligned by a modality classifier, inserted with a GRL, enabling adversarial alignment among modalities as in DANN. \cite{Paper_Rayatdoost_23} also introduces a domain and a subject classifier implemented with GRL layers, that perform domain and subject classification and are trained adversarially so that features extracted from topological maps obtained from Power Spectral Density (PSD) features are free of dataset or subject priors.

Multiple works have further added modifications to DANN to make it more applicable for their purpose. Ding et al. extend DANN in \cite{Paper_Ding_51} by designing two label predictors instead of one. Using pre-trained label predictors on the source data, the two fully connected classifiers are tuned by Maximum Classifier Discrepancy (MCD) criterion \cite{saito2018maximum}. First, target outliers are detected and MCD is maximized to achieve broader classification boundaries. After relocating target samples to new boundaries, MCD is minimized for better adaptation. \cite{Paper_Tang_63} extends the concept of DANN by applying the domain discriminator to the conditional features, i.e. multiplying the features by the softmax output, thereby capturing a 2-D matrix (outer product) that can be inputted into a GRL and a domain discriminator. Likewise, in \cite{Paper_Huang_fMRI_25}, Huang et al. explore the same idea, combined with an entropy-based regularizer that adjusts the informativeness of different samples, on fMRI data for ADHD classification. A similar approach is taken in \cite{Paper_Hong_RP38}, where both global (marginal) and local (conditional) domain discriminators are adversarially trained against the main classifier, with a dynamic weight $\omega$ adjusting their importance. Some works also perform DANN on the shallower representation of the data. In \cite{Paper_Cai_RP26}, features from shallow layers are used for domain discrimination, which is trained adversarially to align the marginal distributions, while deeper features are fed into two different classifiers whose prediction difference is aimed to be maximized to detect target samples close to the decision boundary. Similarly, \cite{Paper_Li_30} benefits from an adversarial adaptation by feeding shallower representations to a domain discriminator, as earlier layers typically produce more task-invariant features that reflect the difference in data domains. Additionally, the association strategy computes the probability of transition between source and target domain based on their features in each batch and introduces a loss in these transitions to encourage them to return to the same class. \cite{Paper_Ye_85} integrates the idea of a DANN with an attention mechanism. A graph convolutional neural network (GCNN) is exploited with numerous stacked CNN layers, creating multi-level features from the GCNN and CNNs. The concatenated representations are inputs to separate adversarial domain classifiers, which help extract more domain-invariant features. For the final label predictor, the feature regions are multiplied by attention weights indicating how difficult it was for the classifier to classify the domains in each region. \cite{Paper_Zeng_RP11} combines a GAN with a DANN to seek domain adaptation. First, a GAN is trained to achieve a robust target data generator and an accurate target data discriminator. Then, the closest source samples to the target distribution, as specified by the discriminator, are further augmented with fake target data and are used in the DANN to adversarially train the final fatigue prediction network against a domain classifier. In \cite{Paper_Li_fMRI_19}, the idea of the adversarial domain discriminator is integrated into a federated learning framework that has pre-trained site-specific feature generators that are further trained to confuse the discriminator.

In some cases, a min-max game is performed to separate features in the data. Zhu et al. adversarially train two classifiers on features from an auto-encoder in \cite{Paper_Zhu_69}. After training the auto-encoder and classifiers, their prediction discrepancy is maximized. Following that, in a min-max game, the auto-encoder is optimized to decrease this discrepancy. In \cite{Paper_Jeon_39}, Jeon et al. design a common point-wise convolutional encoder producing class-relevant and class-irrelevant features and a network estimating the mutual information of these two features that are optimized in a min-max manner to guarantee the omission of subject-specific features from the input of the classification network.

An adversarial scheme may also be used to fit the data to certain priors or prototypes. In \cite{Paper_Peng_9}, a Manifold Adversarial Auto-Encoder (MAAE) is developed to fit a manifold prior distribution to the distribution of the auto-encoder latent space. \cite{Paper_Peng_82} also discards the data specific to patients by presuming a Laplace prior distribution on different patients and considering them as real data. Inspired by GANs, the VAE outputs are regarded as fake data and are fed alongside the real data to a discriminator, aiming to deceive the discriminator. In \cite{Paper_Wang_87}, Wang et al. create source and target prototypes and classify samples based on distance from these prototypes using a domain classifier trained in a min-max game with the generator (the symmetric and positive definite matrix network applied to the data covariance matrix).

Some works consider private encoders per domain in adversarial domain alignment. In \cite{Paper_Luo_53}, Luo et al. propose two variants of Wasserstein-distance-based Multi-source Adversarial Domain Adaptation (wMADA) for domain adaptation in vigilance estimation and emotion recognition. The first variant, wMADA-$\alpha$, adversarially trains $k$ different private discriminators on the Wasserstein distance between source and target outputs. In the second variant, wMADA-$\beta$, source features are inputs to a public discriminator as well. Also, Qu et al. utilize private and common feature extractors in source and target domains plus a domain classifier (with GRL unit) in \cite{Paper_Qu_24} to separate sleep-related features from unrelated ones for insomnia detection. A difference loss also forces the two networks to obtain orthogonal features. In order to improve accuracy, reconstruction losses are embedded in the network and the target common classifier's features are fed into an LSTM and then the final classifier.

Adversarial training may also be applied to transform source data into the target distribution. In \cite{Paper_Haung_71}, Huang et al. propose a generator network that attempts to generate samples similar to target data, from source samples by using an adversarial domain discriminator, as in GANs. Overall, the sample data is first transformed to have the target distribution. Finally, an emotion classifier is trained on this data, allowing the target data to be used directly at test time.

Adversarial domain alignment is broadly used in a great number of works, due to its general framework that can be combined with various feature encoding modules; as in arenas other than medical data analysis, adversarial training is a powerful tool for making a backbone feature encoding model more robust. Moreover, another benefit that adversarial methods bring about is that they may be applied to unlabeled source data as well, in an unsupervised manner. Despite the numerous improvements adversarial approaches bring about, using them can also be challenging. First of all, training them can be unstable, as finding an equilibrium between the two modules adversarially trained against each other may not be practical. In other words, these two modules may end up with a suboptimal solution where their performance is not satisfying for the final task. The performance of an adversarial model might be limited by the mode collapse issue, i.e. if there is no proper alignment between features and classes in different domains, the separate design of the task classifier and domain discriminator may degrade the model's performance \cite{hassanpour2022survey}. Additionally, adversarial scenarios demand a larger number of data samples to provide meaningful results in comparison with other methodologies. This issue can be problematic in medical data analysis, where the volume of data and providing neat data is labor-intensive and expensive. Also, adversarial methods have the problem of being time-consuming at the training stage.
\subsubsubsection{Domain Alignment}

A non-adversarial source and target features alignment can be achieved using domain alignment techniques. In this regard, subspace alignment is one of the most straightforward methods. In the context of subspace alignment, the initial concept is to learn a common intermediate representation shared between domains. It has been found that most adaptation approaches in this category start by creating a low-dimensional representation of original data using a variety of deep or non-deep methods and then use distinct objectives such as Kullback–Leibler Divergence (KLD) and MMD to reduce the discrepancy between marginal and conditional distributions in a new subspace, as in \cite{Paper_Shen_11}, \cite{Paper_Zhou_fMRI_26}, \cite{Paper_Bao_91}, and \cite{Paper_Gao_fMRI_5}. In order to achieve this goal, various deep neural network models are used. For example, Hao Chen et al.\cite{Paper_Chen_13} propose a model consisting of an ANN-based common feature extractor and multiple domain-specific feature extractors, with one network per pair of source and target, which is designed to minimize MMD in order to transform each pair into a different subspace. Likewise, Zhao et al.\cite{Paper_Zhao_52}, allocate a model for each pair of source and target. However, they consider the importance of each class as they minimize MMD. Additionally, the paper \cite{Paper_Zhao_52} provides domain-invariant feature extraction modules built on a Common EEGNet-based Network (C-EEGNet)\cite{Lawhern2018EEGNetAC} as well as domain-specific feature extraction in each pair of sources and targets by using $N$ CNN-based subnets (S-CNNs). A novel alignment algorithm called Local Label-based MMD (LLMMD) is proposed in this paper to diminish the discrepancy between source and target domains, which explores local label-based fine-grained structure information across all domains and extracts label-based domain-invariant features. And In \cite{Paper_Shi_fMRI_10}, the Dempster-Shafer (D-S)\cite{dempster2008upper} evidence theory, and rough adjoint inconsistency are applied to derive weight coefficients for each domain. Afterward, the target domain class proportion and optimal coupling distribution set are solved iteratively. Lastly, each source domain is aligned with the target domain and is used to train the final classifier.

In \cite{Paper_Chai_43}, a Subspace Alignment Auto-Encoder-based model (SAAE) is proposed to embed extracted Differential Entropy (DE) features from EEG signals and minimize MMD in an infinite-dimensional Reproducing Kernel Hilbert Space (RKHS). Moreover, Peizhen Peng et al. \cite{Paper_Peng_82} use auto-encoder and MMD on time-frequency images for the mentioned purpose.

In the paper \cite{Paper_Lee_80}, to extract features, Kee et al. use a single-layer Gated Recurrent Unit (GRU) embedded in a semantic manifold and used Multi-Kernel MMD (MK-MMD) as a divergence metric.

Feature extraction can also be conducted using tensors; Mu Shen et al. propose a tensor-based alignment model in \cite{Paper_Shen_RP32}. This model uses Tucker decomposition to tensorize EEG channel data. As a result of tensor network summation, features of training and testing tensor samples are derived from corresponding subspace matrices. A Deep Domain Adaptation Network (DDAN) is proposed by Wenlong Hang et al. \cite{Paper_Hang_RP2} that employs a CNN to automatically detect features, MMD to minimize distribution discrepancy, and a Center-based Discriminative Feature Learning (CDFL) method to force the deep features closer to their respective class centers and to make the inter-class centers more distinguishable. As well, in \cite{Paper_Meng_RP20}, a novel domain adaptation method, Deep Subdomain Associate Adaptation Network (DSAAN), is described that combines the advantages of both subdomain adaptation and associate loop calculation. This model uses a ResNet \cite{He2016DeepRL} to extract features.

It is also possible to extract meaning from EEG signals by using GNNs; Feng Kuang et al. \cite{Paper_Kuang_81} offer a Multi-Spatial Domain Adaptive Network (MSDAN). Through MSDAN, the original EEG data is mapped into multiple graph-based spaces, and the distribution of the source and target domains in those spaces is narrowed by the use of MMD.

Attention mechanisms can also be useful to solve excessive alignment problems as well. As an example, in \cite{Paper_Ning_77}, a CBAM-based module was designed to extract the common features of the source and target. The MMD in RKHS is also used to align the distribution of the two domains. In this article, to overcome the excessive alignment problem in which the samples of the two domains are mixed, and the categories within each domain cannot be distinguished well, the few-shot learning module is introduced to retain the domain-specific information. And, \cite{Paper_Chen_90} suggests CS-DASA, which learns the common features from multi-frame EEG images using the convLSTM. Also, to adapt the source and target domains, the model uses a subject-specific module using 2D-CNN with MK-MMD loss in the RKHS. Furthermore, a subject-to-subject spatial attention mechanism focused on the discriminative spatial features from the target image data is used. 

Other classic machine learning techniques are similarly useful for reducing dimension and finding shared subspaces. As a dimensionality reduction technique, Transfer Component Analysis (TCA) \cite{TCA_2010} aims to minimize distribution discrepancies by learning a set of transfer components. In \cite{Paper_Liu_RP9} and \cite{Paper_He_70}, a transfer learning-enabled classifier consisting of a TCA is implemented to mitigate the mismatch among distributions. It anticipates a projection to a latent subspace where the projected source and target data achieve a reduced MMD in RKHS. Similarly, Yueying Zhou et al. \cite{Paper_Zhou_65} use TCA, Joint Distribution Adaptation (JDA)\cite{long2013transfer}, Balanced Domain Adaptation (BDA)\cite{wang2018stratified}, and Transfer Joint Matching (TJM) \cite{long2014transfer} with MMD distance measure to adapt the domains. \cite{Paper_Wang_50} uses a JDA-based adaptation module that joints the marginal distribution alignment and conditional distribution alignment to minimize the data distance between the source and the target domains with MMD measure. Very similarly, Transport-Based Joint Distribution Alignment (T-JDA) blocks are proposed in \cite{Paper_Zhang_fMRI_21} that can propagate features or labels from source to target by minimizing the global transportation cost between the empirical joint distribution of a pair of source and target domains. An Independent Component Analysis (ICA) \cite{comon1994independent} method is employed to determine the independent components of unlabeled and labeled EEG signals in \cite{Paper_Qu_RP29}. In this work, the energy features of ICs are extracted as the source and target domains. As a final step, the marginal distributions of the source subspace base vectors are aligned with the base vectors of the target subspace using linear mappings.

Alternatively, another category of methods assumes that there exists a manifold of transformations between the source and target domains; this manifold consists of a space of parameters where each point generates a possible domain. For instance, Wen Zhang and Dongrui Wu \cite{Paper_Zhang_56} propose a Manifold Embedded Knowledge Transfer (MEKT) approach by aligning the covariance matrices of the EEG trials in the Riemannian manifold, extracting tangent features, and then performing domain adaptation by minimizing the joint probability distribution shift between the source and target domains. Jiang et al. \cite{Paper_Jiang_RP33} have also proposed a Kernel-based Riemannian Manifold Domain Adaptation technique (KMDA) in which the covariance matrices are aligned in the Riemannian manifold and then mapped to a high dimensional space by a log-Euclidean metric Gaussian kernel, which is then reduced by MMD.

Despite the fact that domain alignment techniques are powerful and widely used for domain adaptation issues, these techniques need a significant amount of parameters for adaption \cite{chen2022multi} and the constraints they are subject to could cause a distortion of semantic feature structures and a loss of class discriminability \cite{ge2022domain}.
\subsubsubsection{Instance Alignment}
Target domains can also be aligned with source domains at the instance level. In some studies, pairs of source and target data samples are directly guided to become closer to each other.

Chambon et al. \cite{Paper_Chambon_41} find the optimal transport from the source domain to the target domain and train the model on transported source samples.
From a representation learning aspect, Lee et al. \cite{Paper_Lee_fMRI_22} and Wang et al. \cite{Paper_Wang_fMRI_20} try to represent each sample of every source domain by a shared transformation and a low-rank transformation of target samples and use it as the new representation for source samples. The shared transformation further generates a new representation for target samples. Similarly, in \cite{Paper_Lee_6}, Lee et al. follow a contrastive approach by decreasing the distance between pairs of samples in the same class and different subjects compared to samples with different classes and the same subject.

It is also common to represent target samples or predictions based on their similarity to source samples. For instance, in \cite{Paper_Zhao_10}, after training source-shared and source-specific encoders and decoders and a target-specific encoder, the final prediction on target data results as a combination of target model prediction and source model predictions, weighted by their feature similarity to the target sample.
Likewise, Li et al. \cite{Paper_Li_14} duplicate the batch normalization layer for each source. In the test phase, for new target samples, the average of batch normalization branches is computed and further weighted by layer statistics similarities of the target and each one of the source domains.
From a slightly different viewpoint, Lin et al. \cite{Paper_Lin_RP19}, train task and subject predictor networks, and select samples from most similar subjects to train the model on a new domain. 
Moreover, in \cite{Paper_Wang_49}, using direct transfer accuracy, Wang et al. select only related source domains (subjects) to be used for the main adaptation module.

There are also studies conducted on aligning source and target samples based on their discriminative statistics.
For example, Tao et al. \cite{Paper_Tao_15}  align kernel-based classifiers for each domain to match the label structure of the samples and minimize the distribution of source and target domains.
In a similar manner, Shen et al. \cite{Paper_Shen_79}, learn a transformation on samples for each source domain, so that their covariance matrix becomes as close as possible to the covariance matrix of the target domain samples.

As a more comprehensive form of domain alignment, instance alignment attempts to align domains at the sample level. Although this extension allows for more accurate adaptation when it is successful, it can also fail more frequently. Instance alignment requires that the target domain data can be represented by the source domain or vice versa and can be adversely affected by outliers. Additionally, similarity-based methods are not able to adapt to target spaces that differ significantly from source spaces.
\subsubsubsection{Classifier Alignment}
Classifiers trained on features extracted from different sources may result in misaligned predictions for target samples close to the domain boundaries in a multi-source setting. By minimizing specific classifier costs, the classifiers can be better aligned, resulting in more accurate and generalized models. As an example, Zhang et al. \cite{Paper_Zhang_fMRI_21} adapt multi-source domains to a single target and penalize decision inconsistency among diverse classifiers trained on paired joint distribution aligned features by minimizing the consistency loss between classifiers trained on source-target domain pairs. In addition, in \cite{Paper_Chen_13}, a discrepancy loss metric is introduced to achieve convergence of predictions from $N$ classifiers trained on $N$ domain-adapted sources. Moreover, \cite{Paper_Zhao_52} integrates the probability distribution from $N$ classifiers by the weighted mechanism. Each classifier's prediction probability distribution is used to calculate the weights; the global optimization strategy can also remove the negative impact of significant individual differences.

Unlike the described methods, classifier alignment is used with a different purpose by Xia et al. in \cite{Paper_Xia_73}. In their presented model, to have a more robust target classifier, different perturbation of target data is fed into some auxiliary classifiers, which are aligned to each other and to the fixed source classifier using consistency regularization loss.

Classifier alignment techniques are practical, especially when the target domain samples are at the decision boundary. In this case, the variance of predictions will be high, which will significantly impact the results. While these techniques can provide high-accuracy results, they require the development of a module to extract source and target common features for the classifier's input.

\subsubsection{Data Manipulation}
In this group of methods, data is changed and manipulated for adaptation purposes. The major sub-type in these approaches is preprocessing, given that solutions to domain adaptation can be injected into data preprocessing steps.

For example, in \cite{Paper_Albuquerque_40}, the authors showed that feature normalization had a relevant effect on the conditional shift, and by performing z-score normalization as preprocessing, the conditional and marginal shifts could be reduced. Another approach that can be categorized as preprocessing is subject clustering. Concerning this, Li et al. \cite{Paper_Liu_RP46} propose Domain Adaptation with the Subject Clustering (DASC) which clusters the subjects according to their inter-subject similarity of emotion-specific EEG activities and only uses the source cluster that matched the target better for adaptation to the target. Moreover, the optimal subjects from the selected cluster, with positive transfer impacts, are used to classify the emotional situation of the target subject.

Applying this approach as a data manipulation method provides independence from the necessity of training and makes it directly applicable. Also, preprocessing is integrable with other DA methods and this increases our options. On the contrary, information loss and error propagation through the whole pipeline are disadvantages of this approach.
\subsubsection{Feature Disentanglement}

One of the recently popular techniques in DA is to disentangle input data into domain-specific and domain-invariant features. By defining appropriate objectives for each disentangled part, domain-specific information can be removed from the data, and thus, domain-invariant features can be used in a new domain for prediction.

Jeon et al. \cite{Paper_Jeon_39} force class-invariant and class-relevant features to contain the least common information by minimizing their mutual information.
In \cite{Paper_Liu_48}, aside from guiding each part to predict its own information, Liu et al. use adversarial training to prevent them from estimating each other's information. Furthermore, the separated parts are used to reconstruct the original data in order to ensure that there is a minimum loss of information during disentanglement.

In the context of multi-source domain adaptation, Zhao et al. \cite{Paper_Zhao_10} disentangle data from each source into domain-private and shared features and then, reconstruct the original data via a shared decoder. The new target domain uses a private encoder that is trained with the reconstruction objective. During the inference phase, data from the target domain can be classified using both private and shared encoders.

When data samples are inherently a combination of multiple parts mixed together, the disentanglement technique is useful. Brain signals are a complex mixture of the brain's response to various stimuli in the environment. Usually, only one of them (class-relevant response) is intended to be analyzed in DA for medical image analysis. Consequently, disentanglement is a very intuitive and natural way of dealing with complex brain data. Nonetheless, It is important to note that disentanglement requires a significant amount of data to work well, due to the need to extract irrelevant features as well.
\subsubsection{Pseudo-label training}

The generation of pseudo-labels from the source model is a common DA approach in medical image analysis. In this approach, a model is trained on source domain data, then its predictions on target domain data are considered pseudo-labels for the model, and they are exploited to adjust the model to the target domain.

Since pseudo-labels are noisy and not completely accurate, some studies apply them iteratively, and as the model produces better predictions, it uses its more accurate pseudo-labels from the previous iteration. This strategy can be used to transform any transfer learning method that uses target domain labels into an unsupervised domain adaptation one.

Some examples of this type of iterative self-supervision are the works of Zhang et al.  \cite{Paper_Zhang_56}, Eldele et al. \cite{Paper_Eldele_78},  Shi et al. \cite{Paper_Shi_fMRI_17}, Jiang et al. \cite{Paper_Jiang_RP33}, Shen et al. \cite{Paper_Shen_11}, and Wang et al. \cite{Paper_Wang_50}. These works, while aligning source and target samples, generate pseudo-labels at each iteration that is used for the next step. Moreover, Edele et al. tackle the cold start problem by ignoring the target classification loss in the initial iterations. Shi et al. perform JDA iteratively and enhance pseudo labels using label propagation algorithm proposed in \cite{label_propagation}. The idea to enhance the pseudo-labels is also proposed by Han et al. \cite{Paper_Han_fMRI_12}, where they apply a single step of K-Means algorithm to the features of each pseudo-label to form more coherent and less uncertain labels. Wang et al. 
 come up with the aforementioned iterative self-supervision from an objective that is optimized with respect to both representation parameters and target domain labels.
 
In some studies, pseudo-labels are directly treated as target data labels, along with other objectives imposed to guide the model in the correct direction.
Guarneros et al. \cite{Paper_JimnezGuarneros_76} retrain their source model on target data using pseudo-labels with an additional loss term controlling the uncertainty and increasing the diversity of predictions.
Zhao et al. \cite{Paper_Zhao_26} and Heremans et al. \cite{Paper_Heremans_RP43} use pseudo-labels for target classification alongside their adversarial objective.
Tao et al. \cite{Paper_Tao_15} also generate pseudo-labels and consider the target domain as one of the sources and adapt all of them together.

Pseudo-labels may also be used indirectly, for an objective other than classification primarily. 
In \cite{Paper_Hong_RP38}, Hong et al. use pseudo-labels to approximate the class probability of target samples for conditional discrimination between samples of source and target. 
Additionally, in \cite{Paper_Zhou_65}, Zhou et al. utilize JDA as one of their experimented methods which requires target data labels (derived from pseudo-labels) to calculate class conditional distributions. 
Moreover, Meng et al. \cite{Paper_Meng_RP20} use pseudo-labels to partition target domain samples into subdomains and estimate the similarity of samples in source and target domains.

As a result of pseudo-label-based DA, useful information held by the source model is retained during the transition to the target domain. Despite its ability to prevent the forgetting of source domain information, it may not be able to eliminate the bias associated with the source domain. In particular, the performance of this approach is relatively weak when the difference between the source and target domains is so significant that iterative fine-tuning or additional objectives are not capable of correctly adapting the model. Moreover, convergence is a critical concern when using iterative methods, since negative feedback in the self-supervision process may prevent the model from eventually reaching its optimal state.

\subsubsection{Hybrid Methods}

There are some studies that combine multiple domain adaptation approaches. Particularly, as pseudo-labeling can be conducted independently, it can be easily integrated with other adaptation methods. Thus, as discovered in this study, the most common combination is pseudo-labeling and domain alignment (\cite{Paper_Jiang_RP33}, \cite{Paper_Meng_RP20}, \cite{Paper_Zhou_65}, \cite{Paper_JimnezGuarneros_76}, \cite{Paper_Wang_50}, \cite{Paper_Shen_11}, \cite{Paper_Zhang_56}, \cite{Paper_Han_fMRI_12}). Pseudo-labeling is also used alongside adversarial feature alignment (\cite{Paper_Heremans_RP43}, \cite{Paper_Hong_RP38}, \cite{Paper_Eldele_78}, \cite{Paper_Zhao_26}). In these combinations, the pseudo-labels generated from the source model help align the source and target domains more accurately.

Furthermore, classifier alignment is mostly accompanied by domain alignment (\cite{Paper_Chen_13}, \cite{Paper_Zhao_52}, \cite{Paper_Zhang_fMRI_21}). There are also other combinations in the literature. Bao et al. \cite{Paper_Bao_91} and Peng et al. \cite{Paper_Peng_82} try to adapt their model using adversarial feature alignment and domain distribution alignment together. Liu et al. \cite{Paper_Liu_48} use adversarial training to train their disentangled feature extraction model. Zhao et al. \cite{Paper_Zhao_10} incorporate disentanglement along with instance alignment. In \cite{Paper_Wang_49}, Wang et al. utilize adversarial training in conjunction with instance alignment. Additionally, in \cite{Paper_Tao_15} pseudo-labels are used to employ instance alignment.
\subsection{DG Approaches}

\begin{figure*}[t]
\centerline{\includegraphics[width=\textwidth]{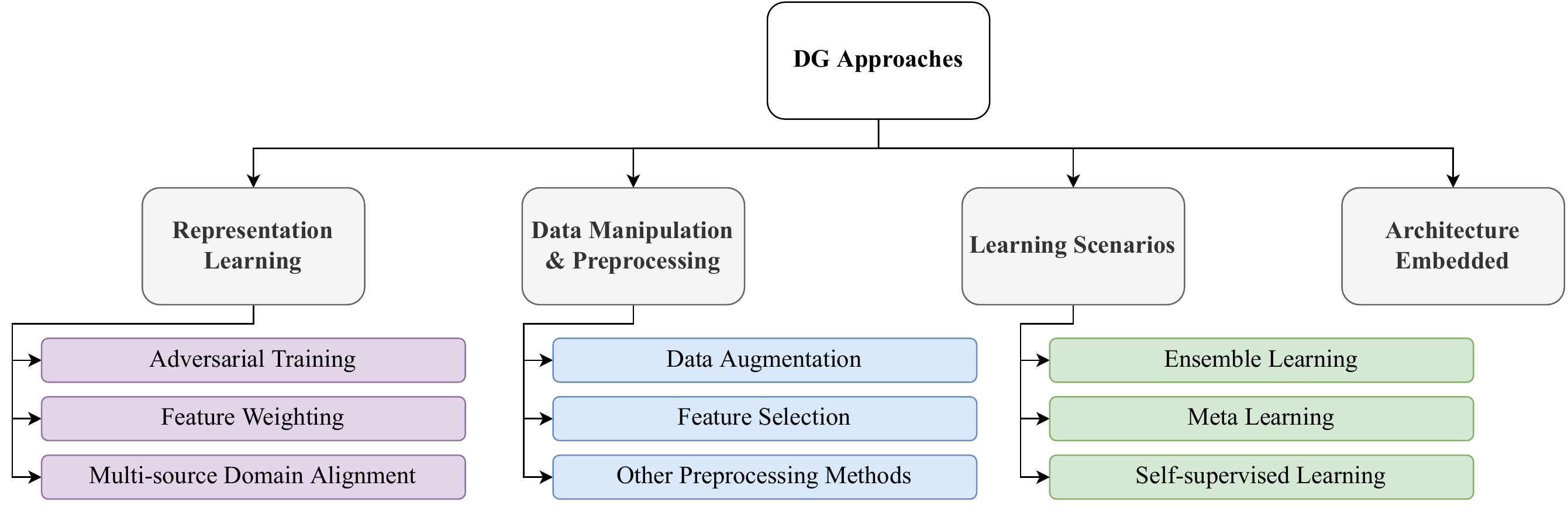}}
\caption{Hierarchy of Domain Generalization approaches in functional medical imaging}
\label{fig:dg_approaches_hierarchy}
\end{figure*}

Various DG methods have been suggested to process functional brain signals. These approaches include representation learning, data manipulation \& preprocessing, learning scenarios, and embedded architectures, which can also be merged to enhance performance on various tasks. Our designed approach hierarchy for these methods based on their approaches and the motivations behind their design is shown in Figure \ref{fig:dg_approaches_hierarchy}. In the remainder of this section, we explain DG-specific approaches as well as approaches presented as part of a domain adaptation method where adaptation techniques are improved via generalization ideas. A precise summary of these explanations is also provided in Table \ref{fmri-papers-table}.
\subsubsection{Representation Learning}

An overall strategy for domain generalization is representation learning, aiming to ensure that the learned representations are domain-invariant, meaning that they do not contain much knowledge about the domain from which they originate. In other words, feature extraction is guaranteed to be generalizable to all domains as long as they include similar semantics about the studied task. In the following, the representation learning methods are grouped into adversarial training,  domain Alignment, and feature weighting. Most of these ideas are designed for multi-source DG scenarios.
\subsubsubsection{Adversarial Training}

Learning a generalizable representation needed for domain generalization may be achieved through adversarial training. It should be noted that most methods that use an adversarial setting for such a purpose have multi-source setups. Derived from DANN, in most of these methods, common feature extraction is applied to different sources; afterward, a domain classifier, as a discriminator, tries to distinguish domains. By min-max training, the discriminator will fail to identify the domain from the extracted features, meaning that features are global among different source domains. Hence, these features can make the main model generalizable.

Many works employ variations of the above strategy for gaining a domain-generalizable feature extraction model. \cite{Paper_Bethge_67} and \cite{Paper_zdenizci_RP16} propose an adversary network that tries to output the source domain identity (i.e., the given dataset or subject) from the encoded features. Training this domain discriminator adversarially with the main classifier results in a generalized pipeline where the learned features are domain-invariant and, hence, accurately classifiable. Similarly, in \cite{Paper_Jia_12}, a domain discriminator is used and is designed to fail at distinguishing the incoming domain while maintaining task-relevant parts in the extracted features obtained via spatial and temporal graph convolutions. In \cite{Paper_Hagad_17}, Hagad et al. employ a DANN, consisting of a domain and emotion classifier, alongside a beta-VAE \cite{higgins2016beta}, treating each of the multiple sources as a single domain and feeding the DANN with outputs of bi-lateral convolutional on the concatenated VAE outputs of the two hemispheres. \cite{Paper_Albuquerque_33} proposes a generalization approach that theoretically guarantees a generalization bound on unseen domains. From the practical viewpoint, they implement this method by using one-vs-all classifiers, each of which is responsible for computing a divergence score between every source and all other sources. By adversarially training these classifiers with a feature encoder and a task classifier, they demonstrate the generalizability of their given method for EEG emotion recognition. In \cite{Paper_Han_16}, a domain classifier is used to force encoders in two different branches to extract features that diminish the distinguishability of domains. In \cite{Paper_Ma_35}, Ma et al. suggest that biased network weights in source feature extractors can be regarded as domain-relevant clues and, therefore, incorporate separate encoders for each domain, having shared unbiased weights and specific biased weights for each domain. They further use the encoded features as inputs to a label predictor and an adversarial domain classifier.

Adversarial training may also be useful for enforcing a low-dimensional distribution of the data. Inspired by GANs, Ming et al. propose an adversarial scheme in \cite{Paper_Ming_RP22} to align multiple sources with an artificial empirical distribution in low dimensions. To this end, they design a discriminator considering samples from the artificial domain as “real” data and the encoder’s output as “fake”. These two networks are further adversarially trained so that eventually, data from all sources is mapped in a coherent low-dimensional representation space. In order to avoid a lack of information for further classification tasks due to the artificial distribution, the generator is split into an adapter and a mapper, and the intermediate output from the adapter is used for the final task.

Although adversarial training can remarkably help generalize the representation learning step and may be implemented in integration with various feature encoding methods, they still face challenges such as time-consuming training and mode collapse, as pointed out in the DA section. Also, most notably, being data-hungry, these methods may be hard to obtain their best performance, especially in DG in medical use cases, where data collection is tricky.
\subsubsubsection{Feature Weighting}
Some works apply feature weighting to obtain domain-invariant features from different sources. These methods’ features are associated with some learnable weights during representation learning. Fusing the features concerning these weights will result in a more general representation of features. It is critical to keep in mind that some features might be weighted to zero during the learning process, resulting in some features being omitted.

As an example of this category, \cite{Paper_Cui_34} presents Feature Weighted Episodic Training (FWET) which consists of feature weighting to determine the importance of various features and episodic training for domain generalization. Feature weighting, regarding the different significance of various brain areas, assigns a weight to each feature. Episodic training is also converted from classification to regression and combined with FW for better generalization.

Feature weighting provides the ability to determine different levels of importance for various features. By doing so, we can improve the value of more generalizable features. However, in some cases, it might be better to use a combination of features instead of setting different weights for features.
\subsubsubsection{Multi-source Domain Alignment}
Similar to feature alignment, mentioned in the domain adaptation section, extracted features from multiple source domains can be aligned to remove the shift between various sources, thereby generalizing the final model.

To reach a shared space, some works apply TCA algorithms, RKHS-based approaches, or MMD-based losses. For example, in \cite{Paper_Ayodele_60}, Ayodele et al. utilize a modified version of the Multi TCA algorithm based on an RKHS approach to extract a common subspace of the datasets. Moreover, Bethge et al. \cite{Paper_Bethge_5} design a multi-source learning framework for domain-invariant representation learning, including a private feature encoder per domain and a cross-domain shared classifier, for which an MMD-based domain alignment loss is leveraged across private feature encoders to decrease domain-specific deficiency within the learned representations.

Musellim et al. \cite{Paper_Musellim_7} propose a prototype-based framework that forces cross-instance style in-variance in the same domain by using subject labels. Also, they use Convolutional Prototype Learning (CPL) as the open-set recognition method for subject classification.

Zhang et al. present a novel Convolutional Recurrent Attention Model (CRAM) in \cite{Paper_Zhang_31} that encodes timepieces extracting the Spatio-temporal information. They apply a recurrent-attention network to explore the temporal dynamics among various time portions and focus on the most discriminative ones.

Common feature extraction may also be done in two steps to align source features. For example, Yousefnezhad et al. \cite{Paper_Yousefnezhad_fMRI_35}, propose a Shared Space Transfer Learning (SSTL) that first finds common features for all subjects in each site and maps them to a site-independent shared space. Next, it uses a scalable optimization procedure that uses a single iteration multi-view approach to extract the common features for each site and then maps them to the site-independent shared space. 
There are also some ideas in this arena that take advantage of graph structures to align sources. For instance, Li et al. \cite{Paper_Li_fMRI_36}, propose a graph decoding model in which a cross-subject graph showing the similarities across subjects is used. By further regularization, developing a kernel-based optimization, which enables the extraction of nonlinear features, would be possible.

In this category, some models attempt to extract features from the intermediate space that can be reconstructed from the original data. It is pertinent to note that in this intermediate space, no information is removed. Accordingly, a Discriminant Autoencoder Network with Sparsity constraint (DANS) is presented in \cite{Paper_Zeng_fMRI_29}, which is developed to learn domain-shared features. The model introduces an optimized discriminant item based on the correlation function in the cost function at the pre-training stage to generate discriminating features for binary classification. Furthermore, Huang et al.\cite{Paper_Huang_fMRI_27} present Manifold-Regularized Multiple Decoders AutoEncoder (MRMD-AE) network that extracts common latent space representations from multiple sources while respecting the individual data geometry by a pre-computed PHATE embedding while maintaining the ability to decode individual raw fMRI signals. Also, the authors of \cite{Paper_Zhang_fMRI_18} have proposed a low-rank subspace built on low-rank representation theory using the multi-source RS-fMRI dataset. They initially encode all domains in a common lower-dimension space. The graph-based data is then loaded into the graph convolution network module, followed by a classification head for autism spectrum disorder diagnosis. In a similar approach, in \cite{Paper_Wang_42}, Wang et al. reduce the feature space dimension to align source data in a lower-dimension space. To this end, they apply plural discrete wavelet transforms and nonlinear analysis to extract representative features. Moreover, they employ a PCA algorithm, along with the feature ranking method of the analysis of variance (ANOVA), to extract features further in five frequency sub-bands based on clinical interest and omit irrelevant features.

Although multi-source domain alignment is an approach that increases model generalization by concentrating on more common features between various sources and constructing shared subspaces, it could also cause a scarcity of discriminative features among different classes.

\subsubsection{Learning Scenarios}
Some methods use learning-based approaches for domain generalization. These methods are categorized into ensemble learning, meta-learning, and self-supervised learning.
\subsubsubsection{Ensemble Learning}
One learning-based idea for generalization is ensemble learning, boosting the final model's performance and accuracy by combining various networks and specifying the main output by majority voting. For example, Li et al. \cite{Paper_Li_66} propose a novel decomposition-based ensemble CNN framework. The outputs are integrated with an ensemble architecture employed in two modes, Train CNNs Together (TT), in which the score after the fully connected layer and before the Softmax layer is averaged and backpropagation is executed on the entire ensemble network, and, Output Fusion (OF), in which the outputs of the Softmax layer are averaged directly. Moreover, in the test phase of \cite{Paper_Zhao_10}, predictions of the shared classifier integrated with those of individual classifiers are ensembled after modulation by similarity weights. As a result, this idea helps increase the generalizability of emotion decoding.

In \cite{Paper_Zhu_RP28}, however, Zhu et al. first evaluate the feasibility of utilizing EEGNet models \cite{eegnet_2018} with various kernel numbers to decode SSVEP in ear-EEG signals. Then, due to the difficulty of separating useful information from background noise caused by weak SSVEP in ear-EEG, they use an ensemble learning strategy to combine EEGNet models with different kernel numbers to enhance the classification of ear-EEG signals. Roots et al. \cite{Paper_Roots_RP6} propose a model called EEGNet Fusion, a multi-branch 2D CNN that utilizes various hyperparameter values for each branch and is more flexible to data from different subjects.

As multiple networks are capable of extracting a wider range of features and processing them in a more varied manner, ensemble learning can significantly improve domain-invariant results. However, this approach cannot reveal the unknown differences between various samples and populations. Also, such models are not easy to interpret.
\subsubsubsection{Meta-Learning}
The main goal of meta-learning is learning to learn, meaning that the model observes how different machine learning methods perform various tasks and uses their meta-data to learn how the learning procedure is performed. For example, Luo et al. propose Pseudo Domain Adaptation via Meta-Learning (PDAML) in \cite{Paper_Luo_22} to reduce the time, cost, and storage usage of their emotion predictor model in the test phase. Firstly, they introduce Pseudo Domain Adaptation (PDA). Also, they use an additive decomposable structure, known as a domain shift governor, and a meta-learning-based approach to make the model fast to generalize to a new domain using the target data.

Some works utilize Model Agnostic Meta-Learning (MAML) \cite{MAML_2017} framework. For instance, Lemkhenter et al. \cite{Paper_Lemkhenter_RP31} introduce a meta-learning method for sleep scoring built on top of MAML, where the model is trained on many subjects with the goal of generalizing to unseen subjects by zero-shot learning. Also, in \cite{Paper_Duan_RP10}, Duan et al. propose Meta-Learning on Constrained transfer Learning (MLCL). The transfer process is quickened by utilizing the MAML algorithm, performed under a novel constrained setting, which preserves adequate flexibility to adapt to a new subject where the number of must-transfer parameters is decreased substantially.

Furthermore, Lee et al. \cite{Paper_Lee_fMRI_22} try to learn the adaptation of feature representations within a meta-learning framework by using an episodic-learning strategy. Regarding episodes of the target task to simulate differences between sites, the modulation network learns different patterns that can cope with various domain shifts.

Using meta-learning leads to more generalization in the model, together with a faster and cheaper training process; because fewer experiments are used in learning and unnecessary ones are removed. However, the rule set utilized in this approach may be incomplete; also, in some of its approaches, there is a limit to the volume of information that meta-features can capture, due to the fact that these features might only capture relations between two attributes or a class and an attribute.
\subsubsubsection{Self-Supervised Learning}
Self-supervised learning is a machine learning method used to extract useful information from data that has not been labeled. Therefore, it is reasonable to address the lack of sufficient labeled data in medical domains using this method. In this area, two general types of self-supervised methods are contrastive and non-contrastive methods. In contrastive methods, the similarity between two augmented versions of a data sample is maximized in a positive pair, whereas the difference between each of these two samples and samples in negative pairs is minimized. On the other hand, in non-contrastive methods, there are no negative pairs, and self-supervised learning is performed only within positive pairs \cite{Tian2021UnderstandingSL}.

Self-supervised contrastive learning can be used to solve domain shift problems, and several novel works have been proposed to perform domain generalization. In \cite{Paper_Shen_19}, Shen et al. propose Contrastive Learning for Inter-Subject Alignment (CLISA), a self-supervised contrastive learning method to address the issue of inter-subject variability. CLISA is grounded on a neuroscientific inspection which assumes that the neural activity state of subjects is similar when they receive indistinguishable stimuli.

Cheng et al. present a subject-aware learning method in \cite{Paper_Cheng_RP8}, which combines adversarial training with self-supervised contrastive learning to reduce the inter-subject variability in bio-signals such as EEG and ECG. With this method, they manage to achieve competitive results in varied kinds of downstream tasks.
In \cite{Paper_Wagh_RP45}, Wagh et al. propose three novel self-supervised pre-text tasks, which exploit known patterns in scalp EEG signals and enable the learning of features that could be transferred to other domains and tasks. In their method, pre-text tasks are designed to examine the spatial similarities between the left and right hemispheres of the brain, the behavior of the brain, and changes related to brain activity.
In \cite{Banluesombatkul2021MetaSleepLearnerAP}, sleep stage classification (sleep scoring) is performed using MetaSleepLearner (MAML), a meta-learning method based on few-shot domain adaptations. The MAML model, however, is vulnerable to overfitting even on datasets with many samples. A self-supervised stage was introduced to MAML by Lemkhenter et al. in \cite{Paper_Lemkhenter_RP31} to solve the over-fitting problem without using newly labeled target data (zero-shot learning).

A similar idea to contrastive self-supervised learning is applied in \cite{Paper_Xia_73}, in which features extracted by an encoder from unlabeled target data were perturbed and then were used as input of a number of classifiers to train a robust and adaptable model for motor imagery classification task in the cross-subject setting. The proposed approach is different from other mentioned self-supervised DG methods, as the pre-training has been done on target data instead of source data.

Self-supervised learning methods have the advantage of reducing the need for labeled data. These types of methods have also shown considerably high performance in different areas. One of the limitations of this method is that it takes time to prepare a proper pre-trained model, and the model also might need additional data sources for pre-training.
\subsubsection{Data Manipulation}

By processing and manipulating the input data, a number of studies have been able to increase the generalizability of their models. Some attempt to augment the input data, mostly through adversarial approaches, while others attempt to eliminate unimportant or redundant data. Furthermore, data normalization has been shown to reduce domain bias in some studies.
\subsubsubsection{Data Augmentation}
In general, more data results in more generalizability because the model can explore a greater proportion of the data space. Consequently, adding new data samples to the available dataset can enhance the generalization capability of the model.
In \cite{Paper_Cheng_RP8}, the authors define augmentations such as channel dropout and temporal cutout and extract features based on contrastive learning. Additionally, subject-invariant features are extracted using adversarial training.
Similarly, the authors of \cite{Paper_Xia_73} enhance the generalization of the source model by using channel dropout in source model training. 
Additionally, Han et al. \cite{Paper_Han_16} equip their model with a set of augmentation functions. Aside from Gaussian noise, scaling, and temporal cutout, they shift the signal's amplitude, roll it in time, and upsample intervals in the signal.

One of the common methods to add new data samples is through adversarial training.
In \cite{Paper_Song_21}, adding new data samples to the dataset is accomplished this way. Additionally, they add constraints, such as covariance matrix alignment, to the training objective to ensure that generated data is similar to the original ones.

A consistency-diversity trade-off always exists in the context of data augmentation. On one hand, it is more likely that augmented samples will be inconsistent with the real distribution of the data if they are too different from the original data samples. On the other hand, if the augmented data are too similar to the original data, it will not allow the model to explore much more of the data space, which makes them ineffective. Despite the fact that appropriate augmentation methods can increase diversity while maintaining consistency, the evaluation of such methods is not straightforward as the actual data space is unknown.
\subsubsubsection{Feature Selection}
Prior to the advent of deep learning, selecting important and essential features from data was one of the most common strategies. This approach is useful even when there are deep models present, as it prevents the model from over-fitting to the data.
In some studies, feature selection is applied as a preprocessing step. In \cite{Paper_Pan_fMRI_16}, graph pooling is used to select important nodes in the brain network. Graph pooling eliminates redundant data by discarding nodes that are close to the average of their neighbors.
In \cite{Paper_Subah_fMRI_14}, to preserve only the important brain regions, Wang et al. used standard brain atlases and significantly reduced the number of features by averaging values in these important brain regions.

Statistical selection of important features has also been addressed in the literature.
Harrison et al. \cite{Paper_Bhaumik_fMRI_23} conduct a t-test for feature selection. As an alternative feature selection method, the method also solves a $l_1+l_2$ regularized classification (also called ElasticNet) and discards features with zero weights.
 Similarly, Harrison et al. \cite{Paper_Yang_RP4} calculate a statistical comparison between the distribution of the values for each of the features, separately for positive and negative data (binary classification), and only retain the features with significant differences.
 
 Manual feature selection has the ability to insert prior or expert knowledge into the system, though in many cases is not available or is very limited. Contrary to this, automatic (learning-based or statistical) feature selection methods can be used in most cases, but their associated criteria are very simple and can only be applied to discard obviously useless features. Information redundancy and sharing are two of the major challenges in feature selection as all of the features may contain some useful information.
\subsubsubsection{Other Preprocessing Methods}
As another data-driven approach for DG, the normalization of data values can help reduce domain biases in the data. Every successful medical image analysis study includes a series of preprocessing steps, so here we only focus on those which are directly aimed at DG.

In \cite{Paper_Fdez_RP5} and \cite{Paper_Liu_48}, each subject's data values were normalized into the $[0, 1]$ interval to remove subject-specific information that affected the scale and position of the data.
Chen et al. \cite{Paper_Chen_90} use the preprocessing proposed in EEGNet in order to start with generalizable features, considering the good generalization power of EEGNet.

Numerous preprocessing methods have been employed to remove data noise and various artifacts available in functional brain signals. When it comes to EEG signals, preprocessing becomes more important since they have a low signal-to-noise ratio, meaning there is a lot of trial-specific, unwanted information. Preprocessing is essential for any kind of adaptation model since the datasets in this field are typically too small for the model to detect and remove data artifacts by itself. Although preprocessing is essential, its performance is always dependent on the prediction model and usually needs other methods as well.

\subsubsection{Architecture Embedded}
Some methods exploit architectures that are naturally capable of more generalizable learning of the task. Also, using particular layers inside the network, such as batch normalization layers, can help reduce the unwanted variability of data within the network. Such ideas have been investigated in multiple works. 
In this category, some methods try to bring extra values to conventional CNNs used for feature extraction. Two different CNN architectures were proposed by Dissanayake et al. in \cite{Paper_Dissanayake_62} for predicting epileptic seizures from EEG signals. As a result of the use of a customized convolutional architecture, it is possible to learn major features from data much more efficiently and robustly. On the CHB-MIT-EEG\cite{Dataset_CHBMIT} dataset, the proposed models performed well when compared to the existing models. In addition, they use interpretability methods to understand how these models work. In \cite{Paper_Cui_RP13}, Cui et al. developed InterpretableCNN, a CNN architecture for driver drowsiness recognition in cross-subject settings, and an interpretation technique to uncover what happens inside the model. Compared with other models, the model achieved competitive results by incorporating separable convolutions to process spatial-temporal aspects of EEG signals. Additionally, the proposed interpretation technique can provide meaningful insights into the model and input data. \cite{Paper_Zhang_fMRI_24} takes advantage of the generalizability of a novel Separated Channel Attention Convolutional Neural network (SC-CNN-Attention), which enables a good performance in the leave-one-site-out scenario. To this end, a separated channel CNN yields temporal features of brain regions, followed by an attention-based network learning temporal dependencies and a fully connected classifier for ADHD diagnosis. Jiang et al.\cite{Paper_Jiang_fMRI_3} proposed 4DResNet, an architecture that combines 4D convolution with 3D attention modules to extract temporo-spatial information from fMRI signals. The attention mechanism improved the framework's ability to recognize distinct features and enhanced its performance. They used their proposed model in different settings, including cross-task and cross-dataset.

In a study published by Jana et al. in \cite{Paper_Jana_RP14}, the capsule network \cite{Sabour2017DynamicRB} is utilized to recognize emotional states across subjects by exploiting spatial and temporal information from EEG signals. To create a spatio-temporal frame group for EEG recordings, spatial frames were stacked with time frames (temporal frames). A particular data-splitting method was also used to make the model perform better on unseen data.

Some methods exploit graph structures to provide a generalizable representation. Self-organizing graph neural networks (SOGNNs) were introduced by Li et al. in \cite{Paper_Li_RP24} for cross-subject emotion recognition on EEG signals. With the help of a self-organized graph construction module, their proposed architecture can dynamically generate specific graph structures for each signal. Cross-subject performance of the model is enhanced by aggregating connections between channels and temporal features. In another work, \cite{Paper_Cao_fMRI_13}, Cao et al. introduced a framework consisting of a 16-layer deep graph convolutional neural network (DeepGCN) with ResNet and DropEdge\cite{Rong2020DropEdgeTD} units for the task of Autism Spectrum Disorder (ASD) diagnosis in a cross-site scenario on the ABIDE I dataset. Based on their experiments, their proposed method is robust to vanishing gradients, over-fitting, and over-smoothing.

In \cite{Paper_Joshi_RP36}, Joshi et al. trained several machine learning methods with a cross-dataset scenario on SEED, DEAP, and IDEA\cite{Joshi2022IDEAID} datasets, using different types of EEG signal features like PSD, Hjorth parameters, and Linear Formulation of Differential Entropy (LF-DE). They found that the bidirectional LSTM model with LF-DE features performed best in inter-dataset mode.

The Sub-epoch-wise Feature Encoder (SEFE), developed by Lee et al. in \cite{Paper_Lee_84}, can be added to well-known deep models for EEG signals to extract temporal information from input data. By using SEFE in DeepConvNet\cite{Schirrmeister2017DeepLW}, ShallowConvNet\cite{Schirrmeister2017DeepLW}, and EEGNet, the performance of the model improves in the task of visual imagery classification in a subject-independent setting.

Some works benefit from the generalization obtained by combining multiple known structures. To detect emotion in Parkinson's disease patients, Dar et al. combined a one-dimensional Convolutional Recurrent Neural Network (CRNN) with an Extreme Learning Machine (ELM) classifier and used several preprocessing methods in \cite{Paper_Dar_RP35}. They demonstrated that their proposed framework is reliable in cross-subject and cross-dataset scenarios by testing it on cross-dataset data. In \cite{Paper_Lin_fMRI_33}, a convolutional-recurrent neural network is proposed that is applied in subject-level cross-validation. First, BOLD signals, calculated from covariance and standard deviation from fMRI time series, are calculated and further passed through a network with spatial and temporal convolutions followed by an LSTM, which captures relations between consecutive neighboring windows. Giving this output to the final classifier leads to a generalizable model for Alzheimer's disease classification.

In order to reduce inter-subject variability in EEG signals, Li et al.\cite{Paper_Li_66} employ the Component-Specific Batch Normalization (CSBN) layer in their proposed ensemble model. In \cite{Paper_JimnezGuarneros_76}, Jiménez-Guarneros et al. use Adaptive Batch Normalization (AdaBN) to reduce cross-subject variability and normalize extracted features from different domains.

Huang et al.\cite{Paper_Huang_fMRI_27} propose a Manifold-Regularized Multiple Decoder, AutoEncoder (MRMD-AE) that can extract shared features from a number of different subjects' fMRI data and reconstruct specific data for each subject with its numerous decoders. Furthermore, a special kind of regularization and penalties have been used to extract more precise shared representations. Harrison et al.\cite{Paper_Harrison_fMRI_4} proposed the PROFUMO framework, which can be used to model rfMRI properties in spatial and temporal domains. Furthermore, it can capture differences in levels of activity and generate additional summaries of this kind of data. Zeng et al.\cite{Paper_Zeng_fMRI_29} introduced Discriminant Autoencoder Network with Sparsity Constraint (DANS) to extract site-shared information from functional connectivity MRI (fcMRI) data in the task of schizophrenia classification. The model is capable of learning reliable patterns in the multi-site setting.

DG methods based on an architecture-embedded approach perform well on specific tasks and datasets because of their specialized structure and architecture. However, one of their disadvantages is that they cannot be used for multiple downstream tasks and settings. In the case of having a single specified task for the final performance of the model is the goal, these type of methods is optimal. However, if a general framework is needed for several tasks and datasets, it might be better to use another approach.
\subsubsection{Feature Disentanglement}

Learning disentangled feature representations can also contribute to model generalization. In \cite{Paper_Peng_9}, similarities between subjects are computed based on their samples' feature correlation. Based on these similarities, subjects are then clustered. Lastly, Singular Value Decomposition (SVD) is performed on each cluster separately and then on the entire data to reduce feature dimension.
As discussed in the DA section, the idea of disentanglement is very intuitive due to the mixed nature of brain data. Still, in DG, the same issue of disentanglement exists as in DA, meaning that the disentanglement process requires a lot of data to produce appropriate results.
\subsubsection{Hybrid Methods}

DG approaches can be fused together to produce a more generalizable model. Among different combinations, data augmentation and self-supervised learning have been tried together (\cite{Paper_Xia_73}, \cite{Paper_Cheng_RP8}). In addition, in \cite{Paper_Huang_fMRI_27} and \cite{Paper_Zeng_fMRI_29} the authors not only attempt to match source domain distributions but also account for domain generalization when training their models through regularization. Also, Li et al, \cite{Paper_Li_66} carry out ensemble learning while embedding adaptive batch normalization layers in their model architecture and Lemkhenter et al. \cite{Paper_Lemkhenter_RP31} integrate meta-learning with self-supervised learning.

  \begin{table*}[t]
  \centering
  \fontsize{3.5}{3.5}\selectfont
  {\renewcommand{\arraystretch}{1.1}
  \caption{Summary of papers related to Domain Adaptation and Generalization methods on EEG modality. In the "Single/Multi Source" column, SS represents Single-Source, and MS represents Multi-Source. Also, ER stands for Emotion Recognition, SA for Seizure Analysis, SD for Sleep Diagnosis, AM for Awareness Monitoring, MI for Motor Imagery, VPA for Visual Perception Analysis, MSD for Mental State Diagnosis, HTA for Human Thought Analysis, ND for Neural Decoding, and BEE for Behavioral State Estimation. The state-of-the-art papers in cross-subject and cross-session settings are shown by red and blue crosses in the dataset columns, respectively.}
  \label{eeg-papers-table}
  \resizebox{\textwidth}{!}{%
  \begin{tabular}{|c|c|c|ccccccc|cccccccccc|ccccccc|ccccccccccc|c|c|c|}
  \hline
   &
     &
     &
    \multicolumn{7}{c|}{DA Approaches} &
    \multicolumn{10}{c|}{DG Approaches} &
    \multicolumn{7}{c|}{} &
    \multicolumn{11}{c|}{Architecture} &
     &
     &
     \\ \cline{4-20} \cline{28-38}
   &
     &
     &
    \multicolumn{4}{c|}{Alignment} &
    \multicolumn{1}{c|}{} &
    \multicolumn{1}{c|}{}&
    &
    \multicolumn{3}{c|}{\shortstack{Representation\\ Learning}} &
    \multicolumn{3}{c|}{\shortstack{Data \\ Manipulation}} &
    \multicolumn{3}{c|}{\shortstack{Learning\\ Scenarios}} &
     \textbf{} &
    \multicolumn{7}{c|}{\multirow{-2}{*}{Dataset}} &
    \multicolumn{8}{c|}{Deep} &
    \multicolumn{3}{c|}{Non-Deep} &
     &
     &
     \\ \cline{4-7} \cline{11-19} \cline{21-38}
  \multirow{-3}{*}{Reference} &
    \multirow{-3}{*}{Year} &
    \multirow{-3}{*}{Code} &
     \textbf{\rotatebox[origin=c]{90}{      Adversarial Feature Alignment      }} &
     \textbf{\rotatebox[origin=c]{90}{Domain Alignment}} &
     \textbf{\rotatebox[origin=c]{90}{Instance Alignment}} &
    \multicolumn{1}{c|}{ \textbf{\rotatebox[origin=c]{90}{Classifier Alignment}}} &
     \multicolumn{1}{c|}{\textbf{\rotatebox[origin=c]{90}{Data Manipulation}}} &
     \multicolumn{1}{c|}{\textbf{\rotatebox[origin=c]{90}{Feature Disentanglement}}} &
     \textbf{\rotatebox[origin=c]{90}{Pseudo-label Training}} &
     
     \textbf{\rotatebox[origin=c]{90}{Adversarial Training}} &
     \textbf{\rotatebox[origin=c]{90}{Feature Weighting}} &
    \multicolumn{1}{c|}{ \textbf{\rotatebox[origin=c]{90}{MS Domain Alignment}}} &
     \textbf{\rotatebox[origin=c]{90}{ Augmentation }} &
     \textbf{\rotatebox[origin=c]{90}{Feature Selection}} &
    \multicolumn{1}{c|}{ \textbf{\rotatebox[origin=c]{90}{Other Preprocessing Methods}}} &
     \textbf{\rotatebox[origin=c]{90}{Ensemble Learning}} &
     \textbf{\rotatebox[origin=c]{90}{Meta Learning}} &
    \multicolumn{1}{c|}{ \textbf{\rotatebox[origin=c]{90}{Self-Supervised Learning}}} &
     \textbf{\rotatebox[origin=c]{90}{Architecture Emb.}} &
     \textbf{\rotatebox[origin=c]{90}{SEED}} &
     \textbf{\rotatebox[origin=c]{90}{DEAP}} &
     \textbf{\rotatebox[origin=c]{90}{SEED IV}} &
     \textbf{\rotatebox[origin=c]{90}{SEED VIG}} &
     \textbf{\rotatebox[origin=c]{90}{DREAMER}} &
     \textbf{\rotatebox[origin=c]{90}{BCI}} &
    \multicolumn{1}{l|}{ \textbf{\rotatebox[origin=c]{90}{Others}}} &
     \textbf{\rotatebox[origin=c]{90}{ANN}} &
     \textbf{\rotatebox[origin=c]{90}{CNN}} &
     \textbf{\rotatebox[origin=c]{90}{RNN}} &
     \textbf{\rotatebox[origin=c]{90}{Attention}} &
     \textbf{\rotatebox[origin=c]{90}{Graph}} &
     \textbf{\rotatebox[origin=c]{90}{GAN}} &
     \textbf{\rotatebox[origin=c]{90}{Auto-Encoder}} &
    \multicolumn{1}{c|}{ \textbf{\rotatebox[origin=c]{90}{Capsule Net}}} &
     \textbf{\rotatebox[origin=c]{90}{SVM}} &
     \textbf{\rotatebox[origin=c]{90}{Tensor Net}} &
    \multicolumn{1}{l|}{ \textbf{\rotatebox[origin=c]{90}{Others}}} &
    \multirow{-3}{*}{\shortstack{Multi/Single\\ Source}} &
    \multirow{-3}{*}{Task Categories} &
    \multirow{-3}{*}{Domain (Cross-X)} \\ \hline
  \cite{Paper_Cui_RP13} &
    2022 &
    {\color[HTML]{0000EE} \href{https://github.com/cuijiancorbin/EEG-based-CrossSubject-Driver-Drowsiness-Recognition-with-anInterpretable-CNN}{link}} &
     &
     &
     &
     &
     &
     &
     &
     &
     &
     &
     &
     &
     &
     &
     &
     &
    $\times$ &
     &
     &
     &
     &
     &
     &
    $\times$ &
     &
    $\times$ &
     &
     &
     &
     &
     &
     &
     &
     &
     &
    SS &
    AM &
    Subject \\
  \rowcolor[HTML]{ECF4FF}
  \cite{Paper_Bethge_5} &
    2022 &
    - &
     &
     &
     &
     &
     &
     &
     &
     &
     &
    $\times$ &
     &
     &
     &
     &
     &
     &
     &
    $\times$ &
    $\times$ &
    $\times$ &
     &
    $\times$ &
     &
     &
    $\times$ &
     &
     &
     &
     &
     &
     &
     &
     &
     &
    \cellcolor[HTML]{ECF4FF} &
    MS &
    ER &
    Subject \\
  \cite{Paper_Jana_RP14} &
    2022 &
    - &
     &
     &
     &
      &
     &
      &
      &
     &
     &
     &
     &
     &
     &
     &
     &
     &
    $\times$ &
    \multicolumn{1}{l}{} &
    $\times$ &
    \multicolumn{1}{l}{} &
    \multicolumn{1}{l}{} &
    \multicolumn{1}{l}{} &
    \multicolumn{1}{l}{} &
    \multicolumn{1}{l|}{} &
     &
    $\times$ &
     &
     &
     &
     &
     &
     &
     &
     &
     &
    SS &
    AM &
    Subject \\
  \rowcolor[HTML]{ECF4FF}
  \cite{Paper_Xia_73} &
    2022 &
    {\color[HTML]{0000EE} \href{https://github.com/xkazm/ASFA}{link}} &
     &
     &
     &
    $\times$ &
     &
      &
      &
     &
     &
     &
    $\times$ &
     &
     &
     &
     &
    $\times$ &
     &
     &
     &
     &
     &
     &
     &
     &
    $\times$ &
     &
     &
     &
     &
     &
     &
     &
     &
     &
    \cellcolor[HTML]{ECF4FF} &
    SS &
    MI &
    Subject \\
  \cite{Paper_Joshi_RP36} &
    2022 &
    - &
     &
     &
     &
      &
     &
      &
      &
     &
     &
     &
     &
     &
     &
     &
     &
     &
    $\times$ &
    $\times$ &
    $\times$ &
     &
     &
     &
     &
    $\times$ &
     &
     &
    $\times$ &
     &
     &
     &
     &
     &
     &
     &
     &
    SS &
    ER &
    Dataset \\
  \rowcolor[HTML]{ECF4FF}
  \cite{Paper_Lee_6} &
    2022 &
    {\color[HTML]{0000EE} \href{https://github.com/DeepBCI/Deep-BCI/tree/master/1_Intelligent_BCI/Inter_Subject_Contrastive_Learning_for_EEG}{link}} &
     &
     &
     $\times$ &
      &
     &
      &
      &
     &
     &
     &
     &
     &
     &
     &
     &
     &
     &
     &
     &
     &
     &
     &
     &
    $\times$ &
    $\times$ &
     &
     &
     &
     &
     &
     &
     &
     &
     &
    \cellcolor[HTML]{ECF4FF} &
    MS &
    VPA &
    Subject \\
  \cite{Paper_Jiang_RP33} &
    2022 &
    - &
     &
    $\times$ &
     &
      &
     &
      &
    $\times$ &
     &
     &
     &
     &
     &
     &
     &
     &
     &
     &
     &
     &
     &
     &
     &
    $\times$ &
     &
    \multicolumn{1}{l}{} &
    \multicolumn{1}{l}{} &
    \multicolumn{1}{l}{} &
    \multicolumn{1}{l}{} &
    \multicolumn{1}{l}{} &
    \multicolumn{1}{l}{} &
    \multicolumn{1}{l}{} &
    \multicolumn{1}{l}{} &
    \multicolumn{1}{l}{} &
    \multicolumn{1}{l}{} &
    \multicolumn{1}{l|}{$\times$} &
    SS &
    MI &
    Subject \\
  \rowcolor[HTML]{ECF4FF}
  \cite{Paper_Dar_RP35} &
    2022 &
    - &
     &
     &
     &
      &
     &
      &
      &
     &
     &
     &
     &
     &
     &
     &
     &
     &
    $\times$ &
     &
     &
    $\times$ &
     &
     &
     &
    $\times$ &
    \multicolumn{1}{l}{\cellcolor[HTML]{ECF4FF}} &
    \multicolumn{1}{l}{\cellcolor[HTML]{ECF4FF}$\times$} &
    \multicolumn{1}{l}{\cellcolor[HTML]{ECF4FF}$\times$} &
    \multicolumn{1}{l}{\cellcolor[HTML]{ECF4FF}} &
    \multicolumn{1}{l}{\cellcolor[HTML]{ECF4FF}} &
    \multicolumn{1}{l}{\cellcolor[HTML]{ECF4FF}} &
    \multicolumn{1}{l}{\cellcolor[HTML]{ECF4FF}} &
    \multicolumn{1}{l}{\cellcolor[HTML]{ECF4FF}} &
    \multicolumn{1}{l}{\cellcolor[HTML]{ECF4FF}} &
    \multicolumn{1}{l}{\cellcolor[HTML]{ECF4FF}} &
    \multicolumn{1}{l|}{\cellcolor[HTML]{ECF4FF}$\times$} &
    MS &
    ER &
    Dataset \\
  \cite{Paper_Peng_9} &
    2022 &
    - &
    $\times$ &
     &
     &
      &
     &
      &
      &
     &
     &
    $\times$ &
     &
     &
     &
     &
     &
     &
     &
     &
     &
     &
     &
     &
     &
    $\times$ &
    \multicolumn{1}{l}{$\times$} &
    \multicolumn{1}{l}{} &
    \multicolumn{1}{l}{} &
    \multicolumn{1}{l}{} &
    \multicolumn{1}{l}{} &
    \multicolumn{1}{l}{} &
    \multicolumn{1}{l}{$\times$} &
    \multicolumn{1}{l}{} &
    \multicolumn{1}{l}{} &
    \multicolumn{1}{l}{} &
    \multicolumn{1}{l|}{} &
    MS &
    SA &
    Subject \\
  \rowcolor[HTML]{ECF4FF}
  \cite{Paper_Liu_48} &
    2022 &
    - &
     &
     &
     &
     &
     &
    $\times$ &
     &
     &
     &
     &
     &
     &
    $\times$ &
     &
     &
     &
     &
     &
     &
     &
     &
     &
     &
    $\times$ &
    \multicolumn{1}{l}{\cellcolor[HTML]{ECF4FF}} &
    \multicolumn{1}{l}{\cellcolor[HTML]{ECF4FF}} &
    \multicolumn{1}{l}{\cellcolor[HTML]{ECF4FF}} &
    \multicolumn{1}{l}{\cellcolor[HTML]{ECF4FF}} &
    \multicolumn{1}{l}{\cellcolor[HTML]{ECF4FF}} &
    \multicolumn{1}{l}{\cellcolor[HTML]{ECF4FF}} &
    \multicolumn{1}{l}{\cellcolor[HTML]{ECF4FF}$\times$} &
    \multicolumn{1}{l}{\cellcolor[HTML]{ECF4FF}} &
    \multicolumn{1}{l}{\cellcolor[HTML]{ECF4FF}} &
    \multicolumn{1}{l}{\cellcolor[HTML]{ECF4FF}} &
    \multicolumn{1}{l|}{\cellcolor[HTML]{ECF4FF}} &
    MS &
    MSD &
    Dataset \\
  \cite{Paper_Bethge_67} &
    2022 &
    {\color[HTML]{0000EE} \href{https://github.com/philipph77/acse-framework}{link}} &
     &
     &
     &
      &
     &
      &
      &
    $\times$ &
     &
     &
     &
     &
     &
     &
     &
     &
     &
    $\times$ &
    $\times$ &
    $\times$ &
     &
    $\times$ &
     &
     &
    $\times$ &
    $\times$ &
     &
     &
     &
     &
     &
     &
     &
     &
     &
    MS &
    MSD &
    Dataset \\
  \rowcolor[HTML]{ECF4FF}
  \cite{Paper_Haung_71} &
    2022 &
    - &
    $\times$ &
     &
     &
      &
     &
      &
      &
     &
     &
     &
     &
     &
     &
     &
     &
     &
     &
     &
    $\times$ &
     &
     &
     &
     &
     &
     &
     &
     &
     &
     &
    $\times$ &
     &
     &
     &
     &
    \cellcolor[HTML]{ECF4FF} &
    SS &
    ER &
    Subject \\
  \cite{Paper_Li_66} &
    2022 &
    - &
     &
     &
     &
      &
     &
      &
      &
     &
     &
     &
     &
     &
     &
    $\times$ &
     &
     &
    $\times$  &
     &
     &
     &
     &
     &
     &
    $\times$ &
     &
    $\times$ &
     &
     &
     &
     &
     &
     &
     &
     &
     &
    SS &
    AM &
    Subject \\
  \rowcolor[HTML]{ECF4FF}
  \cite{Paper_Meng_RP20} &
    2022 &
    - &
     &
    $\times$ &
     &
      &
     &
      &
    $\times$ &
     &
     &
     &
     &
     &
     &
     &
     &
     &
     &
    $\times$ &
     &
     &
     &
     &
     &
     &
     &
    $\times$ &
     &
     &
     &
     &
     &
     &
     &
     &
    \cellcolor[HTML]{ECF4FF} &
    SS &
    ER &
    Subject/Session \\
  \cite{Paper_Qu_RP29} &
    2022 &
     - &
     &
    $\times$ &
     &
     &
     &
     &
     &
     &
     &
     &
     &
     &
     &
     &
     &
     &
     &
     &
     &
     &
     &
     &
     &
    $\times$ &
     &
     &
     &
     &
     &
     &
     &
     &
    $\times$ &
     &
     &
    MS &
    MSD &
    Session \\
    \rowcolor[HTML]{ECF4FF}
  \cite{Paper_Lemkhenter_RP31} &
    2022 &
    - &
     &
     &
     &
      &
     &
      &
      &
     &
     &
     &
     &
     &
     &
     &
    $\times$ &
    $\times$ &
     &
     &
     &
     &
     &
     &
     &
    $\times$ &
     &
    $\times$ &
     &
     &
     &
     &
     &
     &
     &
     &
     &
    SS &
    SD &
    Subject/Dataset \\
  \cite{Paper_Heremans_RP43} &
    2022 &
    - &
    $\times$ &
     &
     &
      &
     &
      &
    $\times$ &
     &
     &
     &
     &
     &
     &
     &
     &
     &
     &
     &
     &
     &
     &
     &
     &
    $\times$ &
     &
     &
    $\times$ &
     &
     &
     &
     &
     &
     &
     &
     &
    SS &
    SD &
    Dataset \\
    \rowcolor[HTML]{ECF4FF}
  \cite{Paper_Musellim_7} &
    2022 &
    - &
     &
     &
     &
     &
     &
     &
     &
     &
     &
    $\times$ &
     &
     &
     &
     &
     &
     &
     &
     &
     &
     &
     &
     &
     &
    $\times$ &
     &
    $\times$ &
     &
     &
     &
     &
     &
     &
     &
     &
     &
    SS &
    MI &
    Subject \\
  \cite{Paper_Avramidis_46} &
    2022 &
    {\color[HTML]{0000EE} \href{https://github.com/klean2050/EEG_CrossModal}{link}} &
    $\times$ &
     &
     &
      &
     &
      &
      &
     &
     &
     &
     &
     &
     &
     &
     &
     &
     &
     &
    $\times$ &
     &
     &
     &
     &
     &
     &
     &
    $\times$ &
     &
     &
     &
     &
     &
     &
     &
     &
    SS &
    ER &
    Subject \\
    \rowcolor[HTML]{ECF4FF}
  \cite{Paper_Zhou_65} &
    2022 &
    - &
     &
    $\times$ &
     &
      &
     &
      &
    $\times$  &
     &
     &
     &
     &
     &
     &
     &
     &
     &
     &
     &
     &
     &
     &
     &
     &
    $\times$ &
     &
     &
     &
     &
     &
     &
     &
     &
    $\times$ &
     &
     &
    SS &
    MSD &
    Task/Subject \\
  \cite{Paper_Zhu_69} &
    2022 &
    - &
    $\times$ &
     &
     &
      &
     &
      &
      &
     &
     &
     &
     &
     &
     &
     &
     &
     &
     &
    $\times$ &
     &
     &
     &
     &
     &
     &
     &
    $\times$ &
     &
     &
     &
     &
     &
     &
     &
     &
     &
    SS &
    ER &
    Session \\
    \rowcolor[HTML]{ECF4FF}
  \cite{Paper_He_70} &
    2022 &
    - &
     &
    $\times$ &
     &
      &
     &
      &
      &
     &
     &
     &
     &
     &
     &
     &
     &
     &
     &
    $\times$ &
    $\times$ &
     &
     &
     &
     &
    $\times$ &
     &
     &
     &
     &
     &
     &
     &
     &
    $\times$ &
     &
     &
    SS &
    ER &
    Day \\
  \cite{Paper_Zhao_92} &
    2022 &
    - &
     &
     &
     &
      &
     &
      &
      &
    $\times$ &
     &
     &
     &
     &
     &
     &
     &
     &
     &
     &
     &
     &
     &
     &
     &
    $\times$ &
    $\times$ &
     &
     &
     &
     &
    $\times$ &
     &
     &
     &
     &
     &
    SS &
    SA &
    Subject \\
    \rowcolor[HTML]{ECF4FF}
  \cite{Paper_Fdez_RP5} &
    2021 &
    - &
     &
     &
     &
      &
     &
      &
      &
     &
     &
     &
     &
     &
    $\times$ &
     &
     &
     &
     &
    $\times$ &
     &
     &
     &
     &
     &
     &
    $\times$ &
     &
     &
     &
     &
     &
     &
     &
     &
     &
     &
    MS &
    ER &
    Subject \\
  \cite{Paper_Jia_12} &
    2021 &
    {\color[HTML]{0000EE} \href{https://github.com/ziyujia/mstgcn}{link}} &
     &
     &
     &
      &
     &
      &
      &
    $\times$ &
     &
     &
     &
     &
     &
     &
     &
     &
     &
     &
     &
     &
     &
     &
     &
    $\times$ &
     &
     &
     &
    $\times$ &
    $\times$ &
     &
     &
     &
     &
     &
     &
    SS &
    SD &
    Subject \\
    \rowcolor[HTML]{ECF4FF}
  \cite{Paper_Wang_87} &
    2021 &
    - &
    $\times$ &
     &
     &
      &
     &
      &
      &
     &
     &
     &
     &
     &
     &
     &
     &
     &
     &
     &
    \color[HTML]{FE0000}$\times$ &
     &
     &
    \color[HTML]{FE0000}$\times$ &
     &
     &
     &
    $\times$ &
     &
     &
     &
    $\times$ &
     &
     &
     &
     &
     &
    MS &
    ER &
    Subject \\
  \cite{Paper_Dissanayake_62} &
    2021 &
    - &
     &
     &
     &
      &
     &
      &
      &
     &
     &
     &
     &
     &
     &
     &
     &
     &
    $\times$ &
     &
     &
     &
     &
     &
     &
    $\times$ &
     &
    $\times$ &
     &
     &
     &
     &
     &
     &
     &
     &
     &
    SS &
    SA &
    Subject \\
    \rowcolor[HTML]{ECF4FF}
  \cite{Paper_Zhao_10} &
    2021 &
    - &
     &
     &
    $\times$ &
      &
     &
    $\times$ &
      &
     &
     &
     &
     &
     &
     &
    $\times$ &
     &
     &
     &
    $\times$ &
     &
     &
     &
     &
     &
     &
     &
     &
    $\times$ &
    $\times$ &
     &
     &
     &
     &
     &
     &
     &
    MS &
    ER &
    Subject \\
  \cite{Paper_Bao_91} &
    2021 &
    - &
    $\times$ &
    $\times$ &
     &
     &
     &
     &
     &
     &
     &
     &
     &
     &
     &
     &
     &
     &
     &
    $\times$ &
     &
     &
     &
     &
     &
     &
     &
    $\times$ &
     &
     &
     &
     &
     &
     &
     &
     &
     &
    MS &
    ER &
    Day/Subject \\
    \rowcolor[HTML]{ECF4FF}
  \cite{Paper_Li_RP24} &
    2021 &
    {\color[HTML]{0000EE} \href{https://github.com/tailofcat/SOGNN}{link}} &
     &
     &
     &
      &
     &
      &
      &
     &
     &
     &
     &
     &
     &
     &
     &
     &
    $\times$ &
    $\times$ &
     &
    \color[HTML]{FE0000}$\times$ &
     &
     &
     &
     &
     &
     &
     &
     &
    $\times$ &
     &
     &
     &
     &
     &
     &
    SS &
    ER &
    Subject \\
  \cite{Paper_JimnezGuarneros_76} &
    2021 &
    - &
     &
    $\times$ &
     &
      &
     &
      &
    $\times$ &
     &
     &
     &
     &
     &
     &
     &
     &
     &
    $\times$ &
     &
     &
     &
     &
     &
     &
    $\times$ &
     &
     &
    $\times$ &
     &
     &
     &
     &
     &
     &
     &
     &
    SS &
    HTA &
    Subject \\
    \rowcolor[HTML]{ECF4FF}
  \cite{Paper_Zeng_RP11} &
    2021 &
    - &
    $\times$ &
     &
     &
      &
     &
      &
      &
     &
     &
     &
     &
     &
     &
     &
     &
     &
     &
     &
     &
     &
     &
     &
     &
    $\times$ &
    $\times$ &
     &
     &
     &
     &
     &
     &
     &
     &
     &
     &
    SS &
    FAM &
    Subject \\
  \cite{Paper_Zhu_RP28} &
    2021 &
    - &
     &
     &
     &
      &
     &
      &
      &
     &
     &
     &
     &
     &
     &
    $\times$ &
     &
     &
     &
     &
     &
     &
     &
     &
     &
    $\times$ &
     &
    $\times$ &
     &
     &
     &
     &
     &
     &
     &
     &
     &
    SS &
    VPA &
    Session \\
    \rowcolor[HTML]{ECF4FF}
  \cite{Paper_Hong_RP38} &
    2021 &
    - &
    $\times$ &
     &
     &
      &
     &
      &
    $\times$ &
     &
     &
     &
     &
     &
     &
     &
     &
     &
     &
     &
     &
     &
     &
     &
    \color[HTML]{0000DD}$\times$ &
     &
     &
    $\times$ &
     &
     &
     &
     &
     &
     &
     &
     &
     &
    SS &
    MI &
    Session \\
  \cite{Paper_Rayatdoost_23} &
    2021 &
    {\color[HTML]{0000EE} \href{https://github.com/philipph77/acse-framework}{link}} &
    $\times$ &
     &
     &
      &
     &
      &
      &
     &
     &
     &
     &
     &
     &
     &
     &
     &
     &
     &
     &
     &
     &
     &
     &
    $\times$ &
     &
    $\times$ &
     &
     &
     &
     &
     &
     &
     &
     &
     &
    SS &
    ER &
    Subject/Dataset \\
    \rowcolor[HTML]{ECF4FF}
  \cite{Paper_Chen_13} &
    2021 &
    {\color[HTML]{0000EE} \href{https://github.com/VoiceBeer/MS-MDA}{link}} &
     &
     $\times$ &
     &
     $\times$ &
     &
      &
      &
     &
     &
     &
     &
     &
     &
     &
     &
     &
     &
    \color[HTML]{FE0000}$\times$ &
     &
    $\times$ &
     &
     &
     &
     &
    $\times$ &
     &
     &
     &
     &
     &
     &
     &
     &
     &
     &
    MS &
    ER &
    Subject/Session \\
  \cite{Paper_Li_14} &
    2021 &
    - &
     &
     &
    $\times$ &
      &
     &
      &
      &
     &
     &
     &
     &
     &
     &
     &
     &
     &
     &
     &
     &
     &
     &
     &
     &
    $\times$ &
     &
    $\times$ &
     &
     &
     &
     &
     &
     &
     &
     &
     &
    SS &
    AM &
    Subject \\
    \rowcolor[HTML]{ECF4FF}
  \cite{Paper_Wang_49} &
    2021 &
    - &
    $\times$ &
     &
    $\times$ &
     &
     &
     &
     &
     &
     &
     &
     &
     &
     &
     &
     &
     &
     &
    $\times$ &
     &
     &
     &
     &
     &
     &
     &
    $\times$ &
     &
     &
     &
     &
     &
     &
     &
     &
     &
    SS &
    ER &
    Subject \\
  \cite{Paper_Tao_15} &
    2021 &
    - &
     &
     &
    $\times$ &
      &
     &
      &
    $\times$ &
     &
     &
     &
     &
     &
     &
     &
     &
     &
     &
    $\times$ &
    $\times$ &
     &
     &
     &
     &
     &
    $\times$ &
     &
     &
     &
     &
     &
     &
     &
     &
     &
     &
    MS &
    ER &
    Subject/Dataset \\
    \rowcolor[HTML]{ECF4FF}
  \cite{Paper_Zhao_52} &
    2021 &
    - &
     &
    $\times$ &
     &
    $\times$ &
     &
      &
      &
     &
     &
     &
     &
     &
     &
     &
     &
     &
     &
     &
     &
     &
     &
     &
     &
    $\times$ &
    $\times$ &
     &
     &
     &
     &
     &
     &
     &
     &
     &
    $\times$ &
    MS &
    MSD &
    Subject \\
  \cite{Paper_Shen_79} &
    2021 &
    {\color[HTML]{0000EE} \href{https://github.com/shenmusmart/MSSA-TN}{link}} &
     &
     &
    $\times$ &
      &
     &
      &
      &
     &
     &
     &
     &
     &
     &
     &
     &
     &
     &
     &
     &
     &
     &
     &
     &
    $\times$ &
     &
     &
     &
     &
     &
     &
     &
     &
     &
    $\times$ &
     &
    MS &
    AM &
    Subject \\
    \rowcolor[HTML]{ECF4FF}
  \cite{Paper_He_20} &
    2021 &
    - &
    $\times$ &
     &
     &
      &
     &
      &
      &
     &
     &
     &
     &
     &
     &
     &
     &
     &
     &
     &
    $\times$ &
     &
     &
    $\times$ &
     &
     &
     &
    $\times$ &
     &
     &
     &
     &
     &
     &
     &
     &
     &
    SS &
    ER &
    Subject/Dataset \\
  \cite{Paper_Liu_RP46} &
    2021 &
    - &
     &
     &
     &
     &
    $\times$ &
     &
     &
     &
     &
     &
     &
     &
     &
     &
     &
     &
     &
     &
    $\times$ &
     &
     &
     &
     &
     &
    $\times$ &
     &
     &
     &
     &
     &
     &
     &
     &
     &
     &
    MS &
    ER &
    Subject \\
    \rowcolor[HTML]{ECF4FF}
  \cite{Paper_Han_16} &
    2021 &
    - &
     &
     &
     &
      &
     &
      &
     &
    $\times$ &
     &
     &
    $\times$ &
     &
     &
     &
     &
     &
     &
     &
     &
     &
     &
     &
    $\times$ &
     &
     &
    $\times$ &
     &
     &
     &
     &
     &
     &
     &
     &
     &
    SS &
    MI &
    Subject/Session \\
  \cite{Paper_Hagad_17} &
    2021 &
    - &
     &
     &
     &
      &
     &
      &
      &
    $\times$ &
     &
     &
     &
     &
     &
     &
     &
     &
     &
    $\times$ &
    $\times$ &
     &
     &
     &
     &
     &
     &
     &
     &
     &
     &
     &
    $\times$ &
     &
     &
     &
     &
    SS &
    ER &
    Subject \\
    \rowcolor[HTML]{ECF4FF}
  \cite{Paper_Ding_51} &
    2021 &
    - &
    $\times$ &
     &
     &
     &
     &
     &
     &
     &
     &
     &
     &
     &
     &
     &
     &
     &
     &
    $\times$ &
     &
     &
     &
     &
     &
     &
    $\times$ &
     &
     & 
     &
     &
     &
     &
     &
     &
     &
     &
     SS &
     ER &
     Subject/Phase 
    \\
  \cite{Paper_Shen_19} &
    2021 &
    - &
     &
     &
     &
      &
     &
      &
      &
     &
     &
     &
     &
     &
     &
     &
     &
    $\times$ &
     &
    $\times$ &
     &
     &
     &
     &
     &
    $\times$ &
     &
    $\times$ &
     &
     &
     &
     &
     &
     &
     &
     &
     &
    MS &
    ER &
    Subject \\
  \rowcolor[HTML]{ECF4FF}
  \cite{Paper_Wang_50} &
    2021 &
    - &
     &
    $\times$ &
     &
      &
     &
      &
    $\times$ &
     &
     &
     &
    $\times$ &
     &
    $\times$ &
     &
     &
     &
     &
     &
     &
    $\times$ &
     &
     &
     &
     &
     &
     &
     &
     &
     &
     &
     &
     &
     &
     &
     &
    SS &
    ER &
    Session \\
  \cite{Paper_Lee_80} &
    2021 &
    {\color[HTML]{0000EE} \href{https://github.com/DeepBCI/Deep-BCI}{link}} &
     &
    $\times$ &
     &
      &
     &
      &
      &
     &
     &
     &
     &
     &
     &
     &
     &
     &
     &
     &
     &
     &
     &
     &
     &
    $\times$ &
     &
     &
    $\times$ &
     &
     &
     &
     &
     &
     &
     &
     &
    MS &
    VPA &
    Subject \\
  \rowcolor[HTML]{ECF4FF}
  \cite{Paper_Kuang_81} &
    2021 &
    - &
     &
    $\times$ &
     &
      &
     &
      &
      &
     &
     &
     &
     &
     &
     &
     &
     &
     &
     &
     &
     &
     &
     &
     &
     &
    $\times$ &
     &
     &
     &
     &
    $\times$ &
     &
     &
     &
     &
     &
    \cellcolor[HTML]{ECF4FF} &
    SS &
    ER &
    Subject/Device \\
  \cite{Paper_Peng_82} &
    2021 &
    - &
    $\times$ &
    $\times$ &
     &
      &
     &
      &
      &
     &
     &
     &
     &
     &
     &
     &
     &
     &
     &
     &
     &
     &
     &
     &
     &
    $\times$ &
     &
     &
     &
     &
     &
     &
     &
     &
    $\times$ &
     &
     &
    SS &
    SA &
    Subject \\
  \rowcolor[HTML]{ECF4FF}
  \cite{Paper_Ning_77} &
    2021 &
    - &
     &
    $\times$ &
     &
      &
     &
      &
      &
     &
     &
     &
     &
     &
     &
     &
     &
     &
     &
    $\times$ &
    $\times$ &
     &
     &
     &
     &
     &
     &
    $\times$ &
     &
    $\times$ &
     &
     &
     &
     &
     &
     &
    \cellcolor[HTML]{ECF4FF} &
    SS &
    ER &
    Subject \\
  \cite{Paper_Ye_85} &
    2021 &
    - &
    $\times$ &
     &
     &
      &
     &
      &
      &
     &
     &
     &
     &
     &
     &
     &
     &
     &
     &
    $\times$ &
     &
     &
     &
     &
     &
     &
     &
     &
     &
    $\times$ &
    $\times$ &
     &
     &
     &
     &
     &
     &
    MS &
    ER &
    Subject \\
  \rowcolor[HTML]{ECF4FF}
  \cite{Paper_Cai_RP26} &
    2021 &
    - &
    $\times$ &
     &
     &
      &
     &
      &
      &
     &
     &
     &
     &
     &
     &
     &
     &
     &
     &
    $\times$ &
     &
     &
     &
     &
     &
     &
     &
     &
     &
     &
     &
    $\times$ &
     &
     &
     &
     &
    \cellcolor[HTML]{ECF4FF} &
    SS &
    ER &
    Subject/Session \\
  \cite{Paper_Lee_84} &
    2021 &
    - &
     &
     &
     &
      &
     &
      &
      &
     &
     &
     &
     &
     &
     &
     &
     &
     &
    $\times$ &
     &
     &
     &
     &
     &
     &
    $\times$ &
     &
    $\times$ &
     &
     &
     &
     &
     &
     &
     &
     &
     &
    SS &
    ND &
    Subject \\
  \rowcolor[HTML]{ECF4FF}
  \cite{Paper_Wagh_RP45} &
    2021 &
    - &
     &
     &
     &
      &
     &
      &
      &
     &
     &
     &
     &
     &
     &
     &
     &
    $\times$ &
     &
     &
     &
     &
     &
     &
     &
    $\times$ &
     &
    $\times$ &
     &
     &
     &
     &
     &
     &
     &
     &
    \cellcolor[HTML]{ECF4FF} &
    SS &
    \cellcolor[HTML]{ECF4FF}BEE &
    Subject/Session/Dataset \\
  \cite{Paper_Song_21} &
    2021 &
    - &
     &
     &
     &
      &
     &
      &
      &
     &
     &
     &
    $\times$ &
     &
     &
     &
     &
     &
     &
     &
     &
     &
     &
     &
    \color[HTML]{FE0000}$\times$ &
     &
     &
     &
     &
     &
     &
    $\times$ &
     &
     &
     &
     &
     &
    MS &
    MI &
    Subject \\
  \rowcolor[HTML]{ECF4FF}
  \cite{Paper_Qu_24} &
    2021 &
    - &
    $\times$ &
     &
     &
      &
     &
      &
      &
     &
     &
     &
     &
     &
     &
     &
     &
     &
     &
     &
     &
     &
     &
     &
     &
    $\times$ &
     &
     &
    $\times$ &
     &
     &
     &
     &
     &
     &
     &
     &
    MS &
    SD &
    Dataset \\
  \cite{Paper_Eldele_78} &
    2021 &
    {\color[HTML]{0000EE} \href{https://github.com/emadeldeen24/ADAST}{link}} &
    $\times$ &
     &
     &
      &
     &
      &
    $\times$ &
     &
     &
     &
     &
     &
     &
     &
     &
     &
     &
     &
     &
     &
     &
     &
     &
    $\times$ &
     &
    $\times$ &
     &
    $\times$ &
     &
     &
     &
     &
     &
     &
     &
    SS &
    SD &
    Dataset \\
  \rowcolor[HTML]{ECF4FF}
  \cite{Paper_Lin_RP19} &
    2021 &
    - &
     &
     &
    $\times$ &
      &
     &
      &
      &
     &
     &
     &
     &
     &
     &
     &
     &
     &
     &
    $\times$ &
     &
     &
     &
     &
     &
     &
     &
    $\times$ &
     &
     &
     &
     &
     &
     &
     &
     &
     &
    SS &
    ER &
    Subject \\
  \cite{Paper_Shen_11} &
    2021 &
    - &
     &
    $\times$ &
     &
      &
     &
      &
    $\times$ &
     &
     &
     &
     &
     &
     &
     &
     &
     &
     &
     &
     &
    $\times$ &
     &
     &
     &
     &
     &
     &
     &
     &
     &
     &
     &
     &
     &
     &
    $\times$ &
    SS &
    ER &
    Session/Trial \\
  \rowcolor[HTML]{ECF4FF}
  \cite{Paper_Luo_22} &
    2021 &
    - &
     &
     &
     &
      &
     &
      &
      &
     &
     &
     &
     &
     &
     &
     &
    $\times$ &
     &
     &
    $\times$ &
     &
     &
     &
     &
     &
     &
     &
     &
    $\times$ &
     &
     &
     &
     &
     &
     &
     &
     &
    SS &
    ER &
    Subject \\
  \cite{Paper_Luo_53} &
    2021 &
    - &
    $\times$ &
     &
     &
      &
     &
      &
      &
     &
     &
     &
     &
     &
     &
     &
     &
     &
     &
    $\times$ &
     &
     &
    \color[HTML]{FE0000}$\times$ &
     &
     &
     &
     &
     &
     &
     &
     &
     &
     &
     &
     &
     &
     &
    SS &
    AM &
    Subject \\
  \rowcolor[HTML]{ECF4FF}
  \cite{Paper_Chen_90} &
    2021 &
    - &
     &
    $\times$ &
     &
      &
     &
      &
      &
     &
     &
     &
     &
     &
    $\times$ &
     &
     &
     &
     &
     &
     &
     &
     &
     &
     &
    $\times$ &
     &
    $\times$ &
     &
    $\times$ &
     &
     &
     &
     &
     &
     &
     &
    SS &
    MSD &
    Subject \\
  \cite{Paper_zdenizci_RP16} &
    2020 &
    - &
     &
     &
     &
      &
     &
      &
      &
    $\times$ &
     &
     &
     &
     &
     &
     &
     &
     &
     &
     &
     &
     &
     &
     &
     &
    $\times$ &
     &
    $\times$ &
     &
     &
     &
     &
     &
     &
     &
     &
     &
    SS &
    MI &
    Subject \\
  \rowcolor[HTML]{ECF4FF}
  \cite{Paper_Cheng_RP8} &
    2020 &
    - &
     &
     &
     &
      &
     &
      &
      &
     &
     &
     &
    $\times$ &
     &
     &
     &
     &
    $\times$ &
     &
     &
     &
     &
     &
     &
     &
    $\times$ &
     &
     &
     &
     &
     &
     &
    $\times$ &
     &
     &
     &
    $\times$ &
    SS &
    \cellcolor[HTML]{ECF4FF}ND &
    Subject \\
  \cite{Paper_Roots_RP6} &
    2020 &
    - &
     &
     &
     &
      &
     &
      &
      &
     &
     &
     &
     &
     &
     &
    $\times$ &
     &
     &
     &
     &
     &
     &
     &
     &
     &
    $\times$ &
     &
    $\times$ &
     &
     &
     &
     &
     &
     &
     &
     &
     &
    SS &
    MI &
    Subject \\
  \rowcolor[HTML]{ECF4FF}
  \cite{Paper_Zhao_26} &
    2020 &
    - &
    $\times$ &
     &
     &
      &
     &
      &
    $\times$ &
     &
     &
     &
     &
     &
     &
     &
     &
     &
     &
     &
     &
     &
     &
     &
    $\times$ &
    $\times$ &
     &
    $\times$ &
     &
     &
     &
     &
     &
     &
     &
     &
     &
    SS &
    MI &
    Subject \\
  \cite{Paper_Tang_63} &
    2020 &
    - &
    $\times$ &
     &
     &
      &
     &
      &
      &
     &
     &
     &
     &
     &
     &
     &
     &
     &
     &
     &
     &
     &
     &
     &
     &
    $\times$ &
     &
    $\times$ &
     &
     &
     &
     &
     &
     &
     &
     &
     &
    SS &
    MI &
    Subject \\
  \rowcolor[HTML]{ECF4FF}
  \cite{Paper_Duan_RP10} &
    2020 &
    - &
     &
     &
     &
      &
     &
      &
      &
     &
     &
     &
     &
     &
     &
     &
    $\times$ &
     &
     &
    $\times$ &
    $\times$ &
     &
     &
     &
    $\times$ &
     &
     &
     &
     &
     &
     &
     &
     &
     &
     &
     &
     &
    MS &
    MI &
    Subject \\
  \cite{Paper_Ayodele_60} &
    2020 &
    - &
     &
     &
     &
      &
     &
      &
      &
     &
     &
    $\times$ &
     &
     &
     &
     &
     &
     &
     &
     &
     &
     &
     &
     &
     &
    $\times$ &
     &
    $\times$ &
    $\times$ &
     &
     &
     &
     &
     &
     &
     &
     &
    MS &
    SA &
    Dataset \\
    \rowcolor[HTML]{ECF4FF}
  \cite{Paper_Shen_RP32} &
    2020 &
    - &
     &
    $\times$ &
     &
      &
     &
      &
      &
     &
     &
     &
     &
     &
    $\times$ &
     &
     &
     &
     &
     &
     &
     &
     &
     &
     &
    $\times$ &
     &
     &
     &
     &
     &
     &
     &
     &
     &
    $\times$ &
     &
    SS &
    AM &
    Session \\
  \cite{Paper_Lebedeva_RP44} &
    2020 &
    - &
    $\times$ &
     &
     &
     &
     &
     &
     &
     &
     &
     &
     &
     &
     &
     &
     &
     &
     &
     &
    $\times$ &
     &
     &
     &
     &
     &
     &
     &
     &
     &
     &
     &
     &
     &
    $\times$ &
     &
    $\times$ &
    SS &
    ER &
    Subject/Session/Dataset \\
    \rowcolor[HTML]{ECF4FF}
  \cite{Paper_Li_30} &
    2019 &
    - &
    $\times$ &
     &
     &
      &
     &
      &
      &
     &
     &
     &
     &
     &
     &
     &
     &
     &
     &
     $\times$ &
     $\times$ &
     &
     &
     &
     &
     &
    $\times$ &
     &
     &
     &
     &
     &
     &
     &
     &
     &
     &
    SS &
    ER &
    Subject/Session \\
  \cite{Paper_Zhang_31} &
    2019 &
    {\color[HTML]{0000EE} \href{https://github.com/dalinzhang/CRAM}{link}} &
     &
     &
     &
      &
     &
      &
      &
     &
     &
    $\times$ &
     &
     &
     &
     &
     &
     &
     &
     &
     &
     &
     &
     &
    $\times$ &
     &
     &
    $\times$ &
     &
    $\times$ &
     &
     &
     &
     &
     &
     &
     &
    SS &
    MI &
    Subject \\
    \rowcolor[HTML]{ECF4FF}
  \cite{Paper_Albuquerque_33} &
    2019 &
    {\color[HTML]{0000EE} \href{https://github.com/belaalb/G2DM}{link}} &
     &
     &
     &
      &
     &
      &
      &
    $\times$ &
     &
     &
     &
     &
     &
     &
     &
     &
     &
    $\times$ &
     &
     &
     &
     &
     &
     &
    \multicolumn{1}{l}{} &
    \multicolumn{1}{l}{} &
    \multicolumn{1}{l}{} &
    \multicolumn{1}{l}{} &
    \multicolumn{1}{l}{} &
    \multicolumn{1}{l}{} &
    \multicolumn{1}{l}{} &
    \multicolumn{1}{l}{} &
    \multicolumn{1}{l}{} &
    \multicolumn{1}{l}{} &
    \multicolumn{1}{l|}{} &
    MS &
    ER &
    Subject \\
  \cite{Paper_Li_55} &
    2019 &
    - &
    $\times$ &
     &
     &
      &
     &
      &
      &
     &
     &
     &
     &
     &
     &
     &
     &
     &
     &
    $\times$ &
     &
    \color[HTML]{0000DD}$\times$ &
     &
     &
     &
    $\times$ &
     &
     &
    $\times$ &
     &
     &
     &
     &
     &
     &
     &
     &
    SS &
    ER &
    Subject \\
  \rowcolor[HTML]{ECF4FF}
  \cite{Paper_Yang_RP4} &
    2019 &
    - &
     &
     &
     &
     &
     &
     &
     &
     &
     &
     &
     &
    $\times$ &
     &
     &
     &
     &
     &
    $\times$ &
    $\times$ &
     &
     &
     &
     &
     &
     &
     &
     &
     &
     &
     &
     &
     &
    $\times$ &
     &
     &
    SS &
    ER &
    Subject \\
  \cite{Paper_Zhang_56} &
    2019 &
    {\color[HTML]{0000EE} \href{https://github.com/chamwen/MEKT}{link}} &
     &
    $\times$ &
     &
      &
     &
      &
    $\times$ &
     &
     &
     &
     &
     &
     &
     &
     &
     &
     &
     &
     &
     &
     &
     &
     &
    $\times$ &
     &
     &
     &
     &
     &
     &
     &
     &
     &
     &
     &
    SS &
    MI &
    Subject \\
  \rowcolor[HTML]{ECF4FF}
  \cite{Paper_Cui_34} &
    2019 &
    - &
     &
     &
     &
      &
     &
      &
      &
     &
     $\times$ &
     &
     &
     &
     &
     &
     &
     &
     &
     &
     &
     &
     &
     &
     &
    $\times$ &
    $\times$ &
     &
     &
     &
     &
     &
     &
     &
     &
     &
     &
    MS &
    AM &
    Subject \\
  \cite{Paper_Hang_RP2} &
    2019 &
    - &
     &
    $\times$ &
     &
      &
     &
      &
      &
     &
     &
     &
     &
     &
     &
     &
     &
     &
     &
     &
     &
     &
     &
     &
    $\times$ &
     &
     &
    $\times$ &
     &
     &
     &
     &
     &
     &
     &
     &
     &
    SS &
    MI &
    Subject \\
    \rowcolor[HTML]{ECF4FF}
  \cite{Paper_Ma_35} &
    2019 &
    - &
     &
     &
     &
      &
     &
      &
      &
    $\times$ &
     &
     &
     &
     &
     &
     &
     &
     &
     &
    $\times$ &
     &
     &
    $\times$ &
     &
     &
     &
    $\times$ &
     &
     &
     &
     &
     &
     &
     &
     &
     &
     &
    MS &
    ER &
    Subject \\
  \cite{Paper_Liu_RP9} &
    2019 &
    - &
     &
    $\times$ &
     &
      &
     &
      &
      &
     &
     &
     &
     &
     &
     &
     &
     &
     &
     &
     &
     &
     &
     &
     &
     &
    $\times$ &
     &
    $\times$ &
     &
     &
     &
     &
     &
     &
     &
     &
     &
    SS &
    AM &
    Subject \\
  \rowcolor[HTML]{ECF4FF}
  \cite{Paper_Ming_RP22} &
    2019 &
    - &
     &
     &
     &
      &
     &
      &
      &
    $\times$ &
     &
     &
     &
     &
     &
     &
     &
     &
     &
     &
     &
     &
     &
     &
     &
    $\times$ &
     &
     &
     &
     &
     &
    $\times$ &
     &
     &
     &
     &
     &
    SS &
    AM &
    Subject/Session \\
  \cite{Paper_Jeon_39} &
    2019 &
    {\color[HTML]{0000EE} \href{https://github.com/eunjin93/SICR_BCI}{link}} &
     &
     &
     &
      &
     &
    $\times$ &
      &
     &
     &
     &
     &
     &
     &
     &
     &
     &
     &
     &
     &
     &
     &
     &
     &
    $\times$ &
    $\times$ &
     &
     &
     &
     &
     &
     &
     &
     &
     &
     &
    SS &
    MI &
    Subject \\
  \rowcolor[HTML]{ECF4FF}
  \cite{Paper_Albuquerque_40} &
    2019 &
    - &
     &
     &
     &
     &
    $\times$ &
     &
     &
     &
     &
     &
     &
     &
     &
     &
     &
     &
     &
     &
     &
     &
     &
     &
     &
    $\times$ &
     &
     &
     &
     &
     &
     &
     &
     &
     &
     &
     &
    MS &
    MSD &
    Subject \\
  \cite{Paper_Chambon_41} &
    2018 &
    - &
     &
     &
    $\times$ &
      &
     &
      &
      &
     &
     &
     &
     &
     &
     &
     &
     &
     &
     &
     &
     &
     &
     &
     &
     &
    $\times$ &
     &
    $\times$ &
     &
     &
     &
     &
     &
     &
     &
     &
     &
    SS &
    SD &
    Dataset \\
  \rowcolor[HTML]{ECF4FF}
  \cite{Paper_Wang_42} &
    2017 &
    - &
     &
     &
     &
      &
     &
      &
      &
     &
     &
    $\times$ &
     &
     &
     &
     &
     &
     &
     &
     &
     &
     &
     &
     &
     &
    $\times$ &
     &
     &
     &
     &
     &
     &
     &
     &
    $\times$ &
     &
    $\times$ &
    SS &
    SA &
    Subject \\
  \cite{Paper_Chai_43} &
    2016 &
    - &
     &
    $\times$ &
     &
      &
     &
      &
      &
     &
     &
     &
     &
     &
     &
     &
     &
     &
     &
    $\times$ &
     &
     &
     &
     &
     &
     &
    \multicolumn{1}{l}{} &
    \multicolumn{1}{l}{} &
    \multicolumn{1}{l}{} &
    \multicolumn{1}{l}{} &
    \multicolumn{1}{l}{} &
    \multicolumn{1}{l}{} &
    \multicolumn{1}{l}{$\times$} &
    \multicolumn{1}{l}{} &
    \multicolumn{1}{l}{} &
    \multicolumn{1}{l}{} &
    \multicolumn{1}{l|}{} &
    SS &
    ER &
    Subject/Session \\ \hline
  \end{tabular}
  }%
  }
  \end{table*}

  \begin{table*}[]
  \centering
  \caption{Summary of papers related to Domain Adaptation and Generalization methods on fMRI modality. Note that ASD-D stands for Autism Spectrum Disorder Diagnosis, VPA for Visual Perception Analysis, AD for Anomaly Detection, HTA for Human Thought Analysis, AD-D for Alzheimer’s Disease Diagnosis, ADHD-D for Attention Deficit Hyperactivity Disorder Diagnosis, MSV for Modelling subject variability, WM for Working Memory, and Sc-D for Schizophrenia Diagnosis. The state-of-the-art papers in cross-subject and cross-site settings are shown by red and green crosses in the dataset columns, respectively.}

  \label{fmri-papers-table}
  \resizebox{\textwidth}{!}{%
  \begin{tabular}{|c|c|c|cccccc|cccccc|ccccccc|ccccccccc|c|c|}
  \hline
   &
     &
     &
    \multicolumn{6}{c|}{DA Approaches} &
    \multicolumn{6}{c|}{DG Approaches} &
    \multicolumn{7}{c|}{} &
    \multicolumn{9}{c|}{Architecture} &
     &
     \\ \cline{4-15} \cline{23-31}
   &
     &
     &
    \multicolumn{4}{c|}{Alignment} &
    \multicolumn{1}{l|}{} &
    \multicolumn{1}{l|}{} &
    \multicolumn{2}{c|}{\shortstack{Representation\\ Learning}} &
    \multicolumn{2}{c|}{Preprocessing} &
    \multicolumn{1}{c|}{\shortstack{Learning\\ scenarios}} &
     &
    \multicolumn{7}{c|}{\multirow{-2}{*}{Dataset}} &
    \multicolumn{7}{c|}{Deep} &
    \multicolumn{2}{c|}{Non-Deep} &
     &
     \\ \cline{4-7} \cline{10-14} \cline{16-31}
  \multirow{-3}{*}{Reference} &
    \multirow{-3}{*}{Year} &
    \multirow{-3}{*}{Code} &
    \textbf{\rotatebox[origin=c]{90}{     Adversarial Feature Alignment     }} &
    \textbf{\rotatebox[origin=c]{90}{Domain Alignment}} &
    \textbf{\rotatebox[origin=c]{90}{Instance Alignment}} &
    \multicolumn{1}{l|}{\textbf{\rotatebox[origin=c]{90}{Classifier Alignment}}} &
    \multicolumn{1}{l|}{\textbf{\rotatebox[origin=c]{90}{Pseudo Label Training}}} &
    \multicolumn{1}{l|}{\textbf{\rotatebox[origin=c]{90}{Feature Disentanglement}}} &
    \textbf{\rotatebox[origin=c]{90}{Adversarial Training}} &
    \multicolumn{1}{c|}{\textbf{\rotatebox[origin=c]{90}{MS Domain Alignment}}} &
    \textbf{\rotatebox[origin=c]{90}{Feature Selection}} &
    \multicolumn{1}{c|}{\textbf{\rotatebox[origin=c]{90}{Other Preprocessing Methods}}} &
    \multicolumn{1}{c|}{\textbf{\rotatebox[origin=c]{90}{Meta Learning}}} &
    \multicolumn{1}{l|}{\textbf{\rotatebox[origin=c]{90}{Architecture Emb.}}} &
    \textbf{\rotatebox[origin=c]{90}{ABIDE}} &
    \textbf{\rotatebox[origin=c]{90}{HCP}} &
    \textbf{\rotatebox[origin=c]{90}{ADHD-200}} &
    \textbf{\rotatebox[origin=c]{90}{OpenfMRI}} &
    \textbf{\rotatebox[origin=c]{90}{ADNI}} &
    \textbf{\rotatebox[origin=c]{90}{ABCD}} &
    \multicolumn{1}{l|}{\textbf{\rotatebox[origin=c]{90}{Others}}} &
    \textbf{\rotatebox[origin=c]{90}{ANN}} &
    \textbf{\rotatebox[origin=c]{90}{CNN}} &
    \textbf{\rotatebox[origin=c]{90}{RNN}} &
    \textbf{\rotatebox[origin=c]{90}{Attention}} &
    \textbf{\rotatebox[origin=c]{90}{Graph}} &
    \textbf{\rotatebox[origin=c]{90}{GAN}} &
    \multicolumn{1}{c|}{\textbf{\rotatebox[origin=c]{90}{Auto-Encoder}}} &
    \textbf{\rotatebox[origin=c]{90}{SVM}} &
    \multicolumn{1}{l|}{\textbf{\rotatebox[origin=c]{90}{kNN}}} &
    \multirow{-3}{*}{Task} &
    \multirow{-3}{*}{Domain (Cross-X)} \\ \hline
  \cite{Paper_Wang_fMRI_9} &
    2022 &
    - &
     &
     &
     &
     &
     &
    $\times$ &
     &
     &
     &
     &
     &
     &
    $\times$ &
     &
     &
     &
     &
     &
     &
    $\times$ &
     &
     &
     &
     &
     &
     &
     &
     &
    ASD-D&
    Site \\
  \rowcolor[HTML]{ECF4FF} 
  \cite{Paper_Shi_fMRI_10} &
    2022 &
    - &
     &
    $\times$ &
     &
     &
     &
     &
     &
     &
     &
     &
     &
     &
    $\times$ &
     &
     &
     &
     &
     &
     &
     &
     &
     &
     &
     &
     &
     &
    $\times$ &
    $\times$ &
    ASD-D&
    Site \\
  \cite{Paper_Han_fMRI_12} &
    2022 &
    - &
     &
    $\times$ &
     &
     &
    $\times$ &
     &
     &
     &
     &
     &
     &
     &
    $\times$ &
     &
     &
     &
     &
     &
    $\times$ &
    $\times$ &
     &
     &
     &
     &
     &
     &
     &
     &
    ASD-D&
    Site \\
  \rowcolor[HTML]{ECF4FF} 
  \cite{Paper_Cao_fMRI_13} &
    2021 &
    - &
     &
     &
     &
     &
     &
     &
     &
     &
     &
     &
     &
    $\times$ &
    $\times$ &
     &
     &
     &
     &
     &
     &
     &
     &
     &
     &
    $\times$ &
     &
     &
     &
     &
    ASD-D&
    Site \\
  \cite{Paper_Subah_fMRI_14} &
    2021 &
    - &
     &
     &
     &
     &
     &
     &
     &
     &
    $\times$ &
     &
     &
     &
    \color[HTML]{00FF00}$\times$ &
     &
     &
     &
     &
     &
     &
    $\times$ &
     &
     &
     &
     &
     &
     &
     &
     &
    ASD-D&
    Site \\
  \rowcolor[HTML]{ECF4FF} 
  \cite{Paper_Pominova_fMRI_15} &
    2021 &
    {\color[HTML]{0000EE} \href{https://github.com/kondratevakate/fmri-fader-net}{link}} &
    $\times$ &
     &
     &
     &
     &
     &
     &
     &
     &
     &
     &
     &
    $\times$ &
     &
     &
     &
     &
     &
     &
     &
    $\times$ &
     &
     &
     &
    $\times$ &
    $\times$ &
     &
     &
    ASD-D&
    Site \\
  \cite{Paper_Pan_fMRI_16} &
    2021 &
    {\color[HTML]{0000EE} \href{https://github.com/jhonP-Li/ASD_GP_GCN}{link}} &
     &
     &
     &
     &
     &
     &
     &
     &
    $\times$ &
     &
     &
     &
    $\times$ &
     &
     &
     &
     &
     &
     &
     &
     &
     &
     &
    $\times$ &
     &
     &
     &
     &
    ASD-D&
    Site \\
  \rowcolor[HTML]{ECF4FF} 
  \cite{Paper_Shi_fMRI_17} &
    2021 &
    - &
     &
     &
     &
     &
    $\times$ &
     &
     &
     &
     &
     &
     &
     &
    $\times$ &
     &
     &
     &
     &
     &
     &
     &
     &
     &
     &
     &
     &
     &
    $\times$ &
     &
    ASD-D&
    Site \\
  \cite{Paper_Huang_fMRI_27} &
    2021 &
    - &
     &
     &
     &
     &
     &
     &
     &
    $\times$ &
     &
     &
     &
    \multicolumn{1}{l|}{$\times$} &
     &
     &
     &
     &
     &
     &
    $\times$ &
    $\times$ &
     &
     &
     &
     &
     &
     &
     &
     &
    VPA &
    Subject \\
  \rowcolor[HTML]{ECF4FF} 
  \cite{Paper_Su_fMRI_2} &
    2021 &
    - &
    $\times$ &
     &
     &
     &
     &
     &
     &
     &
     &
     &
     &
     &
     &
    $\times$ &
     &
     &
     &
     &
    $\times$ &
    $\times$ &
     &
     &
     &
     &
     &
     &
     &
     &
    AD &
    Site \\
  \cite{Paper_Jiang_fMRI_3} &
    2021 &
    - &
     &
     &
     &
     &
     &
     &
     &
     &
     &
     &
     &
    $\times$ &
     &
    \color[HTML]{FE0000}$\times$ &
     &
     &
     &
     &
     &
     &
    $\times$ &
     &
    $\times$ &
     &
     &
     &
     &
     &
    HTA &
    Subject/Task/Dataset \\
  \rowcolor[HTML]{ECF4FF} 
  \cite{Paper_Zhang_fMRI_18} &
    2021 &
    - &
     &
     &
     &
     &
     &
     &
     &
    $\times$ &
     &
     &
     &
     &
    $\times$ &
     &
     &
     &
     &
     &
     &
     &
     &
     &
     &
    $\times$ &
     &
     &
     &
     &
    ASD-D&
    Site \\
  \cite{Paper_Lin_fMRI_33} &
    2021 &
    - &
     &
     &
     &
     &
     &
     &
     &
     &
     &
     &
     &
    $\times$ &
     &
     &
     &
     &
    \color[HTML]{FE0000}$\times$ &
     &
     &
     &
    $\times$ &
    $\times$ &
     &
     &
     &
     &
     &
     &
    AD-D &
    Subject \\
    \rowcolor[HTML]{ECF4FF} 
  \cite{Paper_Lee_fMRI_22} &
    2021 &
    - &
     &
     &
     &
     &
     &
     &
     &
     &
     &
     &
    $\times$ &
     &
    $\times$ &
     &
     &
     &
     &
     &
     &
     &
     &
     &
     &
     &
     &
     &
     &
     &
    ASD-D &
    Site \\
  \cite{Paper_Li_fMRI_19} &
    2020 &
    {\color[HTML]{0000EE} \href{https://github.com/xxlya/Fed_ABIDE}{link}} &
    $\times$ &
     &
     &
     &
     &
     &
     &
     &
     &
     &
     &
     &
    $\times$ &
     &
     &
     &
     &
     &
     &
    $\times$ &
     &
     &
     &
     &
     &
     &
     &
     &
    ASD-D&
    Site \\ 
  \rowcolor[HTML]{ECF4FF} 
  \cite{Paper_Wang_fMRI_20} &
    2020 &
    - &
     &
    \multicolumn{1}{l}{} &
    $\times$ &
     &
     &
     &
     &
     &
     &
     &
     &
     &
    $\times$ &
     &
     &
     &
     &
     &
     &
     &
     &
     &
     &
     &
     &
     &
     &
     &
    ASD-D&
    Site \\
  \cite{Paper_Harrison_fMRI_4} &
    2020 &
    {\color[HTML]{0000EE} \href{https://git.fmrib.ox.ac.uk/profumo/PFM_Simulations}{link}} &
     &
     &
     &
     &
     &
     &
     &
     &
     &
     &
     &
    $\times$ &
     &
    $\times$ &
     &
     &
     &
     &
     &
     &
     &
     &
     &
     &
     &
     &
     &
     &
    MSV &
    Subject \\
    \rowcolor[HTML]{ECF4FF}
  \cite{Paper_Zhang_fMRI_24} &
    2020 &
    - &
     &
     &
     &
     &
     &
     &
     &
     &
     &
     &
     &
    $\times$ &
     &
     &
    \color[HTML]{00FF00}$\times$ &
     &
     &
     &
     &
     &
    $\times$ &
     &
    $\times$ &
     &
     &
     &
     &
     &
    ADHD-D &
    Site \\
  \cite{Paper_Zhang_fMRI_21} &
    2020 &
    - &
     &
    $\times$ &
     &
    $\times$ &
     &
     &
     &
     &
     &
     &
     &
     &
    $\times$ &
     &
     &
     &
     &
     &
     &
    $\times$ &
     &
     &
     &
     &
     &
     &
     &
     &
    ASD-D&
    Site \\
  \rowcolor[HTML]{ECF4FF}
  \cite{Paper_Yousefnezhad_fMRI_35} &
    2020 &
    - &
     &
     &
     &
     &
     &
     &
     &
    $\times$ &
     &
     &
     &
     &
     &
     &
     &
     &
     &
     &
    $\times$ &
     &
     &
     &
     &
     &
     &
     &
    $\times$ &
     &
    HTA &
    Site \\
  \cite{Paper_Gao_fMRI_5} &
    2020 &
    - &
     &
    $\times$ &
     &
     &
     &
     &
     &
     &
     &
     &
     &
     &
     &
    $\times$ &
     &
     &
     &
     &
     &
     &
    $\times$ &
     &
     &
     &
     &
     &
     &
     &
    WM / HTA &
    Subject \\ 
  \rowcolor[HTML]{ECF4FF}
  \cite{Paper_Huang_fMRI_25} &
    2020 &
    - &
    $\times$ &
     &
     &
     &
     &
     &
     &
     &
     &
     &
     &
     &
     &
     &
    $\times$ &
     &
     &
     &
     &
    $\times$ &
     &
     &
     &
     &
     &
     &
     &
     &
    ADHD-D &
    Site \\
  \cite{Paper_Li_fMRI_36} &
    2020 &
    - &
     &
     &
     &
     &
     &
     &
     &
    $\times$ &
     &
     &
     &
     &
     &
     &
     &
    \color[HTML]{FE0000}$\times$ &
     &
     &
     &
     &
     &
     &
     &
     &
     &
     &
     &
     &
    HTA &
    Subject \\
  \rowcolor[HTML]{ECF4FF}
  \cite{Paper_Zhou_fMRI_26} &
    2019 &
    {\color[HTML]{0000EE} \href{https://github.com/shuo-zhou/DawfMRI}{link}} &
     &
    $\times$  &
     &
     &
     &
     &
     &
     &
     &
     &
     &
     &
     &
     &
     &
    $\times$ &
     &
     &
     &
    $\times$ &
     &
     &
     &
     &
     &
     &
     &
     &
    HTA &
    Dataset \\
  \cite{Paper_Zeng_fMRI_29} &
    2018 &
    - &
     &
     &
     &
     &
     &
     &
     &
    $\times$ &
     &
     &
     &
    $\times$ &
     &
     &
     &
     &
     &
     &
    $\times$ &
     &
     &
     &
     &
     &
     &
     &
    $\times$ &
     &
    Sc-D &
    Site \\
    \rowcolor[HTML]{ECF4FF}
    \cite{Paper_Bhaumik_fMRI_23} &
    2018 &
    {\color[HTML]{0000EE} \href{https://github.com/shuo-zhou/DawfMRI}{link}} &
     &
     &
     &
     &
     &
     &
     &
     &
    $\times$ &
     &
     &
     &
    $\times$ &
     &
     &
     &
     &
     &
     &
     &
     &
     &
     &
     &
     &
     &
    $\times$ &
     &
    ASD-D&
    Site \\
   \hline
  \end{tabular}%
  }
  \end{table*}


\section{Datasets}

\label{sec:datasets}
In this section, we provide some general information, including attributes and experiment procedures, about the most common datasets used in the papers. Furthermore, we mention some other datasets that have the same data types, which are well-known in other fields.
\subsection{Main EEG Datasets}
This section introduces the most commonly used EEG datasets, which are organized according to their tasks. In Table \ref{eeg-dataset-table}, a summary of these datasets is presented. Also, it is worth mentioning that for the most frequent ones, including SEED, DEAP, SEED-IV, SEED-VIG, BCI Competition IV: Dataset 2a, BCI Competition IV: Dataset 2b, and DREAMER, the state-of-the-art papers are marked in Table \ref{eeg-papers-table}.
\subsubsection{Emotion Recognition (ER)}

\textbf{DEAP}\cite{Dataset_DEAP} is a multi-modal emotion recognition dataset consisting of 32 subjects (16 males and 16 females with ages between 19 and 37, with an average of 26.9). This dataset includes 32-channel EEG signals alongside 13 peripheral physiological signals. Participants rated several music video clips from five different aspects: arousal, valence, like/dislike, dominance, and familiarity.

\textbf{DREAMER}\cite{Dataset_DREAMER} is a multi-modal dataset including EEG and electrocardiogram (ECG) signals from 23 healthy participants. Each subject watched 18 emotional film clips and rated their emotional response based on three concepts: valence, arousal, and dominance. Participants were between 23 and 33 years old (the mean age is 26.6).

\textbf{SEED}\cite{Dataset_SEED} is a dataset consisting of two main sections: SEED\_EEG, which includes EEG data, and SEED\_Multimodal, which consists of EEG and eye movement data. The EEG signals were recorded as 62-channel samples. Generally, there are 15 Chinese subjects in this dataset consisting of seven males and eight females who are 23.27 years old on average. Each subject had been assigned a number from 1 to 15, and only subjects 6, 7, and 15 did not have eye movement data. All subjects underwent three recording sessions with two-week breaks between successive sessions. There were 15 trials per subject in each session, and during each of the trials, 4-minute movie excerpts were used to induce positive, negative, and neutral emotions.

\textbf{SEED-IV}\cite{Dataset_SEEDIV} is an expansion of the original SEED dataset. This dataset has four emotion classes: happy, neutral, sad, and fear. The dataset's subject population is almost identical to the original. Each subject participated in three recording sessions which were held on different days. There were 24 trials per subject in each session, and during each of these trials, the subject watched a 2 minute clip, and their 62-channel EEG signals and eye movement data were collected.

\textbf{THU-EP}\cite{Dataset_THU-EP} is a dataset consisting of 32-channel EEG signals of 80 subjects. This experiment includes 28 video clips made up of 12 clips related to negative feelings (such as anger, disgust, fear, and sadness), 12 clips related to positive emotions (such as amusement, joy, inspiration, and tenderness), and four clips related to neutral feelings. Overall, there are nine emotion classes, and while all other emotions have three clips, neutral has four video clips. These 28 clips were divided into seven 4-clip blocks, and each participant was required to solve 20 arithmetic problems between two successive blocks.

\textbf{MAHNOB-HCI}\cite{Dataset_MAHNOB} is a multi-modal dataset including 32-channel EEG and ECG signals of 30 subjects. Experiments of this dataset had two parts; in the first part, participants watched movie fragments, then they were asked to determine their emotional state after each fragment on a valence and arousal scale. Then, in the second part, movie fragments or images are shown to participants while a tag is displayed at the bottom of the screen. After watching each item, subjects must press a button, green if they agree with the tag; otherwise, they must press the red button.

\subsubsection{Motor Imagery (MI)}

\textbf{BCI Competition IV}\footnote{\href{https://www.bbci.de/competition/iv/}{https://www.bbci.de/competition/iv/}} provides five datasets, two of which are used frequently in the papers: dataset 2a \cite{Dataset_BCI_2a}, and dataset 2b \cite{Dataset_BCI_2b}. In \textbf{dataset 2a}, 22-channel EEG signals were recorded from nine subjects in two sessions on different days. This dataset has four classes, including the imagination of movement of the left hand, right hand, both feet, and tongue. \textbf{Dataset 2b} contains three bipolar-channel EEG signals gathered from 9 subjects with two classes: left hand and right hand.

\textbf{Korea University Dataset(KU)} \cite{Dataset_KU} is a multi-modal dataset comprising 62-channel EEG and 4-channel EMG signals gathered from 54 healthy subjects, 25 females and 29 males, with an age range of 24-35. This dataset has three tasks: ERP, MI, and SSVEP. In ERP, participants had to copy-spell a given sentence; for the MI task, they completed an imagery task of taking the arrow with the right hand, and in SSVEP, four stimuli were presented in four positions, and the subjects were instructed to gaze in the direction of highlighted stimuli. Each of these tasks had both training and test phases. In the training phase, the experimenters recorded EEG data, then used it to build a classifier. During the test phase, they acquired real-time EEG data and decoded it according to the classifier.

\textbf{Gwangju Institute of Science and Technology EEG dataset(GIST)}\cite{Dataset_GIST} is a multi-modal dataset consisting of 62-channel EEG and 2-channel EMG signals acquired from 52 subjects, 19 females and 33 males, 24.8 years old on average. During the MI task in this dataset, the subjects must imagine moving their fingers from the index finger to the little finger. They must imagine a kinesthetic experience rather than a visual experience of movement.

\subsubsection{Sleep Stage Classification (SSC)}
\textbf{Montreal Archive of Sleep Studies(MASS)}\cite{Dataset_MASS}
 includes whole-night recordings of EEG, EOG, EMG, and ECG signals from 200 healthy subjects (97 males and 103 females) with sleep stage labels.
 
 \textbf{ISRUC-Sleep}\cite{Dataset_ISRUC} is a multi-modal dataset containing 6-channel EEG, 2-channel EOG, 3-channel EMG, and 1-channel ECG signals acquired from human adults made up of 3 main groups. The first group consists of 100 subjects, with one recording session for each subject. The second has eight subjects, consisting of two recording sessions for each subject. The third one includes ten healthy subjects who participated in a single recording session. In these experiments, sleep-wake cycles are classified into awake, non-rapid eye movement (NREM), and rapid eye movement (REM) sleep stages. Furthermore, NREM is divided into stages such as N1 (drowsiness sleep), N2 (light sleep), and N3 (deep sleep).

\subsubsection{Situation Awareness Recognition (SAR)}
\textbf{Taiwan Driving Dataset}\cite{Dataset_Taiwan}
is a collection of 32-channel EEG data recorded from 27 subjects who drove on a four-lane highway in a driving simulator. The participant's task was to keep the car moving in the lane, considering that random events induced the car to move out of the lane.

\subsubsection{Seizure Prediction (SP)}
\textbf{CHB-MIT}\cite{Dataset_CHBMIT} is a dataset including EEG signals from subjects with intractable seizures. The dataset consists of 22 subjects (5 males and 17 females) and 969 hours of scalp EEG recordings with 173 seizures of various types (clonic, atonic, and tonic).

\subsubsection{Vigilance Estimation (VE)}
\textbf{SEED-VIG}\cite{Dataset_SEED_VIG} is a multi-modal dataset consisting of 12-channel EEG, 6-channel EEG, 4-channel forehead EEG, and EOG signals of 23 subjects that are made up of 11 males and 12 females with an average age of 23.3 years old. A four-lane highway was shown to subjects controlling the wheel as well as the gas pedal of a vehicle in front of an LCD screen. Some unique situations were also provided for better inducing fatigue. Most experiments were conducted immediately after lunch. The simulated road was straight and monotonous, and the experiment duration was about 2 hours. Finally, from eye tracking data, they acquired labels ranging from 0 for drowsy to 1 for wakeful.

\subsection{Other EEG Datasets}
There are also datasets like Kermanshah University of Medical Sciences Dataset \cite{Dataset_Kermanshah}, WAUC\cite{Dataset_WAUC}, University Hospital Bonn Dataset\cite{Dataset_Bonn}, MPED\cite{Dataset_MPED}, TUSZ\cite{Dataset_TUSZ}, HGD\cite{Dataset_HGD}, Spanish Words Dataset\cite{Dataset_Spanish}, Long Words Dataset\cite{Dataset_Long}, Sleep-EDF\cite{Dataset_Sleep-EDF}, SHHS\cite{Dataset_SHHS}, and Freiburg Hospital Dataset\cite{Dataset_Freiburg} which are not employed as frequently as the main EEG datasets.
\begin{table*}[t]
\centering
\caption{\textbf{Commonest EEG datasets.} Note that for each dataset, information like reference to the paper in which the dataset was first published, number of subjects whose data is available in the dataset, number of all sample trials that are provided in the dataset, number of EEG channels used for signal acquisition, type of task that the dataset is used for, number of different classes covered by the dataset, year in which the dataset was published, availability state of the dataset, and reference to articles that utilize the dataset is provided. Also note that ER stands for Emotion Recognition, SP for Seizure Prediction, SSC for Sleep Stage Classification, SAR for Situation Awareness Recognition, MI for Motor Imagery, and VE for Vigilance Estimation.}
\label{eeg-dataset-table}
\begin{adjustbox}{width=1\textwidth,center=\textwidth}
\begin{tabular}{
|>{\centering\arraybackslash}m{0.2\textwidth}
|>{\centering\arraybackslash}m{0.1\textwidth}
|>{\centering\arraybackslash}m{0.13\textwidth}
|>{\centering\arraybackslash}m{0.1\textwidth}
|>{\centering\arraybackslash}m{0.12\textwidth}
|>{\centering\arraybackslash}m{0.1\textwidth}
|>{\centering\arraybackslash}m{0.1\textwidth}
|>{\centering\arraybackslash}m{0.15\textwidth}
|>{\centering\arraybackslash}m{0.2\textwidth}|
}
\hline

\textbf{Dataset Name} & \textbf{Subject \#} & \textbf{Total Sample Count} & \textbf{Channel \#}  & \textbf{Task} & \textbf{Class \#} & \textbf{Year} & \textbf{Link + Availability} & \textbf{Article References} \\
\hline
SEED \cite{Dataset_SEED} & 15 & 675 & 62 & ER & 3 & 2015 & \href{https://bcmi.sjtu.edu.cn/home/seed/downloads.html#seed-access-anchor}{Access Request} & \cite{Paper_Gu_3} \cite{Paper_Bethge_5} \cite{Paper_Zhao_10} \cite{Paper_Chen_13} \cite{Paper_Tao_15} \cite{Paper_Hagad_17} \cite{Paper_Shen_19} \cite{Paper_Luo_22} \cite{Paper_Li_30} \cite{Paper_Albuquerque_33} \cite{Paper_Ma_35} \cite{Paper_Chai_43} \cite{Paper_Wang_49} \cite{Paper_Ding_51} \cite{Paper_Luo_53} \cite{Paper_Li_55} \cite{Paper_Bethge_67} \cite{Paper_Zhu_69} \cite{Paper_He_70} \cite{Paper_Li_72} \cite{Paper_Ning_77} \cite{Paper_Ye_85} \cite{Paper_Bao_91} \cite{Paper_Yang_RP4} \cite{Paper_Fdez_RP5} \cite{Paper_Duan_RP10} \cite{Paper_Lin_RP19} \cite{Paper_Meng_RP20} \cite{Paper_Li_RP24} \cite{Paper_Cai_RP26} \cite{Paper_Joshi_RP36} \\
\hline
DEAP \cite{Dataset_DEAP} & 32 & 1280 & 32 & ER & 5 & 2012 & \href{https://www.eecs.qmul.ac.uk/mmv/datasets/deap/index.html}{Access Request} & \cite{Paper_Gu_3} \cite{Paper_Bethge_5} \cite{Paper_Tao_15} \cite{Paper_Hagad_17} \cite{Paper_He_20} \cite{Paper_Li_30} \cite{Paper_Avramidis_46} \cite{Paper_Bethge_67} \cite{Paper_He_70} \cite{Paper_Haung_71} \cite{Paper_Ning_77} \cite{Paper_Wang_87} \cite{Paper_Yang_RP4} \cite{Paper_Duan_RP10} \cite{Paper_Jana_RP14} \cite{Paper_Joshi_RP36} \cite{Paper_Lebedeva_RP44} \cite{Paper_Liu_RP46}\\
\hline
DREAMER \cite{Dataset_DREAMER} & 32 & 414 & 14 & ER & 3 & 2017 & \href{https://zenodo.org/record/546113#.Ynq8P5NBxQI}{Access Request} & \cite{Paper_Bethge_5} \cite{Paper_He_20} \cite{Paper_Bethge_67} \cite{Paper_Wang_87}\\
\hline
SEED-IV \cite{Dataset_SEEDIV}  & 15 & 1080 & 62 & ER & 4 & 2019 & \href{https://bcmi.sjtu.edu.cn/home/seed/seed-iv.html}{Access Request} & \cite{Paper_Bethge_5} \cite{Paper_Shen_11} \cite{Paper_Chen_13} \cite{Paper_Wang_50} \cite{Paper_Li_55} \cite{Paper_Bethge_67} \cite{Paper_Li_RP24} \cite{Paper_Dar_RP35}\\
\hline
CHB-MIT \cite{Dataset_CHBMIT} & 22 & 664 & Mostly 23 or 24-26 & SP
 & 4 & 2010 & \href{https://physionet.org/content/chbmit/1.0.0/}{Available} & \cite{Paper_Peng_9} \cite{Paper_Ayodele_60} \cite{Paper_Dissanayake_62} \cite{Paper_Peng_82} \cite{Paper_Zhao_92} \\
\hline
ISRUC-Sleep \cite{Dataset_ISRUC} & 100, 8, 10 & 126 & 6 & SSC & 3 & 2015 & \href{https://sleeptight.isr.uc.pt/?page_id=48}{Available} & \cite{Paper_Jia_12} \cite{Paper_Lemkhenter_RP31}\\
\hline
MASS \cite{Dataset_MASS}& 200 & 200 & 4, 17, 19, 20 & SSC & 5 & 2014 & \href{http://ceams-carsm.ca/en/mass/}{Access Request} & \cite{Paper_Jia_12} \cite{Paper_Qu_24} \cite{Paper_Chambon_41} \cite{Paper_Heremans_RP43}\\
\hline
Taiwan Driving Dataset \cite{Dataset_Taiwan}& 27 & 81576 & 32 & SAR
 & 3 & 2019 & \href{https://doi.org/10.6084/m9.figshare.6427334.v5}{Raw}  \href{https://doi.org/10.6084/m9.figshare.7666055.v3}{Preprocessed} & \cite{Paper_Li_14} \cite{Paper_Li_66} \cite{Paper_Cui_RP13}\\
\hline
BCI Competition IV: dataset 2a \cite{Dataset_BCI_2a} & 9 & 5184 & 22 & MI
 & 4 & 2008 & \href{http://www.bbci.de/competition/iv/#dataset2a}{Available} & \cite{Paper_Han_16} \cite{Paper_Song_21} \cite{Paper_Zhao_26} \cite{Paper_Zhang_31} \cite{Paper_Hang_RP2} \cite{Paper_Duan_RP10} \cite{Paper_Jiang_RP33} \cite{Paper_Hong_RP38}\\
\hline
BCI Competition IV: dataset 2b \cite{Dataset_BCI_2b} & 9 & 6480 & 3 bipolar & MI
 & 2 & 2008 & \href{http://www.bbci.de/competition/iv/#dataset2b}{Available} & \cite{Paper_Zhao_26} \cite{Paper_Hong_RP38}\\
\hline
THU-EP \cite{Dataset_THU-EP}& 80 & 2240 & 32 & ER
 & 9 & 2022 & \href{https://cloud.tsinghua.edu.cn/d/3d176032a5a545c1b927/}{Available} & \cite{Paper_Shen_19}\\
\hline
MAHNOB-HCI \cite{Dataset_MAHNOB}& 27 & 540 & 32 & ER & 2 & 2011 & \href{https://mahnob-db.eu/hci-tagging/}{Access Request} & \cite{Paper_Rayatdoost_23} \cite{Paper_He_70} \\
\hline
SEED-VIG \cite{Dataset_SEED_VIG}& 23 & 885 & 6, 12 & VE & 2 & 2017 & \href{https://bcmi.sjtu.edu.cn/~seed/seed-vig.html}{Access Request} & \cite{Paper_Ma_35} \cite{Paper_Luo_53} \\
\hline
KU \cite{Dataset_KU}& 54 & 5400 & 62 & MI & 2 & 2019 & \href{http://deepbci.korea.ac.kr/opensource/opendb/}{Access Request} & \cite{Paper_Jeon_39} \\
\hline
GIST \cite{Dataset_GIST} & 52 & 5200 $\leq \leq$ 6240 & 62 & MI & 2 & 2017 & \href{http://gigadb.org/dataset/100295}{Available} & \cite{Paper_Jeon_39} \\
\hline
\end{tabular}
\end{adjustbox}
\end{table*}

\subsection{Main fMRI Datasets}

Here are some of the most commonly used fMRI datasets, categorized by their usage in different tasks. An overview of these datasets can be found in Table \ref{fmri-dataset-table}. In addition, the state-of-the-art papers are marked in Table \ref{fmri-papers-table} for the most frequent ones, including ABIDE, HCP, ADNI, ADHD-200, and OpenfMRI: The Ballon analog risk-taking task dataset.

\subsubsection{Autism Spectrum Disorders Identification}
\textbf{Autism Brain Imaging Data Exchange (ABIDE)}\footnote{\href{https://fcon_1000.projects.nitrc.org/indi/abide/}{https://fcon\_1000.projects.nitrc.org/indi/abide/}} contains two main groups: ABIDE I\cite{Dataset_ABIDEI} and ABIDE II\cite{Dataset_ABIDEII}. Each group involves several sites sharing datasets with resting-state functional magnetic resonance imaging (R-fMRI), anatomical, and phenotypic data types. These datasets have the same labels, including Autism Spectrum Disorder (ASD) and Typical Controls (TC); it is worth mentioning that the scanning procedure is almost identical in two groups of ABIDE across different sites. Although it varies a bit in various places, subjects were told to relax and move a bit in most cases. At some sites, subjects are instructed to look at a plain picture; at other sites, subjects are instructed to close their eyes. To recognize each ABIDE group more clearly, it is necessary to consider their differences. While ABIDE I data is gathered from 17 international sites and yields an 1112-member dataset comprising 539 subjects with ASD and 573 subjects with TC problems aged from 7 to 64 years and 14.7 years old as median, ABIDE II data is assembled from 19 sites and yields a 1114-member dataset including 521 subjects with ASD and 593 subjects with TC problems aged between 5 and 64 years old.

\subsubsection{Decoding Cognitive States}
\textbf{Human Connectome Project (HCP)\cite{Dataset_HCP}} is a dataset consisting of 1206 adult subjects who are healthy and aged between 22 and 35 years old from families with and without siblings. This dataset has five data types: structural MRI, R-fMRI, task fMRI (T-fMRI), diffusion MRI (dMRI), and MEG. During the experiments, participants were asked to follow these instructions. Firstly, process different types of information like words, images, voices, or letters. Secondly, utilize various thinking skills like memory, language generation, and decision-making. Finally, respond in various ways, such as shouting the answer or pressing some buttons. Exploration of the heritability of different brain circuits is possible in this dataset because its subjects consist of twins and non-twin siblings. Besides, genotyping of all subjects will provide genome-wide association studies (GWAS) to assess genetic influences on brain circuitry.

\textbf{OpenfMRI}\footnote{\href{https://openfmri.org/}{https://openfmri.org/}} database is a repository for human brain imaging data gathered using MRI and EEG techniques. This database provides several datasets, one of which is used frequently in the papers: Balloon analog risk-taking task dataset\cite{Dataset_OpenfMRIBalloon}. The Ballon analog risk-taking task dataset contains fMRI data from 16 right-handed and healthy subjects. During the experiment, subjects must inflate simulated balloons, and for each successive pump during a particular trial, monetary rewards were assigned to them. The number of trials in this experiment varied for each subject because the task was self-paced. There were 10-minute scanning runs unless the subject ran out of balloons.

\subsubsection{ADHD Classification}
\textbf{ADHD-200}\cite{Dataset_ADHD} is a dataset with the R-fMRI type of data acquired from 8 independent sites and composed of 973 subjects, including 176 participants as the test dataset and 776 participants as the training dataset, which is comprised of 491 ordinary individuals and 285 participants with ADHD aged between 7 and 21 years old. This dataset has some adjoining phenotypic data, including diagnostic status, dimensional ADHD symptom measures, age, sex, intelligence quotient (IQ), and lifetime medication status.

\subsubsection{Alzheimer’s disease (AD) classification}
\textbf{Alzheimer’s Disease Neuroimaging Initiative (ADNI)}\footnote{\href{https://adni.loni.usc.edu/}{https://adni.loni.usc.edu/}} provides two primary datasets: ADNI 1\cite{Dataset_ADNII} and ADNI 2\cite{Dataset_ADNIII}. In ADNI 1, 1.5T T1-weighted structural MRI data were acquired from 819 subjects, including 192 AD, 229 cognitively normal (CN), and 398 MCI participants. However, ADNI 2 added 782 participants to the 819 recruited by ADNI 1. Also, ADNI-2 added a cohort clinically evaluated as cognitively normal but with subjective memory complaints (SMC).

\subsection{Other fMRI Datasets}
There are also datasets like BOLD5000\cite{Dataset_Bold5000}, the Sherlock movie-watching dataset\cite{Dataset_Sherlock}, Auditory oddball task to the FBIRN phase II dataset\cite{Dataset_AuditoryOddball}, Individual Brain Charting (IBC) project\cite{Dataset_IBC}, and some other OpenfMRI datasets such as Probabilistic classification, Deterministic classification, Rhyme judgment, Mixed-gambles task, Stop signal task: Letter classification, Stop signal task: Letter naming, Stop signal task: Pseudoword naming, Simon task, Flanker task, DS105WB, DS105ROI, DS011, DS001, DS232, Raider movie, and Raider image which are not frequently used in papers like primary fMRI datasets.

\begin{table*}[]
    \centering
    \begin{adjustbox}{width=1\textwidth,center=\textwidth}
    \begin{tabular}{
    |>{\centering\arraybackslash}m{0.2\textwidth}
    |>{\centering\arraybackslash}m{0.1\textwidth}
    |>{\centering\arraybackslash}m{0.12\textwidth}
    |>{\centering\arraybackslash}m{0.1\textwidth}
    |>{\centering\arraybackslash}m{0.1\textwidth}
    |>{\centering\arraybackslash}m{0.15\textwidth}
    |>{\centering\arraybackslash}m{0.23\textwidth}
    |>{\centering\arraybackslash}m{0.2\textwidth}|
    }
    \hline
    \textbf{Dataset Name}                                                                    & \textbf{Subject \#} & \textbf{Task} & \textbf{Class \#} & \textbf{Year} & \textbf{Link + Availablity} & \textbf{Extra Information} & \textbf{Article References} \\ \hline
    ABIDE I\cite{Dataset_ABIDEI}                                                                                  & 1112                & ASDI          & 2                 & 2012          & \href{http://fcon_1000.projects.nitrc.org/indi/req_access.htm}{Access Request} & Site \#: 17            & \cite{Paper_Wang_fMRI_9} \cite{Paper_Shi_fMRI_10} \cite{Paper_Han_fMRI_12} \cite{Paper_Cao_fMRI_13} \cite{Paper_Subah_fMRI_14} \cite{Paper_Pan_fMRI_16} \cite{Paper_Shi_fMRI_17} \cite{Paper_Zhang_fMRI_18} \cite{Paper_Li_fMRI_19} \cite{Paper_Wang_fMRI_20} \cite{Paper_Zhang_fMRI_21} \cite{Paper_Lee_fMRI_22} \cite{Paper_Bhaumik_fMRI_23} \\ \hline
    ABIDE II\cite{Dataset_ABIDEII}                                                                                 & 1114                & ASDI          & 2                 & 2017          & \href{http://fcon_1000.projects.nitrc.org/indi/req_access.html}{Access Request}& Site \#: 19              & \cite{Paper_Wang_fMRI_9}\cite{Paper_Pominova_fMRI_15}                            \\ \hline
    HCP\cite{Dataset_HCP}                                                                                      & 1206                & DCS           & 2                 & 2013          & \href{https://humanconnectome.org/study/hcp-young-adult}{Available} &         \textbf{--}          & \cite{Paper_Su_fMRI_2} \cite{Paper_Jiang_fMRI_3} \cite{Paper_Harrison_fMRI_4} \cite{Paper_Gao_fMRI_5} \\ \hline
    OpenfMRI: Balloon Analog Risk-Taking Task\cite{Dataset_OpenfMRIBalloon}    & 16                  & DCS           &        2           & 2016          & \href{https://legacy.openfmri.org/dataset/ds000001/}{Available}&         \textbf{--}            & \cite{Paper_Li_fMRI_36}\cite{Paper_Zhou_fMRI_26}                          \\ \hline
    ADHD-200\cite{Dataset_ADHD}                                                                                 & 973                 & ADHDC         &         3          & 2012          & \href{http://fcon_1000.projects.nitrc.org/indi/req_access.html}{Access Request} & Site \#: 8             & \cite{Paper_Zhang_fMRI_24} \cite{Paper_Huang_fMRI_25} \\ \hline
    ADNI I\cite{Dataset_ADNII}                                                                                   & 819                 & ADC           &           3        & 2010          & \href{https://ida.loni.usc.edu/collaboration/access/appLicense.jsp}{Access Request} &      \textbf{--}       & \cite{Paper_Lin_fMRI_33} \\ \hline
    ADNI II\cite{Dataset_ADNIII}                                                                                  & 1601                 & ADC           &        4           & 2015          & \href{https://ida.loni.usc.edu/collaboration/access/appLicense.jsp}{Access Request} &     \textbf{--}        &   \cite{Paper_Lin_fMRI_33} \\ \hline
    \end{tabular}
    \end{adjustbox}
    \caption{\textbf{Commonest fMRI datasets.} Note that in this table, for each dataset, information like the reference to the paper in which the dataset was first published, the number of subjects whose data is available in the dataset, the type of task that the dataset is used for, the number of different classes covered by the dataset, year in which the dataset was published, availability state of the dataset, some extra information like the number of sites from which data was gathered or the number of total sample counts, and reference to articles that utilize the dataset is provided. Also note that ASDI stands for Autism Spectrum Disorders Identification, DCS for the Decoding of Cognitive States, ADHDC for ADHD Classification, and ADC for Alzheimer’s Disease Classification.}
    \label{fmri-dataset-table}
    \end{table*} 


\section{Future Directions}
\label{sec:futuredirection}
In this section, we address the attention-worthy tracks that deserve to be investigated more deeply in the literature of domain adaptation and generalization in functional medical data in the future. These topics are presented as follows: 
\subsection{Interpretation}\label{future_direction_interpretation}
Interpretability of deep neural networks, specifically in the health and medical domains, has always been a limiting factor for use cases requiring trust in the obtained results. This has attracted the attention of experts to factors and causalities that are unknown. Interpretability and visualization play a critical role in the generalization and adaptation of methods for health-related data analysis. In spite of the fact that only a few interpretable DG/DA models have been proposed for the analysis of medical data, most existing approaches remain black boxes. By making a DG/DA model interpretable, it may be possible to determine the influential aspects and features for each domain, select common elements, eliminate noise, and finally, improve its reliability for health professionals. Hence, developing effective strategies in this field is essential, as they have significant real-world implications.

\subsection{Multi-Source Processing}\label{future_direction_multisource}
Using a multi-source approach, one assumes that the source data originates from multiple domains, and a model learns more stable patterns across the source domains when it distinguishes between source domains (such as subjects or sites), resulting in a more stable pattern across source domains and a better ability to generalize to previously unseen target spaces. For example, in real-world scenarios involving medical image analysis, we are not able to access the target domain during training, so we need to extract shared information from multiple sources to achieve more general results and perform better. Although there are a variety of models in this category, such as in \cite{Paper_Chen_13} and \cite{Paper_Gu_3}, it remains difficult to determine how to effectively incorporate differences between source domains during the training phase of DG/DA approaches proposed for medical data.

\subsection{Incremental and Online Learning} \label{future_direction_incremental}
A limited number of data sources are available during the training process for real-world issues, which means that when a new source, sample, or target becomes available over time, the model needs to be trained based on limited data and then be improved accordingly; as an example, brain data processing can take on data from a new subject, a new center, or a new trial. As a result, if the planned model is able to learn from consistently added data over time, the model will be more accurate and provide significant benefits, such as the ability to adapt to time-related changes. Taking these observations into account in practical use cases, it is necessary to establish online and incremental strategies to address the mentioned scenario in a timely manner.
\subsection{Privacy Preserving} \label{future_direction_privacy}
A noticeable amount of work has recently focused on preserving information privacy for DG/DA in deep neural networks, in which the network is adapted or generalized without any prior knowledge about the data sources used for the training of the model. In such cases, only the architecture of the network (or, in a more extreme case, a black-box query system) is available, so many common DG/DA approaches that rely on moderating data from the source domains fail. Although these privacy-preserving ideas have been widely applied to various fields, such as computer vision and natural language processing, they have not yet been used in medical data analysis to the greatest extent possible. It is worth noting that the importance of privacy in medical fields of study is undeniable. There are serious concerns in some cases about the confidentiality of patients' information that necessitates keeping the signals recorded from people private. Accessing people's clinical information also has the risk of identifying population traits. Furthermore, many medical centers might not be willing to share their data with others so as to keep the methodologies and sources safe from potential rivals. Lastly, the exclusion of data from adaptation and generalization approaches should be carefully considered so that the performance of the models in medical tasks is not severely compromised since medical tasks are more vital than usual vision-based tasks because they involve people's lives and health conditions. Still, to the best of our knowledge, so far, only three works, one on EEG data (\cite{Paper_Xia_73}) and two on fMRI data (\cite{Paper_Han_fMRI_12} and \cite{Paper_Li_fMRI_19}), have studied source-free domain adaptation on EEG data. Thus, there are increasing demands for research dedicated to pursuing DG/DA while preserving data privacy.
\subsection{Disentangled Representation Learning}
Disentanglement in representation learning refers to the process of learning different features from deep networks, each one being narrowly separable from the other, and often supervised by specific semantics within the data. These techniques have been extensively used for disentangling features in images and videos. Evidently, this area can also be examined in the context of medical data, in which a variety of clues relating to subjects and tasks are intertwined with the given data signals. Particularly, suppose that task-specific and subject-specific features can be appropriately separated from each other in functional images. In that case, DG/DA can be achieved efficiently by focusing on task-specific features that can be obtained by designing approaches for omitting the influence of features varying among subjects. The disentanglement of features in brain data, however, is more complex than the disentanglement of features in images or videos, which are relatively understandable and easily visualizable, making it possible to distinguish features more meaningfully. The amount of knowledge we have about functional brain data, however, is limited, and more research is necessary to understand them. Therefore, defining separate features in them is much more challenging, and they cannot be validated effectively. In this study, we have noted several works that address disentanglement for DG/DA on functional medical data(\cite{Paper_Jeon_39}, \cite{Paper_Liu_48}, \cite{Paper_Zhao_10}, and \cite{Paper_Peng_9}) \cite{Paper_wang_multi-site_2022}. Nevertheless, a more profound investigation of disentanglement for solving the problem of domain shift in functional medical data remains a challenging problem that needs to be addressed in the future.
\subsection{Multi-Modal Medical Data Processing}\label{future_direction_multi_modal}
Medical data, especially brain signals, can be processed multi-modally so that the model's performance improves. We can gain a more substantial amount of information if we process multiple types of brain signals at the same time. In addition, we can extract useful information from each of them since each type of brain signal has its own properties, advantages, and disadvantages. Because brain signals contain noise and artifacts, they are inherently a mixture of unwanted information. Although EEG signals have a high temporal resolution, they lack spatial resolution. However, fMRI signals have a high spatial resolution but cannot accurately determine the timestamp of observations. Consequently, processing EEG/fMRI data at the same time allows us to utilize EEG's superior spatial resolution and fMRI's high temporal resolution. This allows us to make more accurate predictions. The dual-accurate property can assist in tasks such as human thought detection and processing, which require both temporal accuracy (people's thoughts change instantly, so immediate detection is sometimes essential) and spatial accuracy (thinking is a complex process and spatial information is required to cover all aspects).

Furthermore, different types of signals are produced by different physical and biological processes. In contrast to common pairs of modality such as image and text, which contain much more common information (e.g. you can estimate text based on its corresponding image), using medical images together can provide more information.

There are challenges with the pairwise disparity between different types of brain signals. Due to their completely different structures, the efficient fusion of these signals requires carefully designed architectures.

There are also measurement costs associated with multi-modal medical data processing. Each signal has its own limitations and costs, so if we measure two signals together, we are complying with all of those limitations and costs. As an example, in contrast to fMRI data, EEG signals can be measured while moving under controlled conditions, like driving. Adding fMRI to EEG limits our ability to study driving-related tasks and limits our experiments on steady-state scenarios.
\subsection{Federated Learning of Medical Data }\label{future_direction_federated_learning}
There are numerous medical datasets available for research, but the amount of data generated by medical institutions and hospitals is well beyond what academics can access. Even regardless of privacy issues, the resources required to process, store, maintain, and update the dataset make it impractical to gather all data continuously measured at medical sites. A high-performance system that works well in real-world applications can be obtained by processing this data using federated learning. Local sources (e.g., hospitals) will not be concerned about their private patient information or statistics being exposed since a federated learning framework denies direct access to local data. Furthermore, each source can train and prepare its model using a cheap computer system, update its data and model according to its schedule, and use a high-performing model trained based on information from multiple sources at virtually no extra cost.

Using a federated learning approach to analyze medical data is also motivated by the diversity of signals and images generated by different physical and biological measurement processes and devices. Brand-new measurement technologies and more accurate devices are continuously being developed in order to improve the usable information extracted from medical data. Consequently, it's challenging to design a centralized system that handles all types and variations of medical data.

Federated learning architectures, with flexible source updates and inter-source information flow, can significantly help with efficient and high-performance medical data analysis.

\subsection{Cross-Modality Generalization and Adaptation}\label{future_direction_cross_modality}
As previously explained in detail, numerous definitions of domains, identified by subjects, sessions, datasets, and many more, are studied in generalization and adaptation. However, presuming domains as modalities and investigating cross-modality cases have not yet been explored in this era. In this case, it is worth noting that the data recorded via different neuroimaging techniques may differ from each other since they reflect the behavior and functionality of different sections of the brain. The incomplete knowledge of the actual brain mechanisms implies that sometimes, one type of data might be a better choice than the other to be used in a fixed manner for a specific task; for example, a model trained on EEG data might need to be applied to fMRI data for the same task. Additionally, it should be noted that operating imaging equipment is typically time-consuming and costly, and that the data collected from multiple centers might inevitably have varying data types. It might not be affordable for them to change the equipment to meet the requirements of the data type specified for the model being used. Consequently, domain shift within different modalities is a worthy and useful problem that has to be explored deeply in medical data analysis. Of course, serious challenges in aligning data representations from different modalities need to be tackled for cross-modality studies. For this purpose to be achieved in the future, it will be necessary to develop a more precise understanding of the meaning behind captured signals, along with more interpretable and disentangled representation learning techniques.


\section{Conclusion}
\label{sec:conclusion}
Several methods have been developed in the literature to allow medical image analysis models trained on one or more source domains to adapt to unidentified Out-Of-Distribution (OOD) target data and train generalizable models. There have been many methods developed covering various applications. This area of research has recently received considerable attention and is an essential component of the deployment of many machine-learning algorithms. We presented a systematic and comprehensive review of domain adaptation and generalization methods for functional medical images, particularly functional brain images. In addition to the fundamentals of OOD generalization and adaptation, we presented the major approaches, architectures, and essential datasets used in the field of medical image analysis. Our analysis addresses a number of potential outstanding issues and promising directions for future research to advance the field, which was discussed in the section \ref{sec:futuredirection}.

\section{Acknowledgements}
This paper and the research behind it would not have been possible without the help of Saba Hashemi and Amirreza Soleymanbeigi, who helped us with the preparation and review of fMRI-related studies.

\bibliographystyle{unsrtnat}
\bibliography{references}

\begin{thebibliography}{282}
\providecommand{\natexlab}[1]{#1}
\providecommand{\url}[1]{\texttt{#1}}
\expandafter\ifx\csname urlstyle\endcsname\relax
  \providecommand{\doi}[1]{doi: #1}\else
  \providecommand{\doi}{doi: \begingroup \urlstyle{rm}\Url}\fi

\bibitem[Wang et~al.(2022{\natexlab{a}})Wang, Lan, Liu, Ouyang, Qin, Lu, Chen,
  Zeng, and Yu]{domaingeneralizationsurvey2022_second}
Jindong Wang, Cuiling Lan, Chang Liu, Yidong Ouyang, Tao Qin, Wang Lu, Yiqiang
  Chen, Wenjun Zeng, and Philip Yu.
\newblock Generalizing to unseen domains: A survey on domain generalization.
\newblock \emph{IEEE Transactions on Knowledge and Data Engineering},
  2022{\natexlab{a}}.

\bibitem[Zhou et~al.(2022{\natexlab{a}})Zhou, Liu, Qiao, Xiang, and
  Loy]{domaingeneralizationsurvey2022}
Kaiyang Zhou, Ziwei Liu, Yu~Qiao, Tao Xiang, and Chen~Change Loy.
\newblock Domain generalization: A survey.
\newblock \emph{IEEE Transactions on Pattern Analysis and Machine
  Intelligence}, 2022{\natexlab{a}}.

\bibitem[Farahani et~al.(2021)Farahani, Voghoei, Rasheed, and
  Arabnia]{domainadatationsurvey}
Abolfazl Farahani, Sahar Voghoei, Khaled Rasheed, and Hamid~R Arabnia.
\newblock A brief review of domain adaptation.
\newblock \emph{Advances in data science and information engineering}, pages
  877--894, 2021.

\bibitem[Liu et~al.(2022{\natexlab{a}})Liu, Yoo, Xing, Oh, El~Fakhri, Kang,
  Woo, et~al.]{domainadaptationsurvey2}
Xiaofeng Liu, Chaehwa Yoo, Fangxu Xing, Hyejin Oh, Georges El~Fakhri, Je-Won
  Kang, Jonghye Woo, et~al.
\newblock Deep unsupervised domain adaptation: a review of recent advances and
  perspectives.
\newblock \emph{APSIPA Transactions on Signal and Information Processing},
  11\penalty0 (1), 2022{\natexlab{a}}.

\bibitem[Guan and Liu(2021)]{domainadaptationmedical}
Hao Guan and Mingxia Liu.
\newblock Domain adaptation for medical image analysis: a survey.
\newblock \emph{IEEE Transactions on Biomedical Engineering}, 69\penalty0
  (3):\penalty0 1173--1185, 2021.

\bibitem[Wu et~al.(2019)Wu, Zhou, Zhao, Yue, and Keutzer]{bg_da_wu_2019}
Bichen Wu, Xuanyu Zhou, Sicheng Zhao, Xiangyu Yue, and Kurt Keutzer.
\newblock Squeezesegv2: Improved model structure and unsupervised domain
  adaptation for road-object segmentation from a lidar point cloud.
\newblock \emph{2019 International Conference on Robotics and Automation
  (ICRA)}, pages 4376--4382, 2019.

\bibitem[Zhao et~al.(2020{\natexlab{a}})Zhao, Li, Reed, Xu, and
  Keutzer]{bg_da_Zhao_2020}
Sicheng Zhao, Bo~Li, Colorado Reed, Pengfei Xu, and Kurt Keutzer.
\newblock Multi-source domain adaptation in the deep learning era: A systematic
  survey.
\newblock \emph{ArXiv}, abs/2002.12169, 2020{\natexlab{a}}.

\bibitem[Pan and Yang(2010)]{bg_da_Pan_2010}
Sinno~Jialin Pan and Qiang Yang.
\newblock A survey on transfer learning.
\newblock \emph{IEEE Transactions on Knowledge and Data Engineering},
  22:\penalty0 1345--1359, 2010.

\bibitem[Zhao et~al.(2020{\natexlab{b}})Zhao, Li, Xu, and Keutzer]{multi2020}
Sicheng Zhao, Bo~Li, Pengfei Xu, and Kurt Keutzer.
\newblock Multi-source domain adaptation in the deep learning era: A systematic
  survey.
\newblock \emph{arXiv preprint arXiv:2002.12169}, 2020{\natexlab{b}}.

\bibitem[He and Wu(2020)]{bg_da_He_2020}
He~He and Dongrui Wu.
\newblock Different set domain adaptation for brain-computer interfaces: A
  label alignment approach.
\newblock \emph{IEEE Transactions on Neural Systems and Rehabilitation
  Engineering}, 28:\penalty0 1091--1108, 2020.

\bibitem[Busto and Gall(2017)]{bg_da_Busto_2017}
Pau~Panareda Busto and Juergen Gall.
\newblock Open set domain adaptation.
\newblock In \emph{2017 IEEE International Conference on Computer Vision
  (ICCV)}, pages 754--763, 2017.
\newblock \doi{10.1109/ICCV.2017.88}.

\bibitem[Saito et~al.(2018{\natexlab{a}})Saito, Yamamoto, Ushiku, and
  Harada]{bg_da_Saito_2018}
Kuniaki Saito, Shohei Yamamoto, Y.~Ushiku, and Tatsuya Harada.
\newblock Open set domain adaptation by backpropagation.
\newblock \emph{ArXiv}, abs/1804.10427, 2018{\natexlab{a}}.

\bibitem[You et~al.(2019)You, Long, Cao, Wang, and Jordan]{bg_da_You_2019}
Kaichao You, Mingsheng Long, Zhangjie Cao, Jianmin Wang, and Michael~I. Jordan.
\newblock Universal domain adaptation.
\newblock \emph{2019 IEEE/CVF Conference on Computer Vision and Pattern
  Recognition (CVPR)}, pages 2715--2724, 2019.

\bibitem[Havaei et~al.(2017)Havaei, Davy, Warde-Farley, Biard, Courville,
  Bengio, Pal, Jodoin, and Larochelle]{bg_med_havaei_brain_2017}
Mohammad Havaei, Axel Davy, David Warde-Farley, Antoine Biard, Aaron Courville,
  Yoshua Bengio, Chris Pal, Pierre-Marc Jodoin, and Hugo Larochelle.
\newblock Brain tumor segmentation with {Deep} {Neural} {Networks}.
\newblock \emph{Medical Image Analysis}, 35:\penalty0 18--31, January 2017.
\newblock ISSN 13618415.
\newblock \doi{10.1016/j.media.2016.05.004}.
\newblock URL
  \url{https://linkinghub.elsevier.com/retrieve/pii/S1361841516300330}.

\bibitem[Ottom et~al.(2022)Ottom, Rahman, and Dinov]{bg_med_ottom_znet:_2022}
Mohammad~Ashraf Ottom, Hanif~Abdul Rahman, and Ivo~D. Dinov.
\newblock Znet: {Deep} {Learning} {Approach} for {2D} {MRI} {Brain} {Tumor}
  {Segmentation}.
\newblock \emph{IEEE Journal of Translational Engineering in Health and
  Medicine}, 10:\penalty0 1--8, 2022.
\newblock ISSN 2168-2372.
\newblock \doi{10.1109/JTEHM.2022.3176737}.
\newblock URL \url{https://ieeexplore.ieee.org/document/9779760/}.

\bibitem[Jiang et~al.(2021{\natexlab{a}})Jiang, Zhai, and
  Kong]{bg_med_jiang_novel_2021}
Min Jiang, Fuhao Zhai, and Jun Kong.
\newblock A novel deep learning model {DDU}-net using edge features to enhance
  brain tumor segmentation on {MR} images.
\newblock \emph{Artificial Intelligence in Medicine}, 121:\penalty0 102180,
  November 2021{\natexlab{a}}.
\newblock ISSN 09333657.
\newblock \doi{10.1016/j.artmed.2021.102180}.
\newblock URL
  \url{https://linkinghub.elsevier.com/retrieve/pii/S0933365721001731}.

\bibitem[Khan et~al.(2021)Khan, Khan, Harouni, Abbasi, Iqbal, and
  Mehmood]{bg_med_khan_brain_2021}
Amjad~Rehman Khan, Siraj Khan, Majid Harouni, Rashid Abbasi, Sajid Iqbal, and
  Zahid Mehmood.
\newblock Brain tumor segmentation using {K}‐means clustering and deep
  learning with synthetic data augmentation for classification.
\newblock \emph{Microscopy Research and Technique}, 84\penalty0 (7):\penalty0
  1389--1399, July 2021.
\newblock ISSN 1059-910X, 1097-0029.
\newblock \doi{10.1002/jemt.23694}.
\newblock URL \url{https://onlinelibrary.wiley.com/doi/10.1002/jemt.23694}.

\bibitem[Ullah et~al.(2022)Ullah, Usman, Jeon, and
  Gwak]{bg_med_ullah_cascade_2022}
Zahid Ullah, Muhammad Usman, Moongu Jeon, and Jeonghwan Gwak.
\newblock Cascade multiscale residual attention {CNNs} with adaptive {ROI} for
  automatic brain tumor segmentation.
\newblock \emph{Information Sciences}, 608:\penalty0 1541--1556, August 2022.
\newblock ISSN 00200255.
\newblock \doi{10.1016/j.ins.2022.07.044}.
\newblock URL
  \url{https://linkinghub.elsevier.com/retrieve/pii/S0020025522007332}.

\bibitem[Jiang et~al.(2022{\natexlab{a}})Jiang, Zhang, Lin, Dong, Cheng, and
  Liang]{bg_med_jiang_swinbts:_2022}
Yun Jiang, Yuan Zhang, Xin Lin, Jinkun Dong, Tongtong Cheng, and Jing Liang.
\newblock {SwinBTS}: {A} {Method} for {3D} {Multimodal} {Brain} {Tumor}
  {Segmentation} {Using} {Swin} {Transformer}.
\newblock \emph{Brain Sciences}, 12\penalty0 (6):\penalty0 797, June
  2022{\natexlab{a}}.
\newblock ISSN 2076-3425.
\newblock \doi{10.3390/brainsci12060797}.
\newblock URL \url{https://www.mdpi.com/2076-3425/12/6/797}.

\bibitem[Han et~al.(2017)Han, Kang, Jeong, Park, Kim, Bang, and
  Seong]{bg_med_breast_han_deep_2017}
Seokmin Han, Ho-Kyung Kang, Ja-Yeon Jeong, Moon-Ho Park, Wonsik Kim, Won-Chul
  Bang, and Yeong-Kyeong Seong.
\newblock A deep learning framework for supporting the classification of breast
  lesions in ultrasound images.
\newblock \emph{Physics in Medicine \& Biology}, 62\penalty0 (19):\penalty0
  7714--7728, September 2017.
\newblock ISSN 1361-6560.
\newblock \doi{10.1088/1361-6560/aa82ec}.
\newblock URL
  \url{https://iopscience.iop.org/article/10.1088/1361-6560/aa82ec}.

\bibitem[Laar et~al.(2013{\natexlab{a}})Laar, Brugman, Nijboer, Poel, Nijholt,
  and Miller]{bg_med_breast_laar_brainbrush_2013}
van~de Laar, Bram, Ivo Brugman, Femke Nijboer, Mannes Poel, Anton Nijholt, and
  Leslie Miller.
\newblock \emph{{BrainBrush}, a multimodal application for creative
  expressivity}.
\newblock IARIA XPS Press, February 2013{\natexlab{a}}.
\newblock ISBN 9781612082509.
\newblock OCLC: 6893370982.

\bibitem[Nahid et~al.(2018)Nahid, Mehrabi, and
  Kong]{bg_med_breast_nahid_histopathological_2018}
Abdullah-Al Nahid, Mohamad~Ali Mehrabi, and Yinan Kong.
\newblock Histopathological {Breast} {Cancer} {Image} {Classification} by
  {Deep} {Neural} {Network} {Techniques} {Guided} by {Local} {Clustering}.
\newblock \emph{BioMed Research International}, 2018:\penalty0 1--20, 2018.
\newblock ISSN 2314-6133, 2314-6141.
\newblock \doi{10.1155/2018/2362108}.
\newblock URL \url{https://www.hindawi.com/journals/bmri/2018/2362108/}.

\bibitem[Rasti et~al.(2017)Rasti, Teshnehlab, and
  Phung]{bg_med_breast_rasti_breast_2017}
Reza Rasti, Mohammad Teshnehlab, and Son~Lam Phung.
\newblock Breast cancer diagnosis in {DCE}-{MRI} using mixture ensemble of
  convolutional neural networks.
\newblock \emph{Pattern Recognition}, 72:\penalty0 381--390, December 2017.
\newblock ISSN 00313203.
\newblock \doi{10.1016/j.patcog.2017.08.004}.
\newblock URL
  \url{https://linkinghub.elsevier.com/retrieve/pii/S0031320317303084}.

\bibitem[Samala et~al.(2018)Samala, Chan, Hadjiiski, Helvie, Richter, and
  Cha]{bg_med_breast_samala_evolutionary_2018}
Ravi~K Samala, Heang-Ping Chan, Lubomir~M Hadjiiski, Mark~A Helvie, Caleb
  Richter, and Kenny Cha.
\newblock Evolutionary pruning of transfer learned deep convolutional neural
  network for breast cancer diagnosis in digital breast tomosynthesis.
\newblock \emph{Physics in Medicine \& Biology}, 63\penalty0 (9):\penalty0
  095005, May 2018.
\newblock ISSN 1361-6560.
\newblock \doi{10.1088/1361-6560/aabb5b}.
\newblock URL
  \url{https://iopscience.iop.org/article/10.1088/1361-6560/aabb5b}.

\bibitem[Ben-Cohen et~al.(2018)Ben-Cohen, Klang, Kerpel, Konen, Amitai, and
  Greenspan]{bg_med_abd_ben-cohen_fully_2018}
Avi Ben-Cohen, Eyal Klang, Ariel Kerpel, Eli Konen, Michal~Marianne Amitai, and
  Hayit Greenspan.
\newblock Fully convolutional network and sparsity-based dictionary learning
  for liver lesion detection in {CT} examinations.
\newblock \emph{Neurocomputing}, 275:\penalty0 1585--1594, January 2018.
\newblock ISSN 09252312.
\newblock \doi{10.1016/j.neucom.2017.10.001}.
\newblock URL
  \url{https://linkinghub.elsevier.com/retrieve/pii/S0925231217316259}.

\bibitem[Causey et~al.(2021)Causey, Stubblefield, Qualls, Fowler, Cai, Walker,
  Guan, and Huang]{bg_med_abd_causey_ensemble_2021}
Jason Causey, Jonathan Stubblefield, Jake Qualls, Jennifer Fowler, Lingrui Cai,
  Karl Walker, Yuanfang Guan, and Xiuzhen Huang.
\newblock An {Ensemble} of {U}-{Net} {Models} for {Kidney} {Tumor}
  {Segmentation} with {CT} images.
\newblock \emph{IEEE/ACM Transactions on Computational Biology and
  Bioinformatics}, pages 1--1, 2021.
\newblock ISSN 1545-5963, 1557-9964, 2374-0043.
\newblock \doi{10.1109/TCBB.2021.3085608}.
\newblock URL \url{https://ieeexplore.ieee.org/document/9444778/}.

\bibitem[Hou et~al.(2020)Hou, Xie, Li, Wang, Lv, Xie, and
  Nan]{bg_med_abd_hou_triple-stage_2020}
Xiaoshuai Hou, Chunmei Xie, Fengyi Li, Jiaping Wang, Chuanfeng Lv, Guotong Xie,
  and Yang Nan.
\newblock A {Triple}-{Stage} {Self}-{Guided} {Network} for {Kidney} {Tumor}
  {Segmentation}.
\newblock In \emph{2020 {IEEE} 17th {International} {Symposium} on {Biomedical}
  {Imaging} ({ISBI})}, pages 341--344, Iowa City, IA, USA, April 2020. IEEE.
\newblock ISBN 9781538693308.
\newblock \doi{10.1109/ISBI45749.2020.9098609}.
\newblock URL \url{https://ieeexplore.ieee.org/document/9098609/}.

\bibitem[Anirudh et~al.(2016)Anirudh, Thiagarajan, Bremer, and
  Kim]{bg_med_abd_lung_2016}
Rushil Anirudh, Jayaraman~J. Thiagarajan, Timo Bremer, and Hyojin Kim.
\newblock Lung nodule detection using {3D} convolutional neural networks
  trained on weakly labeled data.
\newblock page 978532, San Diego, California, United States, March 2016.
\newblock \doi{10.1117/12.2214876}.
\newblock URL
  \url{http://proceedings.spiedigitallibrary.org/proceeding.aspx?doi=10.1117/12.2214876}.

\bibitem[Sarker et~al.(2017)Sarker, Shuvo, Hossain, and
  Hasan]{bg_med_abd_segmentation_2017}
Prionjit Sarker, Md. Maruf~Hossain Shuvo, Zakir Hossain, and Sabbir Hasan.
\newblock Segmentation and classification of lung tumor from {3D} {CT} image
  using {K}-means clustering algorithm.
\newblock In \emph{2017 4th {International} {Conference} on {Advances} in
  {Electrical} {Engineering} ({ICAEE})}, pages 731--736, Dhaka, September 2017.
  IEEE.
\newblock ISBN 9781538608692.
\newblock \doi{10.1109/ICAEE.2017.8255451}.
\newblock URL \url{http://ieeexplore.ieee.org/document/8255451/}.

\bibitem[Wei et~al.(2019)Wei, Wang, Zhao, Zhang, Yuan, and
  Li]{bg_med_abd_wei_two-phase_2019}
Hao Wei, Qin Wang, Weibing Zhao, Minqing Zhang, Kun Yuan, and Zhen Li.
\newblock Two-phase {Framework} for {Automatic} {Kidney} and {Kidney} {Tumor}
  {Segmentation}.
\newblock In \emph{Submissions to the 2019 {Kidney} {Tumor} {Segmentation}
  {Challenge}: {KiTS19}}. University of Minnesota Libraries Publishing, 2019.
\newblock \doi{10.24926/548719.043}.
\newblock URL
  \url{https://kits.lib.umn.edu/two-phase-framework-for-automatic-kidn\\ey-and-kidney-tumor-segmentation/}.

\bibitem[Wu et~al.(2022)Wu, Furuzuki, Li, Kamiya, Mabu, Tanabe, Ito, and
  Kido]{bg_med_abd_wu_segmentation_2022}
Jiaqi Wu, Muki Furuzuki, Guangxu Li, Tohru Kamiya, Shingo Mabu, Masahiro
  Tanabe, Katsuyoshi Ito, and Shoji Kido.
\newblock Segmentation of liver tumors in multiphase computed tomography images
  using hybrid method.
\newblock \emph{Computers \& Electrical Engineering}, 97:\penalty0 107626,
  January 2022.
\newblock ISSN 00457906.
\newblock \doi{10.1016/j.compeleceng.2021.107626}.
\newblock URL
  \url{https://linkinghub.elsevier.com/retrieve/pii/S0045790621005565}.

\bibitem[Kiranyaz et~al.(2016)Kiranyaz, Ince, and
  Gabbouj]{bg_med_heart_kiranyaz_real-time_2016}
Serkan Kiranyaz, Turker Ince, and Moncef Gabbouj.
\newblock Real-{Time} {Patient}-{Specific} {ECG} {Classification} by 1-{D}
  {Convolutional} {Neural} {Networks}.
\newblock \emph{IEEE Transactions on Biomedical Engineering}, 63\penalty0
  (3):\penalty0 664--675, March 2016.
\newblock ISSN 0018-9294, 1558-2531.
\newblock \doi{10.1109/TBME.2015.2468589}.
\newblock URL \url{http://ieeexplore.ieee.org/document/7202837/}.

\bibitem[Kutlu et~al.(2016)Kutlu, Allahverdi, and
  Altan]{bg_med_heart_kutlu_multistage_2016}
Yakup Kutlu, Novruz Allahverdi, and Gokhan Altan.
\newblock A {Multistage} {Deep} {Belief} {Networks} {Application} on
  {Arrhythmia} {Classification}.
\newblock \emph{International Journal of Intelligent Systems and Applications
  in Engineering}, 4\penalty0 (Special Issue-1):\penalty0 222--228, December
  2016.
\newblock ISSN 2147-6799.
\newblock \doi{10.18201/ijisae.2016SpecialIssue-146978}.
\newblock URL \url{https://ijisae.org/IJISAE/article/view/964}.

\bibitem[Lih et~al.(2020)Lih, Jahmunah, San, Ciaccio, Yamakawa, Tanabe,
  Kobayashi, Faust, and Acharya]{bg_med_heart_lih_comprehensive_2020}
Oh~Shu Lih, V~Jahmunah, Tan~Ru San, Edward~J Ciaccio, Toshitaka Yamakawa,
  Masayuki Tanabe, Makiko Kobayashi, Oliver Faust, and U~Rajendra Acharya.
\newblock Comprehensive electrocardiographic diagnosis based on deep learning.
\newblock \emph{Artificial Intelligence in Medicine}, 103:\penalty0 101789,
  March 2020.
\newblock ISSN 09333657.
\newblock \doi{10.1016/j.artmed.2019.101789}.
\newblock URL
  \url{https://linkinghub.elsevier.com/retrieve/pii/S0933365719309030}.

\bibitem[Liu et~al.(2018)Liu, Huang, Chang, Wang, and
  He]{bg_med_heart_liu_multiple-feature-branch_2018}
Wenhan Liu, Qijun Huang, Sheng Chang, Hao Wang, and Jin He.
\newblock Multiple-feature-branch convolutional neural network for myocardial
  infarction diagnosis using electrocardiogram.
\newblock \emph{Biomedical Signal Processing and Control}, 45:\penalty0 22--32,
  August 2018.
\newblock ISSN 17468094.
\newblock \doi{10.1016/j.bspc.2018.05.013}.
\newblock URL
  \url{https://linkinghub.elsevier.com/retrieve/pii/S1746809418301150}.

\bibitem[Tan et~al.(2018)Tan, Hagiwara, Pang, Lim, Oh, Adam, Tan, Chen, and
  Acharya]{bg_med_heart_tan_application_2018}
Jen~Hong Tan, Yuki Hagiwara, Winnie Pang, Ivy Lim, Shu~Lih Oh, Muhammad Adam,
  Ru~San Tan, Ming Chen, and U.~Rajendra Acharya.
\newblock Application of stacked convolutional and long short-term memory
  network for accurate identification of {CAD} {ECG} signals.
\newblock \emph{Computers in Biology and Medicine}, 94:\penalty0 19--26, March
  2018.
\newblock ISSN 00104825.
\newblock \doi{10.1016/j.compbiomed.2017.12.023}.
\newblock URL
  \url{https://linkinghub.elsevier.com/retrieve/pii/S0010482517304201}.

\bibitem[Amin et~al.(2019)Amin, Alsulaiman, Muhammad, Mekhtiche, and
  Shamim~Hossain]{bg_med_mi_amin_deep_2019}
Syed~Umar Amin, Mansour Alsulaiman, Ghulam Muhammad, Mohamed~Amine Mekhtiche,
  and M.~Shamim~Hossain.
\newblock Deep {Learning} for {EEG} motor imagery classification based on
  multi-layer {CNNs} feature fusion.
\newblock \emph{Future Generation Computer Systems}, 101:\penalty0 542--554,
  December 2019.
\newblock ISSN 0167739X.
\newblock \doi{10.1016/j.future.2019.06.027}.
\newblock URL
  \url{https://linkinghub.elsevier.com/retrieve/pii/S0167739X19306077}.

\bibitem[Hassanpour et~al.(2019)Hassanpour, Moradikia, Adeli, Khayami, and
  Shamsinejadbabaki]{bg_med_mi_hassanpour_novel_2019}
Ahmad Hassanpour, Majid Moradikia, Hojjat Adeli, Seyed~Raouf Khayami, and
  Pirooz Shamsinejadbabaki.
\newblock A novel end‐to‐end deep learning scheme for classifying
  multi‐class motor imagery electroencephalography signals.
\newblock \emph{Expert Systems}, 36\penalty0 (6), December 2019.
\newblock ISSN 0266-4720, 1468-0394.
\newblock \doi{10.1111/exsy.12494}.
\newblock URL \url{https://onlinelibrary.wiley.com/doi/10.1111/exsy.12494}.

\bibitem[Li et~al.(2019)Li, Zhang, Zhang, Lei, Cui, and
  Guo]{bg_med_mi_li_channel-projection_2019}
Yang Li, Xian-Rui Zhang, Bin Zhang, Meng-Ying Lei, Wei-Gang Cui, and Yu-Zhu
  Guo.
\newblock A {Channel}-{Projection} {Mixed}-{Scale} {Convolutional} {Neural}
  {Network} for {Motor} {Imagery} {EEG} {Decoding}.
\newblock \emph{IEEE Transactions on Neural Systems and Rehabilitation
  Engineering}, 27\penalty0 (6):\penalty0 1170--1180, June 2019.
\newblock ISSN 1534-4320, 1558-0210.
\newblock \doi{10.1109/TNSRE.2019.2915621}.
\newblock URL \url{https://ieeexplore.ieee.org/document/8709723/}.

\bibitem[Ortiz-Echeverri et~al.(2019)Ortiz-Echeverri, Salazar-Colores,
  Rodríguez-Reséndiz, and Gómez-Loenzo]{bg_med_mi_ortiz-echeverri_new_2019}
César~J. Ortiz-Echeverri, Sebastián Salazar-Colores, Juvenal
  Rodríguez-Reséndiz, and Roberto~A. Gómez-Loenzo.
\newblock A {New} {Approach} for {Motor} {Imagery} {Classification} {Based} on
  {Sorted} {Blind} {Source} {Separation}, {Continuous} {Wavelet} {Transform},
  and {Convolutional} {Neural} {Network}.
\newblock \emph{Sensors}, 19\penalty0 (20):\penalty0 4541, October 2019.
\newblock ISSN 1424-8220.
\newblock \doi{10.3390/s19204541}.
\newblock URL \url{https://www.mdpi.com/1424-8220/19/20/4541}.

\bibitem[Zhang et~al.(2020{\natexlab{a}})Zhang, Yao, Chen, Wang, Chang, and
  Liu]{bg_med_mi_zhang_making_2020}
Dalin Zhang, Lina Yao, Kaixuan Chen, Sen Wang, Xiaojun Chang, and Yunhao Liu.
\newblock Making {Sense} of {Spatio}-{Temporal} {Preserving} {Representations}
  for {EEG}-{Based} {Human} {Intention} {Recognition}.
\newblock \emph{IEEE Transactions on Cybernetics}, 50\penalty0 (7):\penalty0
  3033--3044, July 2020{\natexlab{a}}.
\newblock ISSN 2168-2267, 2168-2275.
\newblock \doi{10.1109/TCYB.2019.2905157}.
\newblock URL \url{https://ieeexplore.ieee.org/document/8698218/}.

\bibitem[Zhang et~al.(2019{\natexlab{a}})Zhang, Duan, Sole-Casals,
  Dinares-Ferran, Cichocki, Yang, and Sun]{bg_med_mi_zhang_novel_2019}
Zhiwen Zhang, Feng Duan, Jordi Sole-Casals, Josep Dinares-Ferran, Andrzej
  Cichocki, Zhenglu Yang, and Zhe Sun.
\newblock A {Novel} {Deep} {Learning} {Approach} {With} {Data} {Augmentation}
  to {Classify} {Motor} {Imagery} {Signals}.
\newblock \emph{IEEE Access}, 7:\penalty0 15945--15954, 2019{\natexlab{a}}.
\newblock ISSN 2169-3536.
\newblock \doi{10.1109/ACCESS.2019.2895133}.
\newblock URL \url{https://ieeexplore.ieee.org/document/8630915/}.

\bibitem[Arasteh et~al.(2021)Arasteh, Mahdizadeh, Mirian, Lee, and
  McKeown]{bg_med_parkinson_arasteh_deep_2021}
Emad Arasteh, Ailar Mahdizadeh, Maryam Mirian, Soojin Lee, and Martin McKeown.
\newblock Deep {Transfer} {Learning} for {Parkinson}’s {Disease} {Monitoring}
  by {Image}-{Based} {Representation} of {Resting}-{State} {EEG} {Using}
  {Directional} {Connectivity}.
\newblock \emph{Algorithms}, 15\penalty0 (1):\penalty0 5, December 2021.
\newblock ISSN 1999-4893.
\newblock \doi{10.3390/a15010005}.
\newblock URL \url{https://www.mdpi.com/1999-4893/15/1/5}.

\bibitem[Hassin-Baer et~al.(2022)Hassin-Baer, Cohen, Israeli-Korn, Yahalom,
  Benizri, Sand, Issachar, Geva, Shani-Hershkovich, and
  Peremen]{bg_med_parkinson_identification_2022}
Sharon Hassin-Baer, Oren~S. Cohen, Simon Israeli-Korn, Gilad Yahalom, Sandra
  Benizri, Daniel Sand, Gil Issachar, Amir~B. Geva, Revital Shani-Hershkovich,
  and Ziv Peremen.
\newblock Identification of an early-stage {Parkinson}’s disease neuromarker
  using event-related potentials, brain network analytics and machine-learning.
\newblock \emph{PLOS ONE}, 17\penalty0 (1):\penalty0 e0261947, January 2022.
\newblock ISSN 1932-6203.
\newblock \doi{10.1371/journal.pone.0261947}.
\newblock URL \url{https://dx.plos.org/10.1371/journal.pone.0261947}.

\bibitem[Lee et~al.(2021{\natexlab{a}})Lee, Hussein, Ward, Jane~Wang, and
  McKeown]{bg_med_parkinson_lee_convolutional-recurrent_2021}
Soojin Lee, Ramy Hussein, Rabab Ward, Z.~Jane~Wang, and Martin~J. McKeown.
\newblock A convolutional-recurrent neural network approach to resting-state
  {EEG} classification in {Parkinson}’s disease.
\newblock \emph{Journal of Neuroscience Methods}, 361:\penalty0 109282,
  September 2021{\natexlab{a}}.
\newblock ISSN 01650270.
\newblock \doi{10.1016/j.jneumeth.2021.109282}.
\newblock URL
  \url{https://linkinghub.elsevier.com/retrieve/pii/S016502702100217X}.

\bibitem[Oh et~al.(2020)Oh, Hagiwara, Raghavendra, Yuvaraj, Arunkumar,
  Murugappan, and Acharya]{bg_med_parkinson_oh_deep_2020}
Shu~Lih Oh, Yuki Hagiwara, U.~Raghavendra, Rajamanickam Yuvaraj, N.~Arunkumar,
  M.~Murugappan, and U.~Rajendra Acharya.
\newblock A deep learning approach for {Parkinson}’s disease diagnosis from
  {EEG} signals.
\newblock \emph{Neural Computing and Applications}, 32\penalty0 (15):\penalty0
  10927--10933, August 2020.
\newblock ISSN 0941-0643, 1433-3058.
\newblock \doi{10.1007/s00521-018-3689-5}.
\newblock URL \url{http://link.springer.com/10.1007/s00521-018-3689-5}.

\bibitem[Bi et~al.(2018{\natexlab{a}})Bi, Shu, Sun, and
  Xu]{bg_med_alzheimer_bi_random_2018}
Xia-an Bi, Qing Shu, Qi~Sun, and Qian Xu.
\newblock Random support vector machine cluster analysis of resting-state
  {fMRI} in {Alzheimer}'s disease.
\newblock \emph{PLOS ONE}, 13\penalty0 (3):\penalty0 e0194479, March
  2018{\natexlab{a}}.
\newblock ISSN 1932-6203.
\newblock \doi{10.1371/journal.pone.0194479}.
\newblock URL \url{https://dx.plos.org/10.1371/journal.pone.0194479}.

\bibitem[Ieracitano et~al.(2019)Ieracitano, Mammone, Bramanti, Hussain, and
  Morabito]{bg_med_alzheimer_ieracitano_convolutional_2019}
Cosimo Ieracitano, Nadia Mammone, Alessia Bramanti, Amir Hussain, and
  Francesco~C. Morabito.
\newblock A {Convolutional} {Neural} {Network} approach for classification of
  dementia stages based on {2D}-spectral representation of {EEG} recordings.
\newblock \emph{Neurocomputing}, 323:\penalty0 96--107, January 2019.
\newblock ISSN 09252312.
\newblock \doi{10.1016/j.neucom.2018.09.071}.
\newblock URL
  \url{https://linkinghub.elsevier.com/retrieve/pii/S0925231218311524}.

\bibitem[Nakamura et~al.(2018)Nakamura, Cuesta, Fernández, Arahata, Iwata,
  Kuratsubo, Bundo, Hattori, Sakurai, Fukuda, Washimi, Endo, Takeda, Diers,
  Bajo, Maestú, Ito, and Kato]{bg_med_alzheimer_nakamura_electromagnetic_2018}
Akinori Nakamura, Pablo Cuesta, Alberto Fernández, Yutaka Arahata, Kaori
  Iwata, Izumi Kuratsubo, Masahiko Bundo, Hideyuki Hattori, Takashi Sakurai,
  Koji Fukuda, Yukihiko Washimi, Hidetoshi Endo, Akinori Takeda, Kersten Diers,
  Ricardo Bajo, Fernando Maestú, Kengo Ito, and Takashi Kato.
\newblock Electromagnetic signatures of the preclinical and prodromal stages of
  {Alzheimer}’s disease.
\newblock \emph{Brain}, 141\penalty0 (5):\penalty0 1470--1485, May 2018.
\newblock ISSN 0006-8950, 1460-2156.
\newblock \doi{10.1093/brain/awy044}.
\newblock URL \url{https://academic.oup.com/brain/article/141/5/1470/4924218}.

\bibitem[Sheng et~al.(2019)Sheng, Wang, Zhang, Liu, Ma, Liu, Shao, and
  Chen]{bg_med_alzheimer_sheng_novel_2019}
Jinhua Sheng, Bocheng Wang, Qiao Zhang, Qingqiang Liu, Yangjie Ma, Weixiang
  Liu, Meiling Shao, and Bin Chen.
\newblock A novel joint {HCPMMP} method for automatically classifying
  {Alzheimer}’s and different stage {MCI} patients.
\newblock \emph{Behavioural Brain Research}, 365:\penalty0 210--221, June 2019.
\newblock ISSN 01664328.
\newblock \doi{10.1016/j.bbr.2019.03.004}.
\newblock URL
  \url{https://linkinghub.elsevier.com/retrieve/pii/S0166432819300324}.

\bibitem[Goshvarpour and
  Goshvarpour(2020)]{bg_med_schizophrenia_goshvarpour_schizophrenia_2020}
Atefeh Goshvarpour and Ateke Goshvarpour.
\newblock Schizophrenia diagnosis using innovative {EEG} feature-level fusion
  schemes.
\newblock \emph{Physical and Engineering Sciences in Medicine}, 43\penalty0
  (1):\penalty0 227--238, March 2020.
\newblock ISSN 2662-4729, 2662-4737.
\newblock \doi{10.1007/s13246-019-00839-1}.
\newblock URL \url{http://link.springer.com/10.1007/s13246-019-00839-1}.

\bibitem[Phang et~al.(2020)Phang, Noman, Hussain, Ting, and
  Ombao]{bg_med_schizophrenia_phang_multi-domain_2020}
Chun-Ren Phang, Fuad Noman, Hadri Hussain, Chee-Ming Ting, and Hernando Ombao.
\newblock A {Multi}-{Domain} {Connectome} {Convolutional} {Neural} {Network}
  for {Identifying} {Schizophrenia} {From} {EEG} {Connectivity} {Patterns}.
\newblock \emph{IEEE Journal of Biomedical and Health Informatics}, 24\penalty0
  (5):\penalty0 1333--1343, May 2020.
\newblock ISSN 2168-2194, 2168-2208.
\newblock \doi{10.1109/JBHI.2019.2941222}.
\newblock URL \url{https://ieeexplore.ieee.org/document/8836535/}.

\bibitem[Santos-Mayo et~al.(2017)Santos-Mayo, San-Jose-Revuelta, and
  Arribas]{bg_med_schizophrenia_santos-mayo_computer-aided_2017}
Lorenzo Santos-Mayo, Luis~M. San-Jose-Revuelta, and Juan~Ignacio Arribas.
\newblock A {Computer}-{Aided} {Diagnosis} {System} {With} {EEG} {Based} on the
  {P3b} {Wave} {During} an {Auditory} {Odd}-{Ball} {Task} in {Schizophrenia}.
\newblock \emph{IEEE Transactions on Biomedical Engineering}, 64\penalty0
  (2):\penalty0 395--407, February 2017.
\newblock ISSN 0018-9294, 1558-2531.
\newblock \doi{10.1109/TBME.2016.2558824}.
\newblock URL \url{http://ieeexplore.ieee.org/document/7460246/}.

\bibitem[Shim et~al.(2016)Shim, Hwang, Kim, Lee, and
  Im]{bg_med_schizophrenia_shim_machine-learning-based_2016}
Miseon Shim, Han-Jeong Hwang, Do-Won Kim, Seung-Hwan Lee, and Chang-Hwan Im.
\newblock Machine-learning-based diagnosis of schizophrenia using combined
  sensor-level and source-level {EEG} features.
\newblock \emph{Schizophrenia Research}, 176\penalty0 (2-3):\penalty0 314--319,
  October 2016.
\newblock ISSN 09209964.
\newblock \doi{10.1016/j.schres.2016.05.007}.
\newblock URL
  \url{https://linkinghub.elsevier.com/retrieve/pii/S0920996416302274}.

\bibitem[Baygin et~al.(2021)Baygin, Dogan, Tuncer, Datta~Barua, Faust,
  Arunkumar, Abdulhay, Emma~Palmer, and
  Rajendra~Acharya]{bg_med_asd_baygin_automated_2021}
Mehmet Baygin, Sengul Dogan, Turker Tuncer, Prabal Datta~Barua, Oliver Faust,
  N.~Arunkumar, Enas~W. Abdulhay, Elizabeth Emma~Palmer, and
  U.~Rajendra~Acharya.
\newblock Automated {ASD} detection using hybrid deep lightweight features
  extracted from {EEG} signals.
\newblock \emph{Computers in Biology and Medicine}, 134:\penalty0 104548, July
  2021.
\newblock ISSN 00104825.
\newblock \doi{10.1016/j.compbiomed.2021.104548}.
\newblock URL
  \url{https://linkinghub.elsevier.com/retrieve/pii/S0010482521003425}.

\bibitem[Bi et~al.(2018{\natexlab{b}})Bi, Wang, Shu, Sun, and
  Xu]{bg_med_asd_bi_classification_2018}
Xia-an Bi, Yang Wang, Qing Shu, Qi~Sun, and Qian Xu.
\newblock Classification of {Autism} {Spectrum} {Disorder} {Using} {Random}
  {Support} {Vector} {Machine} {Cluster}.
\newblock \emph{Frontiers in Genetics}, 9:\penalty0 18, February
  2018{\natexlab{b}}.
\newblock ISSN 1664-8021.
\newblock \doi{10.3389/fgene.2018.00018}.
\newblock URL
  \url{http://journal.frontiersin.org/article/10.3389/fgene.2018.00018/full}.

\bibitem[Ranjani and Supraja(2021)]{bg_med_asd_ranjani_classifying_2021}
M~Ranjani and P~Supraja.
\newblock Classifying the {Autism} and {Epilepsy} {Disorder} {Based} on {EEG}
  {Signal} {Using} {Deep} {Convolutional} {Neural} {Network} ({DCNN}).
\newblock In \emph{2021 {International} {Conference} on {Advance} {Computing}
  and {Innovative} {Technologies} in {Engineering} ({ICACITE})}, pages
  880--886, Greater Noida, India, March 2021. IEEE.
\newblock ISBN 9781728177410.
\newblock \doi{10.1109/ICACITE51222.2021.9404634}.
\newblock URL \url{https://ieeexplore.ieee.org/document/9404634/}.

\bibitem[Tawhid et~al.(2021)Tawhid, Siuly, Wang, Whittaker, Wang, and
  Zhang]{bg_med_asd_tawhid_spectrogram_2021}
Md. Nurul~Ahad Tawhid, Siuly Siuly, Hua Wang, Frank Whittaker, Kate Wang, and
  Yanchun Zhang.
\newblock A spectrogram image based intelligent technique for automatic
  detection of autism spectrum disorder from {EEG}.
\newblock \emph{PLOS ONE}, 16\penalty0 (6):\penalty0 e0253094, June 2021.
\newblock ISSN 1932-6203.
\newblock \doi{10.1371/journal.pone.0253094}.
\newblock URL \url{https://dx.plos.org/10.1371/journal.pone.0253094}.

\bibitem[Islam et~al.(2021)Islam, Islam, Rahman, Mondal, Singha, Ahmad, Awal,
  Islam, and Moni]{bg_med_emotion_islam_eeg_2021}
Md.~Rabiul Islam, Md.~Milon Islam, Md.~Mustafizur Rahman, Chayan Mondal,
  Suvojit~Kumar Singha, Mohiuddin Ahmad, Abdul Awal, Md.~Saiful Islam, and
  Mohammad~Ali Moni.
\newblock {EEG} {Channel} {Correlation} {Based} {Model} for {Emotion}
  {Recognition}.
\newblock \emph{Computers in Biology and Medicine}, 136:\penalty0 104757,
  September 2021.
\newblock ISSN 00104825.
\newblock \doi{10.1016/j.compbiomed.2021.104757}.
\newblock URL
  \url{https://linkinghub.elsevier.com/retrieve/pii/S0010482521005515}.

\bibitem[Liu et~al.(2020)Liu, Ding, Li, Cheng, Song, Wan, and
  Chen]{bg_med_emotion_liu_multi-channel_2020}
Yu~Liu, Yufeng Ding, Chang Li, Juan Cheng, Rencheng Song, Feng Wan, and Xun
  Chen.
\newblock Multi-channel {EEG}-based emotion recognition via a multi-level
  features guided capsule network.
\newblock \emph{Computers in Biology and Medicine}, 123:\penalty0 103927,
  August 2020.
\newblock ISSN 00104825.
\newblock \doi{10.1016/j.compbiomed.2020.103927}.
\newblock URL
  \url{https://linkinghub.elsevier.com/retrieve/pii/S0010482520302663}.

\bibitem[Luo et~al.(2020)Luo, Zhu, Wan, and Lu]{bg_med_emotion_luo_data_2020}
Yun Luo, Li-Zhen Zhu, Zi-Yu Wan, and Bao-Liang Lu.
\newblock Data augmentation for enhancing {EEG}-based emotion recognition with
  deep generative models.
\newblock \emph{Journal of Neural Engineering}, 17\penalty0 (5):\penalty0
  056021, October 2020.
\newblock ISSN 1741-2560, 1741-2552.
\newblock \doi{10.1088/1741-2552/abb580}.
\newblock URL
  \url{https://iopscience.iop.org/article/10.1088/1741-2552/abb580}.

\bibitem[Xiao et~al.(2022)Xiao, Shi, Ye, Xu, Chen, and
  Ren]{bg_med_emotion_xiao_4d_2022}
Guowen Xiao, Meng Shi, Mengwen Ye, Bowen Xu, Zhendi Chen, and Quansheng Ren.
\newblock {4D} attention-based neural network for {EEG} emotion recognition.
\newblock \emph{Cognitive Neurodynamics}, 16\penalty0 (4):\penalty0 805--818,
  August 2022.
\newblock ISSN 1871-4080, 1871-4099.
\newblock \doi{10.1007/s11571-021-09751-5}.
\newblock URL \url{https://link.springer.com/10.1007/s11571-021-09751-5}.

\bibitem[Yin et~al.(2021)Yin, Zheng, Hu, Zhang, and
  Cui]{bg_med_emotion_yin_eeg_2021}
Yongqiang Yin, Xiangwei Zheng, Bin Hu, Yuang Zhang, and Xinchun Cui.
\newblock {EEG} emotion recognition using fusion model of graph convolutional
  neural networks and {LSTM}.
\newblock \emph{Applied Soft Computing}, 100:\penalty0 106954, March 2021.
\newblock ISSN 15684946.
\newblock \doi{10.1016/j.asoc.2020.106954}.
\newblock URL
  \url{https://linkinghub.elsevier.com/retrieve/pii/S1568494620308929}.

\bibitem[Zhong et~al.(2022)Zhong, Wang, and
  Miao]{bg_med_emotion_zhong_eeg-based_2022}
Peixiang Zhong, Di~Wang, and Chunyan Miao.
\newblock {EEG}-{Based} {Emotion} {Recognition} {Using} {Regularized} {Graph}
  {Neural} {Networks}.
\newblock \emph{IEEE Transactions on Affective Computing}, 13\penalty0
  (3):\penalty0 1290--1301, July 2022.
\newblock ISSN 1949-3045, 2371-9850.
\newblock \doi{10.1109/TAFFC.2020.2994159}.
\newblock URL \url{https://ieeexplore.ieee.org/document/9091308/}.

\bibitem[Dissanayake et~al.(2022)Dissanayake, Fernando, Denman, Sridharan, and
  Fookes]{bg_med_seizure_dissanayake_geometric_2022}
Theekshana Dissanayake, Tharindu Fernando, Simon Denman, Sridha Sridharan, and
  Clinton Fookes.
\newblock Geometric {Deep} {Learning} for {Subject} {Independent} {Epileptic}
  {Seizure} {Prediction} {Using} {Scalp} {EEG} {Signals}.
\newblock \emph{IEEE Journal of Biomedical and Health Informatics}, 26\penalty0
  (2):\penalty0 527--538, February 2022.
\newblock ISSN 2168-2194, 2168-2208.
\newblock \doi{10.1109/JBHI.2021.3100297}.
\newblock URL \url{https://ieeexplore.ieee.org/document/9497714/}.

\bibitem[Gao et~al.(2022)Gao, Chen, Liu, Liang, Wu, Qian, Xie, and
  Zhang]{bg_med_seizure_gao_pediatric_2022}
Yikai Gao, Xun Chen, Aiping Liu, Deng Liang, Le~Wu, Ruobing Qian, Hongtao Xie,
  and Yongdong Zhang.
\newblock Pediatric {Seizure} {Prediction} in {Scalp} {EEG} {Using} a
  {Multi}-{Scale} {Neural} {Network} {With} {Dilated} {Convolutions}.
\newblock \emph{IEEE Journal of Translational Engineering in Health and
  Medicine}, 10:\penalty0 1--9, 2022.
\newblock ISSN 2168-2372.
\newblock \doi{10.1109/JTEHM.2022.3144037}.
\newblock URL \url{https://ieeexplore.ieee.org/document/9684436/}.

\bibitem[Jana and Mukherjee(2021)]{bg_med_seizure_jana_deep_2021}
Ranjan Jana and Imon Mukherjee.
\newblock Deep learning based efficient epileptic seizure prediction with {EEG}
  channel optimization.
\newblock \emph{Biomedical Signal Processing and Control}, 68:\penalty0 102767,
  July 2021.
\newblock ISSN 17468094.
\newblock \doi{10.1016/j.bspc.2021.102767}.
\newblock URL
  \url{https://linkinghub.elsevier.com/retrieve/pii/S1746809421003645}.

\bibitem[Li et~al.(2022{\natexlab{a}})Li, Huang, Song, Qian, Liu, and
  Chen]{bg_med_seizure_li_eeg-based_2022}
Chang Li, Xiaoyang Huang, Rencheng Song, Ruobing Qian, Xiang Liu, and Xun Chen.
\newblock {EEG}-based seizure prediction via {Transformer} guided {CNN}.
\newblock \emph{Measurement}, 203:\penalty0 111948, November
  2022{\natexlab{a}}.
\newblock ISSN 02632241.
\newblock \doi{10.1016/j.measurement.2022.111948}.
\newblock URL
  \url{https://linkinghub.elsevier.com/retrieve/pii/S0263224122011447}.

\bibitem[Priya~Prathaban and
  Balasubramanian(2021)]{bg_med_seizure_priya_prathaban_dynamic_2021}
Banu Priya~Prathaban and Ramachandran Balasubramanian.
\newblock Dynamic learning framework for epileptic seizure prediction using
  sparsity based {EEG} {Reconstruction} with {Optimized} {CNN} classifier.
\newblock \emph{Expert Systems with Applications}, 170:\penalty0 114533, May
  2021.
\newblock ISSN 09574174.
\newblock \doi{10.1016/j.eswa.2020.114533}.
\newblock URL
  \url{https://linkinghub.elsevier.com/retrieve/pii/S0957417420311775}.

\bibitem[Ra et~al.(2021)Ra, Li, and Li]{bg_med_seizure_ra_novel_2021}
Jee~S. Ra, Tianning Li, and Yan Li.
\newblock A {Novel} {Permutation} {Entropy}-{Based} {EEG} {Channel} {Selection}
  for {Improving} {Epileptic} {Seizure} {Prediction}.
\newblock \emph{Sensors}, 21\penalty0 (23):\penalty0 7972, November 2021.
\newblock ISSN 1424-8220.
\newblock \doi{10.3390/s21237972}.
\newblock URL \url{https://www.mdpi.com/1424-8220/21/23/7972}.

\bibitem[Yang et~al.(2021)Yang, Zhao, Sun, Lu, and
  Ma]{bg_med_seizure_yang_effective_2021}
Xinwu Yang, Jiaqi Zhao, Qi~Sun, Jianbo Lu, and Xu~Ma.
\newblock An {Effective} {Dual} {Self}-{Attention} {Residual} {Network} for
  {Seizure} {Prediction}.
\newblock \emph{IEEE Transactions on Neural Systems and Rehabilitation
  Engineering}, 29:\penalty0 1604--1613, 2021.
\newblock ISSN 1534-4320, 1558-0210.
\newblock \doi{10.1109/TNSRE.2021.3103210}.
\newblock URL \url{https://ieeexplore.ieee.org/document/9508965/}.

\bibitem[Zhang et~al.(2021{\natexlab{a}})Zhang, Chen, Ranjan, Ke, Tang, and
  Zomaya]{bg_med_seizure_zhang_lightweight_2021}
Shasha Zhang, Dan Chen, Rajiv Ranjan, Hengjin Ke, Yunbo Tang, and Albert~Y.
  Zomaya.
\newblock A lightweight solution to epileptic seizure prediction based on {EEG}
  synchronization measurement.
\newblock \emph{The Journal of Supercomputing}, 77\penalty0 (4):\penalty0
  3914--3932, April 2021{\natexlab{a}}.
\newblock ISSN 0920-8542, 1573-0484.
\newblock \doi{10.1007/s11227-020-03426-4}.
\newblock URL \url{https://link.springer.com/10.1007/s11227-020-03426-4}.

\bibitem[Barua et~al.(2019)Barua, Ahmed, Ahlström, and
  Begum]{bg_med_fatigue_barua_automatic_2019}
Shaibal Barua, Mobyen~Uddin Ahmed, Christer Ahlström, and Shahina Begum.
\newblock Automatic driver sleepiness detection using {EEG}, {EOG} and
  contextual information.
\newblock \emph{Expert Systems with Applications}, 115:\penalty0 121--135,
  January 2019.
\newblock ISSN 09574174.
\newblock \doi{10.1016/j.eswa.2018.07.054}.
\newblock URL
  \url{https://linkinghub.elsevier.com/retrieve/pii/S0957417418304792}.

\bibitem[Wang et~al.(2018{\natexlab{a}})Wang, Li, and
  Liu]{bg_med_fatigue_wang_analysis_2018}
Qingjun Wang, Yibo Li, and Xueping Liu.
\newblock Analysis of {Feature} {Fatigue} {EEG} {Signals} {Based} on {Wavelet}
  {Entropy}.
\newblock \emph{International Journal of Pattern Recognition and Artificial
  Intelligence}, 32\penalty0 (08):\penalty0 1854023, August 2018{\natexlab{a}}.
\newblock ISSN 0218-0014, 1793-6381.
\newblock \doi{10.1142/S021800141854023X}.
\newblock URL
  \url{https://www.worldscientific.com/doi/abs/10.1142/S021800141854023X}.

\bibitem[Xiong et~al.(2016)Xiong, Gao, Yang, Yu, and
  Huang]{bg_med_fatigue_xiong_classifying_2016}
Yijun Xiong, Junfeng Gao, Yong Yang, Xiaolin Yu, and Wentao Huang.
\newblock Classifying {Driving} {Fatigue} {Based} on {Combined} {Entropy}
  {Measure} {Using} {EEG} {Signals}.
\newblock \emph{International Journal of Control and Automation}, 9\penalty0
  (3):\penalty0 329--338, March 2016.
\newblock ISSN 20054297, 20054297.
\newblock \doi{10.14257/ijca.2016.9.3.30}.
\newblock URL \url{http://article.nadiapub.com/IJCA/vol9_no3/30.pdf}.

\bibitem[Zhang et~al.(2017)Zhang, Li, Liu, Zhang, Wang, Luo, Zhou, Zhu, Salman,
  Hu, and Wang]{bg_med_fatigue_zhang_design_2017}
Xiaoliang Zhang, Jiali Li, Yugang Liu, Zutao Zhang, Zhuojun Wang, Dianyuan Luo,
  Xiang Zhou, Miankuan Zhu, Waleed Salman, Guangdi Hu, and Chunbai Wang.
\newblock Design of a {Fatigue} {Detection} {System} for {High}-{Speed}
  {Trains} {Based} on {Driver} {Vigilance} {Using} a {Wireless} {Wearable}
  {EEG}.
\newblock \emph{Sensors}, 17\penalty0 (3):\penalty0 486, March 2017.
\newblock ISSN 1424-8220.
\newblock \doi{10.3390/s17030486}.
\newblock URL \url{http://www.mdpi.com/1424-8220/17/3/486}.

\bibitem[Cheng et~al.(2022)Cheng, Wei, Du, Qiu, Tian, Ma, and
  He]{bg_med_vigilance_cheng_vigilancenet:_2022}
Xinyu Cheng, Wei Wei, Changde Du, Shuang Qiu, Sanli Tian, Xiaojun Ma, and
  Huiguang He.
\newblock {VigilanceNet}: {Decouple} {Intra}- and {Inter}-{Modality} {Learning}
  for {Multimodal} {Vigilance} {Estimation} in {RSVP}-{Based} {BCI}.
\newblock In \emph{Proceedings of the 30th {ACM} {International} {Conference}
  on {Multimedia}}, pages 209--217, Lisboa Portugal, October 2022. ACM.
\newblock ISBN 9781450392037.
\newblock \doi{10.1145/3503161.3548367}.
\newblock URL \url{https://dl.acm.org/doi/10.1145/3503161.3548367}.

\bibitem[Luo and
  Lu(2021{\natexlab{a}})]{bg_med_vigilance_luo_wasserstein-distance-based_2021}
Yun Luo and Bao-Liang Lu.
\newblock Wasserstein-{Distance}-{Based} {Multi}-{Source} {Adversarial}
  {Domain} {Adaptation} for {Emotion} {Recognition} and {Vigilance}
  {Estimation}.
\newblock In \emph{2021 {IEEE} {International} {Conference} on {Bioinformatics}
  and {Biomedicine} ({BIBM})}, pages 1424--1428, Houston, TX, USA, December
  2021{\natexlab{a}}. IEEE.
\newblock ISBN 9781665401265.
\newblock \doi{10.1109/BIBM52615.2021.9669383}.
\newblock URL \url{https://ieeexplore.ieee.org/document/9669383/}.

\bibitem[Song et~al.(2021{\natexlab{a}})Song, Zhou, and
  Wang]{bg_med_vigilance_song_deep_2021}
Kuiyong Song, Lianke Zhou, and Hongbin Wang.
\newblock Deep {Coupling} {Recurrent} {Auto}-{Encoder} with {Multi}-{Modal}
  {EEG} and {EOG} for {Vigilance} {Estimation}.
\newblock \emph{Entropy}, 23\penalty0 (10):\penalty0 1316, October
  2021{\natexlab{a}}.
\newblock ISSN 1099-4300.
\newblock \doi{10.3390/e23101316}.
\newblock URL \url{https://www.mdpi.com/1099-4300/23/10/1316}.

\bibitem[Wang et~al.(2021{\natexlab{a}})Wang, Qiu, Wei, Zhang, He, Xu, and
  Ming]{bg_med_vigilance_wang_vigilance_2021}
Kangning Wang, Shuang Qiu, Wei Wei, Chuncheng Zhang, Huiguang He, Minpeng Xu,
  and Dong Ming.
\newblock Vigilance {Estimating} in {SSVEP}-{Based} {BCI} {Using} {Multimodal}
  {Signals}.
\newblock In \emph{2021 43rd {Annual} {International} {Conference} of the
  {IEEE} {Engineering} in {Medicine} \& {Biology} {Society} ({EMBC})}, pages
  5974--5978, Mexico, November 2021{\natexlab{a}}. IEEE.
\newblock ISBN 9781728111797.
\newblock \doi{10.1109/EMBC46164.2021.9629736}.
\newblock URL \url{https://ieeexplore.ieee.org/document/9629736/}.

\bibitem[Fu et~al.(2022)Fu, Huang, Zhang, Wang, and
  Zhang]{bg_med_sleep_fu_deep_2022}
Ziyang Fu, Chen Huang, Li~Zhang, Shihui Wang, and Yan Zhang.
\newblock Deep {Learning} {Model} of {Sleep} {EEG} {Signal} by {Using}
  {Bidirectional} {Recurrent} {Neural} {Network} {Encoding} and {Decoding}.
\newblock \emph{Electronics}, 11\penalty0 (17):\penalty0 2644, August 2022.
\newblock ISSN 2079-9292.
\newblock \doi{10.3390/electronics11172644}.
\newblock URL \url{https://www.mdpi.com/2079-9292/11/17/2644}.

\bibitem[Phan et~al.(2022)Phan, Mikkelsen, Chen, Koch, Mertins, and
  De~Vos]{bg_med_sleep_phan_sleeptransformer:_2022}
Huy Phan, Kaare Mikkelsen, Oliver~Y. Chen, Philipp Koch, Alfred Mertins, and
  Maarten De~Vos.
\newblock {SleepTransformer}: {Automatic} {Sleep} {Staging} {With}
  {Interpretability} and {Uncertainty} {Quantification}.
\newblock \emph{IEEE Transactions on Biomedical Engineering}, 69\penalty0
  (8):\penalty0 2456--2467, August 2022.
\newblock ISSN 0018-9294, 1558-2531.
\newblock \doi{10.1109/TBME.2022.3147187}.
\newblock URL \url{https://ieeexplore.ieee.org/document/9697331/}.

\bibitem[Sudhakar et~al.(2021)Sudhakar, Hari~Krishnan, Krishnamoorthy,
  Janney~J, Pradeepa, and Raghavi]{bg_med_sleep_sudhakar_sleep_2021}
T.~Sudhakar, G.~Hari~Krishnan, N.~R. Krishnamoorthy, Bethanney Janney~J,
  M.~Pradeepa, and J.~P. Raghavi.
\newblock Sleep {Disorder} {Diagnosis} using {EEG} based {Deep} {Learning}
  {Techniques}.
\newblock In \emph{2021 {Seventh} {International} conference on {Bio}
  {Signals}, {Images}, and {Instrumentation} ({ICBSII})}, pages 1--4, Chennai,
  India, March 2021. IEEE.
\newblock ISBN 9781665441261.
\newblock \doi{10.1109/ICBSII51839.2021.9445158}.
\newblock URL \url{https://ieeexplore.ieee.org/document/9445158/}.

\bibitem[Wang et~al.(2022{\natexlab{b}})Wang, Xiao, Fang, Li, Wang, and
  Zhao]{bg_med_sleep_wang_bi_2022}
Yao Wang, Zhuangwen Xiao, Shuaiwen Fang, Weiming Li, Jinhai Wang, and Xiaoyun
  Zhao.
\newblock {BI} - {Directional} long short-term memory for automatic detection
  of sleep apnea events based on single channel {EEG} signal.
\newblock \emph{Computers in Biology and Medicine}, 142:\penalty0 105211, March
  2022{\natexlab{b}}.
\newblock ISSN 00104825.
\newblock \doi{10.1016/j.compbiomed.2022.105211}.
\newblock URL
  \url{https://linkinghub.elsevier.com/retrieve/pii/S0010482522000038}.

\bibitem[Bhuvaneshwari et~al.(2023)Bhuvaneshwari, Grace Mary~Kanaga, and
  Anitha]{bg_med_ssvep_bhuvaneshwari_bio-inspired_2023}
M.~Bhuvaneshwari, E.~Grace Mary~Kanaga, and J.~Anitha.
\newblock Bio-inspired {Red} {Fox}-{Sine} cosine optimization for the feature
  selection of {SSVEP}-based {EEG} signals for {BCI} applications.
\newblock \emph{Biomedical Signal Processing and Control}, 80:\penalty0 104245,
  February 2023.
\newblock ISSN 17468094.
\newblock \doi{10.1016/j.bspc.2022.104245}.
\newblock URL
  \url{https://linkinghub.elsevier.com/retrieve/pii/S1746809422006991}.

\bibitem[Karunasena et~al.(2021)Karunasena, Ariyarathna, Ranaweera,
  Wijayakulasooriya, Kim, and
  Dassanayake]{bg_med_ssvep_karunasena_single-channel_2021}
Sanduni~P. Karunasena, Darshana~C. Ariyarathna, Ruwan Ranaweera, Janaka
  Wijayakulasooriya, Kwangtaek Kim, and Tharaka Dassanayake.
\newblock Single-{Channel} {EEG} {SSVEP}-based {BCI} for {Robot} {Arm}
  {Control}.
\newblock In \emph{2021 {IEEE} {Sensors} {Applications} {Symposium} ({SAS})},
  pages 1--6, Sundsvall, Sweden, August 2021. IEEE.
\newblock ISBN 9781728194318.
\newblock \doi{10.1109/SAS51076.2021.9530189}.
\newblock URL \url{https://ieeexplore.ieee.org/document/9530189/}.

\bibitem[Liu et~al.(2022{\natexlab{b}})Liu, Chen, Li, Wang, Gao, and
  Gao]{bg_med_ssvep_liu_align_2022}
Bingchuan Liu, Xiaogang Chen, Xiang Li, Yijun Wang, Xiaorong Gao, and Shangkai
  Gao.
\newblock Align and {Pool} for {EEG} {Headset} {Domain} {Adaptation} ({ALPHA})
  to {Facilitate} {Dry} {Electrode} {Based} {SSVEP}-{BCI}.
\newblock \emph{IEEE Transactions on Biomedical Engineering}, 69\penalty0
  (2):\penalty0 795--806, February 2022{\natexlab{b}}.
\newblock ISSN 0018-9294, 1558-2531.
\newblock \doi{10.1109/TBME.2021.3105331}.
\newblock URL \url{https://ieeexplore.ieee.org/document/9516951/}.

\bibitem[Sun et~al.(2022)Sun, Ding, Liu, Zheng, Chen, Hui, Na, Wang, and
  Fan]{bg_med_ssvep_sun_cross-subject_2022}
Ying Sun, Wenzheng Ding, Xiaolin Liu, Dezhi Zheng, Xinlei Chen, Qianxin Hui,
  Rui Na, Shuai Wang, and Shangchun Fan.
\newblock Cross-subject fusion based on time-weighting canonical correlation
  analysis in {SSVEP}-{BCIs}.
\newblock \emph{Measurement}, 199:\penalty0 111524, August 2022.
\newblock ISSN 02632241.
\newblock \doi{10.1016/j.measurement.2022.111524}.
\newblock URL
  \url{https://linkinghub.elsevier.com/retrieve/pii/S026322412200745X}.

\bibitem[Tabanfar et~al.(2023)Tabanfar, Ghassemi, and
  Hassan~Moradi]{bg_med_ssvep_tabanfar_subject-independent_2023}
Zahra Tabanfar, Farnaz Ghassemi, and Mohammad Hassan~Moradi.
\newblock A subject-independent {SSVEP}-based {BCI} target detection system
  based on fuzzy ordering of {EEG} task-related components.
\newblock \emph{Biomedical Signal Processing and Control}, 79:\penalty0 104171,
  January 2023.
\newblock ISSN 17468094.
\newblock \doi{10.1016/j.bspc.2022.104171}.
\newblock URL
  \url{https://linkinghub.elsevier.com/retrieve/pii/S1746809422006255}.

\bibitem[Yan et~al.(2022)Yan, Wu, Du, and
  Xu]{bg_med_ssvep_yan_cross-subject_2022}
Wenqiang Yan, Yongcheng Wu, Chenghang Du, and Guanghua Xu.
\newblock Cross-subject spatial filter transfer method for {SSVEP}-{EEG}
  feature recognition.
\newblock \emph{Journal of Neural Engineering}, 19\penalty0 (3):\penalty0
  036008, June 2022.
\newblock ISSN 1741-2560, 1741-2552.
\newblock \doi{10.1088/1741-2552/ac6b57}.
\newblock URL
  \url{https://iopscience.iop.org/article/10.1088/1741-2552/ac6b57}.

\bibitem[Zhu et~al.(2021{\natexlab{a}})Zhu, Li, Lu, and
  Li]{bg_med_ssvep_zhu_eegnet_2021}
Yuanlu Zhu, Ying Li, Jinling Lu, and Pengcheng Li.
\newblock {EEGNet} {With} {Ensemble} {Learning} to {Improve} the
  {Cross}-{Session} {Classification} of {SSVEP} {Based} {BCI} {From}
  {Ear}-{EEG}.
\newblock \emph{IEEE Access}, 9:\penalty0 15295--15303, 2021{\natexlab{a}}.
\newblock ISSN 2169-3536.
\newblock \doi{10.1109/ACCESS.2021.3052656}.
\newblock URL \url{https://ieeexplore.ieee.org/document/9328251/}.

\bibitem[Bagchi and Bathula(2021)]{bg_med_vpa_bagchi_adequately_2021}
Subhranil Bagchi and Deepti~R. Bathula.
\newblock Adequately {Wide} {1D} {CNN} facilitates improved {EEG} based
  {Visual} {Object} {Recognition}.
\newblock In \emph{2021 29th {European} {Signal} {Processing} {Conference}
  ({EUSIPCO})}, pages 1276--1280, Dublin, Ireland, August 2021. IEEE.
\newblock ISBN 9789082797060.
\newblock \doi{10.23919/EUSIPCO54536.2021.9615945}.
\newblock URL \url{https://ieeexplore.ieee.org/document/9615945/}.

\bibitem[Bagchi and Bathula(2022)]{bg_med_vpa_bagchi_eeg-convtransformer_2022}
Subhranil Bagchi and Deepti~R. Bathula.
\newblock {EEG}-{ConvTransformer} for single-trial {EEG}-based visual stimulus
  classification.
\newblock \emph{Pattern Recognition}, 129:\penalty0 108757, September 2022.
\newblock ISSN 00313203.
\newblock \doi{10.1016/j.patcog.2022.108757}.
\newblock URL
  \url{https://linkinghub.elsevier.com/retrieve/pii/S0031320322002382}.

\bibitem[Kumari et~al.(2022)Kumari, Anwar, and
  Bhattacharjee]{bg_med_vpa_kumari_automated_2022}
Nandini Kumari, Shamama Anwar, and Vandana Bhattacharjee.
\newblock Automated visual stimuli evoked multi-channel {EEG} signal
  classification using {EEGCapsNet}.
\newblock \emph{Pattern Recognition Letters}, 153:\penalty0 29--35, January
  2022.
\newblock ISSN 01678655.
\newblock \doi{10.1016/j.patrec.2021.11.019}.
\newblock URL
  \url{https://linkinghub.elsevier.com/retrieve/pii/S0167865521004104}.

\bibitem[Lee et~al.(2022{\natexlab{a}})Lee, Hwang, Lee, Shin, Jeon, and
  Byun]{bg_med_vpa_lee_inter-subject_2022}
Pilhyeon Lee, Sunhee Hwang, Jewook Lee, Minjung Shin, Seogkyu Jeon, and Hyeran
  Byun.
\newblock Inter-subject {Contrastive} {Learning} for {Subject} {Adaptive}
  {EEG}-based {Visual} {Recognition}.
\newblock In \emph{2022 10th {International} {Winter} {Conference} on
  {Brain}-{Computer} {Interface} ({BCI})}, pages 1--6, Gangwon-do, Korea,
  Republic of, February 2022{\natexlab{a}}. IEEE.
\newblock ISBN 9781665413374.
\newblock \doi{10.1109/BCI53720.2022.9734886}.
\newblock URL \url{https://ieeexplore.ieee.org/document/9734886/}.

\bibitem[Zheng and Chen(2021)]{bg_med_vpa_zheng_attention-based_2021}
Xiao Zheng and Wanzhong Chen.
\newblock An {Attention}-based {Bi}-{LSTM} {Method} for {Visual} {Object}
  {Classification} via {EEG}.
\newblock \emph{Biomedical Signal Processing and Control}, 63:\penalty0 102174,
  January 2021.
\newblock ISSN 17468094.
\newblock \doi{10.1016/j.bspc.2020.102174}.
\newblock URL
  \url{https://linkinghub.elsevier.com/retrieve/pii/S174680942030313X}.

\bibitem[Laar et~al.(2013{\natexlab{b}})Laar, Brugman, Nijboer, Poel, Nijholt,
  and Miller]{bg_med_thought_laar_brainbrush_2013}
van~de Laar, Bram, Ivo Brugman, Femke Nijboer, Mannes Poel, Anton Nijholt, and
  Leslie Miller.
\newblock \emph{{BrainBrush}, a multimodal application for creative
  expressivity}.
\newblock IARIA XPS Press, February 2013{\natexlab{b}}.
\newblock ISBN 9781612082509.
\newblock OCLC: 6893370982.

\bibitem[Agarwal and
  Kumar(2022)]{bg_med_imagined_speech_agarwal_electroencephalographybased_2022}
Prabhakar Agarwal and Sandeep Kumar.
\newblock Electroencephalography‐based imagined speech recognition using deep
  long short‐term memory network.
\newblock \emph{ETRI Journal}, 44\penalty0 (4):\penalty0 672--685, August 2022.
\newblock ISSN 1225-6463, 2233-7326.
\newblock \doi{10.4218/etrij.2021-0118}.
\newblock URL
  \url{https://onlinelibrary.wiley.com/doi/10.4218/etrij.2021-0118}.

\bibitem[Kumar and Scheme(2021)]{bg_med_imagined_speech_kumar_deep_2021}
Pradeep Kumar and Erik Scheme.
\newblock A {Deep} {Spatio}-{Temporal} {Model} for {EEG}-{Based} {Imagined}
  {Speech} {Recognition}.
\newblock In \emph{{ICASSP} 2021 - 2021 {IEEE} {International} {Conference} on
  {Acoustics}, {Speech} and {Signal} {Processing} ({ICASSP})}, pages 995--999,
  Toronto, ON, Canada, June 2021. IEEE.
\newblock ISBN 9781728176055.
\newblock \doi{10.1109/ICASSP39728.2021.9413989}.
\newblock URL \url{https://ieeexplore.ieee.org/document/9413989/}.

\bibitem[Li et~al.(2021{\natexlab{a}})Li, Chao, Li, Fu, Ji, Wu, and
  Shi]{bg_med_imagined_speech_li_decoding_2021}
Fu~Li, Weibing Chao, Yang Li, Boxun Fu, Youshuo Ji, Hao Wu, and Guangming Shi.
\newblock Decoding imagined speech from {EEG} signals using hybrid-scale
  spatial-temporal dilated convolution network.
\newblock \emph{Journal of Neural Engineering}, 18\penalty0 (4):\penalty0
  0460c4, August 2021{\natexlab{a}}.
\newblock ISSN 1741-2560, 1741-2552.
\newblock \doi{10.1088/1741-2552/ac13c0}.
\newblock URL
  \url{https://iopscience.iop.org/article/10.1088/1741-2552/ac13c0}.

\bibitem[Sabour et~al.(2017{\natexlab{a}})Sabour, Frosst, and
  Hinton]{sabour2017dynamic}
Sara Sabour, Nicholas Frosst, and Geoffrey~E Hinton.
\newblock Dynamic routing between capsules.
\newblock \emph{Advances in neural information processing systems}, 30,
  2017{\natexlab{a}}.

\bibitem[Ganin et~al.(2016)Ganin, Ustinova, Ajakan, Germain, Larochelle,
  Laviolette, Marchand, and Lempitsky]{ganin2016domain}
Yaroslav Ganin, Evgeniya Ustinova, Hana Ajakan, Pascal Germain, Hugo
  Larochelle, Fran{\c{c}}ois Laviolette, Mario Marchand, and Victor Lempitsky.
\newblock Domain-adversarial training of neural networks.
\newblock \emph{The journal of machine learning research}, 17\penalty0
  (1):\penalty0 2096--2030, 2016.

\bibitem[Zhao et~al.(2021{\natexlab{a}})Zhao, Zheng, Ma, Li, and
  Zheng]{Paper_Zhao_26}
He~Zhao, Qingqing Zheng, Kai Ma, Huiqi Li, and Yefeng Zheng.
\newblock Deep representation-based domain adaptation for nonstationary eeg
  classification.
\newblock \emph{IEEE Transactions on Neural Networks and Learning Systems},
  32:\penalty0 535--545, 2021{\natexlab{a}}.

\bibitem[Lebedeva(2020)]{Paper_Lebedeva_RP44}
E.~Yu. Lebedeva.
\newblock Data augmentation for domain-adversarial training in eeg-based
  emotion recognition.
\newblock 2020.

\bibitem[Ganin and Lempitsky(2015)]{ganin2015unsupervised}
Yaroslav Ganin and Victor Lempitsky.
\newblock Unsupervised domain adaptation by backpropagation.
\newblock In \emph{International conference on machine learning}, pages
  1180--1189. PMLR, 2015.

\bibitem[Su et~al.(2021)Su, Shen, Peng, and Hu]{Paper_Su_fMRI_2}
Jianpo Su, Hui Shen, Limin Peng, and Dewen Hu.
\newblock Few-shot domain-adaptive anomaly detection for cross-site brain
  images.
\newblock \emph{IEEE Transactions on Pattern Analysis and Machine
  Intelligence}, PP:\penalty0 1--1, 2021.

\bibitem[Heremans et~al.(2022)Heremans, Phan, Borz{\'e}e, Buyse, Testelmans,
  and Vos]{Paper_Heremans_RP43}
Elisabeth Roxane~Marie Heremans, Huy~P Phan, Pascal Borz{\'e}e, Bertien Buyse,
  Dries Testelmans, and Maarten~De Vos.
\newblock From unsupervised to semi-supervised adversarial domain adaptation in
  electroencephalography-based sleep staging.
\newblock \emph{Journal of Neural Engineering}, 19, 2022.

\bibitem[Zhao et~al.(2022)Zhao, Zhang, Zhang, Xiao, Wang, Xu, and
  Zheng]{Paper_Zhao_92}
Yanna Zhao, Gaobo Zhang, Yongfeng Zhang, Tiantian Xiao, Ziwei Wang, Fangzhou
  Xu, and Yuanjie Zheng.
\newblock Multi-view cross-subject seizure detection with information
  bottleneck attribution.
\newblock \emph{Journal of Neural Engineering}, 19, 2022.

\bibitem[He et~al.(2022{\natexlab{a}})He, Zhong, and Pan]{Paper_He_20}
Zhipeng He, Yongshi Zhong, and Jiahui Pan.
\newblock An adversarial discriminative temporal convolutional network for
  eeg-based cross-domain emotion recognition.
\newblock \emph{Computers in biology and medicine}, page 105048,
  2022{\natexlab{a}}.

\bibitem[Liu et~al.(2022{\natexlab{c}})Liu, Li, Yao, J.M.Monaghan, and
  McAlpine]{Paper_Liu_48}
Zhe Liu, Yun Li, L.~Yao, Jessica J.M.Monaghan, and David McAlpine.
\newblock Disentangled and side-aware unsupervised domain adaptation for
  cross-dataset subjective tinnitus diagnosis.
\newblock \emph{ArXiv}, abs/2205.03230, 2022{\natexlab{c}}.

\bibitem[Wang et~al.(2021{\natexlab{b}})Wang, Zhang, Xu, Ping, and
  Chu]{Paper_Wang_49}
Fei Wang, Weiwei Zhang, Zongfeng Xu, Jingyu Ping, and Hao Chu.
\newblock A deep multi-source adaptation transfer network for cross-subject
  electroencephalogram emotion recognition.
\newblock \emph{Neural Comput. Appl.}, 33:\penalty0 9061--9073,
  2021{\natexlab{b}}.

\bibitem[Pominova et~al.(2021{\natexlab{a}})Pominova, Kondrateva, Sharaev,
  Bernstein, and Burnaev]{Paper_Pominova_fMRI_15}
Marina Pominova, Ekaterina Kondrateva, M.~Sharaev, Alexander~V. Bernstein, and
  Evgeny Burnaev.
\newblock Fader networks for domain adaptation on fmri: Abide-ii study.
\newblock In \emph{International Conference on Machine Vision},
  2021{\natexlab{a}}.

\bibitem[Pominova et~al.(2021{\natexlab{b}})Pominova, Kondrateva, Sharaev,
  Bernstein, and Burnaev]{pominova2021fader}
Marina Pominova, Ekaterina Kondrateva, Maxim Sharaev, Alexander Bernstein, and
  Evegeny Burnaev.
\newblock Fader networks for domain adaptation on fmri: abide-ii study.
\newblock In \emph{Thirteenth International Conference on Machine Vision},
  volume 11605, pages 570--577. SPIE, 2021{\natexlab{b}}.

\bibitem[Li et~al.(2021{\natexlab{b}})Li, Wang, Zheng, Zong, Qi, Cui, Zhang,
  and Song]{Paper_Li_55}
Y.~Li, Lei Wang, Wenming Zheng, Yuan Zong, Lei Qi, Zhen Cui, Tong Zhang, and
  Tengfei Song.
\newblock A novel bi-hemispheric discrepancy model for eeg emotion recognition.
\newblock \emph{IEEE Transactions on Cognitive and Developmental Systems},
  13:\penalty0 354--367, 2021{\natexlab{b}}.

\bibitem[Eldele et~al.(2021)Eldele, Ragab, Chen, Wu, Kwoh, Li, and
  Guan]{Paper_Eldele_78}
Emadeldeen Eldele, Mohamed Ragab, Zhenghua Chen, Min Wu, C.~Kwoh, Xiaoli Li,
  and Cuntai Guan.
\newblock Adast: Attentive cross-domain eeg-based sleep staging framework with
  iterative self-training.
\newblock 2021.

\bibitem[Bao et~al.(2020)Bao, Zhuang, Tong, Yan, Shu, Wang, Zeng, and
  Shen]{Paper_Bao_91}
Guangcheng Bao, Ning Zhuang, Li~Tong, Bin Yan, Jun Shu, Linyuan Wang, Ying
  Zeng, and Zhichong Shen.
\newblock Two-level domain adaptation neural network for eeg-based emotion
  recognition.
\newblock \emph{Frontiers in Human Neuroscience}, 14, 2020.

\bibitem[Gretton et~al.(2012)Gretton, Borgwardt, Rasch, Sch{\"o}lkopf, and
  Smola]{gretton2012kernel}
Arthur Gretton, Karsten~M Borgwardt, Malte~J Rasch, Bernhard Sch{\"o}lkopf, and
  Alexander Smola.
\newblock A kernel two-sample test.
\newblock \emph{The Journal of Machine Learning Research}, 13\penalty0
  (1):\penalty0 723--773, 2012.

\bibitem[Avramidis et~al.(2022)Avramidis, Garoufis, Zlatintsi, and
  Maragos]{Paper_Avramidis_46}
Kleanthis Avramidis, Christos Garoufis, Athanasia Zlatintsi, and Petros
  Maragos.
\newblock Enhancing affective representations of music-induced eeg through
  multimodal supervision and latent domain adaptation.
\newblock In \emph{ICASSP}, 2022.

\bibitem[Rayatdoost et~al.(2021)Rayatdoost, Yin, Rudrauf, and
  Soleymani]{Paper_Rayatdoost_23}
Soheil Rayatdoost, Yufeng Yin, David Rudrauf, and M.~Soleymani.
\newblock Subject-invariant eeg representation learning for emotion
  recognition.
\newblock \emph{ICASSP 2021 - 2021 IEEE International Conference on Acoustics,
  Speech and Signal Processing (ICASSP)}, pages 3955--3959, 2021.

\bibitem[Ding et~al.(2021)Ding, Kimura, ichi Fukui, and Numao]{Paper_Ding_51}
Ke~Ding, Tsukasa Kimura, Ken ichi Fukui, and Masayuki Numao.
\newblock Eeg emotion enhancement using task-specific domain adversarial neural
  network.
\newblock \emph{2021 International Joint Conference on Neural Networks
  (IJCNN)}, pages 1--8, 2021.

\bibitem[Saito et~al.(2018{\natexlab{b}})Saito, Watanabe, Ushiku, and
  Harada]{saito2018maximum}
Kuniaki Saito, Kohei Watanabe, Yoshitaka Ushiku, and Tatsuya Harada.
\newblock Maximum classifier discrepancy for unsupervised domain adaptation.
\newblock In \emph{Proceedings of the IEEE conference on computer vision and
  pattern recognition}, pages 3723--3732, 2018{\natexlab{b}}.

\bibitem[Tang and Zhang(2020)]{Paper_Tang_63}
Xingliang Tang and Xianrui Zhang.
\newblock Conditional adversarial domain adaptation neural network for motor
  imagery eeg decoding.
\newblock \emph{Entropy}, 22, 2020.

\bibitem[Huang et~al.(2020)Huang, Hsieh, Yang, and Lee]{Paper_Huang_fMRI_25}
Ya-Lin Huang, Wan-Ting Hsieh, Hao-Chun Yang, and Chi-Chun Lee.
\newblock Conditional domain adversarial transfer for robust cross-site adhd
  classification using functional mri.
\newblock \emph{ICASSP 2020 - 2020 IEEE International Conference on Acoustics,
  Speech and Signal Processing (ICASSP)}, pages 1190--1194, 2020.

\bibitem[Hong et~al.(2021)Hong, Zheng, Liu, Chen, Ma, Gao, and
  Zheng]{Paper_Hong_RP38}
Xiaolin Hong, Qingqing Zheng, Luyan Liu, Peiyin Chen, Kai Ma, Zhongke Gao, and
  Yefeng Zheng.
\newblock Dynamic joint domain adaptation network for motor imagery
  classification.
\newblock \emph{IEEE Transactions on Neural Systems and Rehabilitation
  Engineering}, 29:\penalty0 556--565, 2021.

\bibitem[Cai et~al.(2021)Cai, Guo, Yang, Chen, and Xu]{Paper_Cai_RP26}
Ziliang Cai, Miaomiao Guo, Xinsheng Yang, Xintong Chen, and Guizhi Xu.
\newblock [cross-subject electroencephalogram emotion recognition based on
  maximum classifier discrepancy].
\newblock \emph{Sheng wu yi xue gong cheng xue za zhi = Journal of biomedical
  engineering = Shengwu yixue gongchengxue zazhi}, 38 3:\penalty0 455--462,
  2021.

\bibitem[Li et~al.(2020{\natexlab{a}})Li, Qiu, Du, Wang, and He]{Paper_Li_30}
Jinpeng Li, Shuang Qiu, Changde Du, Yixin Wang, and Huiguang He.
\newblock Domain adaptation for eeg emotion recognition based on latent
  representation similarity.
\newblock \emph{IEEE Transactions on Cognitive and Developmental Systems},
  12\penalty0 (2):\penalty0 344--353, 2020{\natexlab{a}}.
\newblock \doi{10.1109/TCDS.2019.2949306}.

\bibitem[Ye et~al.(2021)Ye, Zhu, Li, Pan, and He]{Paper_Ye_85}
Yalan Ye, Xinyue Zhu, Yunxia Li, Tongjie Pan, and Wenwen He.
\newblock Cross-subject eeg-based emotion recognition using adversarial domain
  adaption with attention mechanism.
\newblock \emph{2021 43rd Annual International Conference of the IEEE
  Engineering in Medicine \& Biology Society (EMBC)}, pages 1140--1144, 2021.

\bibitem[Zeng et~al.(2021)Zeng, Li, Borghini, Zhao, Aric{\'o}, Flumeri,
  Sciaraffa, Zakaria, Kong, and Babiloni]{Paper_Zeng_RP11}
Hong Zeng, Xiufeng Li, Gianluca Borghini, Yue Zhao, Pietro Aric{\'o},
  Gianluca~Di Flumeri, Nicolina Sciaraffa, Wael Zakaria, Wanzeng Kong, and
  Fabio Babiloni.
\newblock An eeg-based transfer learning method for cross-subject fatigue
  mental state prediction.
\newblock \emph{Sensors (Basel, Switzerland)}, 21, 2021.

\bibitem[Li et~al.(2020{\natexlab{b}})Li, Gu, Dvornek, Staib, Ventola, and
  Duncan]{Paper_Li_fMRI_19}
Xiaoxiao Li, Yufeng Gu, Nicha~C. Dvornek, Lawrence~H. Staib, Pamela Ventola,
  and James~S. Duncan.
\newblock Multi-site fmri analysis using privacy-preserving federated learning
  and domain adaptation: Abide results.
\newblock \emph{Medical image analysis}, 65:\penalty0 101765,
  2020{\natexlab{b}}.

\bibitem[Zhu et~al.(2022)Zhu, Ye, Lu, Li, and Wu]{Paper_Zhu_69}
Xinyue Zhu, Yalan Ye, Li~Lu, Yunxia Li, and Haohui Wu.
\newblock Cross-session eeg-based emotion recognition via maximizing domain
  discrepancy.
\newblock \emph{2022 3rd International Conference on Electronic Communication
  and Artificial Intelligence (IWECAI)}, pages 568--572, 2022.

\bibitem[Jeon et~al.(2021)Jeon, Ko, Yoon, and Suk]{Paper_Jeon_39}
Eunjin Jeon, Wonjun Ko, Jee~Seok Yoon, and Heung-Il Suk.
\newblock Mutual information-driven subject-invariant and class-relevant deep
  representation learning in bci.
\newblock \emph{IEEE transactions on neural networks and learning systems}, PP,
  2021.

\bibitem[Peng et~al.(2022)Peng, Xie, Zhang, Zhang, Yang, and Wei]{Paper_Peng_9}
Peizhen Peng, Liping Xie, Kanjian Zhang, Jinxia Zhang, Lu~Yang, and Haikun Wei.
\newblock Domain adaptation for epileptic eeg classification using adversarial
  learning and riemannian manifold.
\newblock \emph{Biomed. Signal Process. Control.}, 75:\penalty0 103555, 2022.

\bibitem[Peng et~al.(2021)Peng, Song, Yang, and Wei]{Paper_Peng_82}
Peizhen Peng, Yang Song, Lu~Yang, and Haikun Wei.
\newblock Seizure prediction in eeg signals using stft and domain adaptation.
\newblock \emph{Frontiers in Neuroscience}, 15, 2021.

\bibitem[Wang et~al.(2021{\natexlab{c}})Wang, Qiu, Ma, and He]{Paper_Wang_87}
Yixin Wang, Shuang Qiu, Xuelin Ma, and Huiguang He.
\newblock A prototype-based spd matrix network for domain adaptation eeg
  emotion recognition.
\newblock \emph{Pattern Recognit.}, 110:\penalty0 107626, 2021{\natexlab{c}}.

\bibitem[Luo and Lu(2021{\natexlab{b}})]{Paper_Luo_53}
Yun Luo and Bao-Liang Lu.
\newblock Wasserstein-distance-based multi-source adversarial domain adaptation
  for emotion recognition and vigilance estimation.
\newblock \emph{2021 IEEE International Conference on Bioinformatics and
  Biomedicine (BIBM)}, pages 1424--1428, 2021{\natexlab{b}}.

\bibitem[Qu et~al.(2021)Qu, Kao, Hong, Chi, Grunstein, Gordon, and
  Wang]{Paper_Qu_24}
Wei Qu, Chien-Hui~Tancy Kao, Hong Hong, Z.~Chi, Ronald~R. Grunstein,
  Christopher Gordon, and Zhiyong Wang.
\newblock Single-channel eeg based insomnia detection with domain adaptation.
\newblock \emph{Computers in biology and medicine}, 139:\penalty0 104989, 2021.

\bibitem[Huang et~al.(2022)Huang, Zhou, and Jiang]{Paper_Haung_71}
Dongmin Huang, Sijin Zhou, and Dazhi Jiang.
\newblock Generator-based domain adaptation method with knowledge free for
  cross-subject eeg emotion recognition.
\newblock \emph{Cogn. Comput.}, 14:\penalty0 1316--1327, 2022.

\bibitem[HassanPour~Zonoozi and Seydi(2022)]{hassanpour2022survey}
Mahta HassanPour~Zonoozi and Vahid Seydi.
\newblock A survey on adversarial domain adaptation.
\newblock \emph{Neural Processing Letters}, pages 1--41, 2022.

\bibitem[Shen et~al.(2022)Shen, Peng, Dai, Lu, and Kong]{Paper_Shen_11}
Fangyao Shen, Yong Peng, Guojun Dai, Baoliang Lu, and Wanzeng Kong.
\newblock Coupled projection transfer metric learning for cross-session emotion
  recognition from eeg.
\newblock \emph{Syst.}, 10:\penalty0 47, 2022.

\bibitem[Zhou et~al.(2018)Zhou, Cox, and Lu]{Paper_Zhou_fMRI_26}
S.~Zhou, Christopher~R. Cox, and Haiping Lu.
\newblock Improving whole-brain neural decoding of fmri with domain adaptation.
\newblock \emph{bioRxiv}, 2018.

\bibitem[Gao et~al.(2020)Gao, Zhang, Cao, Guo, and cai Zhang]{Paper_Gao_fMRI_5}
Yufei Gao, Yameng Zhang, Zhiyuan Cao, Xiaojuan Guo, and Jia cai Zhang.
\newblock Decoding brain states from fmri signals by using unsupervised domain
  adaptation.
\newblock \emph{IEEE Journal of Biomedical and Health Informatics},
  24:\penalty0 1677--1685, 2020.

\bibitem[Chen et~al.(2021{\natexlab{a}})Chen, Jin, Li, Fan, Li, and
  He]{Paper_Chen_13}
Hao Chen, Ming Jin, Zhunan Li, Cunhang Fan, Jinpeng Li, and Huiguang He.
\newblock Ms-mda: Multisource marginal distribution adaptation for
  cross-subject and cross-session eeg emotion recognition.
\newblock \emph{Frontiers in Neuroscience}, 15, 12 2021{\natexlab{a}}.
\newblock \doi{10.3389/fnins.2021.778488}.

\bibitem[Zhao et~al.(2021{\natexlab{b}})Zhao, Dai, Borghini, Zhang, Li, Zhang,
  Aric{\'o}, di~Flumeri, Babiloni, and Zeng]{Paper_Zhao_52}
Yue Zhao, Guojun Dai, Gianluca Borghini, Jiaming Zhang, Xiufeng Li, Zhenyan
  Zhang, Pietro Aric{\'o}, Gianluca di~Flumeri, Fabio Babiloni, and Hong Zeng.
\newblock Label-based alignment multi-source domain adaptation for
  cross-subject eeg fatigue mental state evaluation.
\newblock \emph{Frontiers in Human Neuroscience}, 15, 2021{\natexlab{b}}.

\bibitem[Lawhern et~al.(2018{\natexlab{a}})Lawhern, Solon, Waytowich, Gordon,
  Hung, and Lance]{Lawhern2018EEGNetAC}
Vernon~J. Lawhern, Amelia~J. Solon, Nicholas~R. Waytowich, Stephen~M. Gordon,
  Chou~Po Hung, and Brent Lance.
\newblock Eegnet: a compact convolutional neural network for eeg-based
  brain–computer interfaces.
\newblock \emph{Journal of Neural Engineering}, 15, 2018{\natexlab{a}}.

\bibitem[Shi et~al.(2022)Shi, wei Xin, and cai Zhang]{Paper_Shi_fMRI_10}
Chunlei Shi, Xian wei Xin, and Jia cai Zhang.
\newblock Domain adaptation based on rough adjoint inconsistency and optimal
  transport for identifying autistic patients.
\newblock \emph{Computer methods and programs in biomedicine}, 215:\penalty0
  106615, 2022.

\bibitem[Dempster(2008)]{dempster2008upper}
Arthur~P Dempster.
\newblock Upper and lower probabilities induced by a multivalued mapping.
\newblock In \emph{Classic works of the Dempster-Shafer theory of belief
  functions}, pages 57--72. Springer, 2008.

\bibitem[Chai et~al.(2016)Chai, Wang, Zhao, Liu, Bai, and Li]{Paper_Chai_43}
Xin Chai, Qisong Wang, Yongping Zhao, Xin Liu, Ou~Bai, and Yongqiang Li.
\newblock Unsupervised domain adaptation techniques based on auto-encoder for
  non-stationary eeg-based emotion recognition.
\newblock \emph{Computers in biology and medicine}, 79:\penalty0 205--214,
  2016.

\bibitem[Lee et~al.(2021{\natexlab{b}})Lee, Hwang, Jeon, and
  Byun]{Paper_Lee_80}
Pilhyeon Lee, Sunhee Hwang, Seogkyu Jeon, and Hyeran Byun.
\newblock Subject adaptive eeg-based visual recognition.
\newblock In \emph{ACPR}, 2021{\natexlab{b}}.

\bibitem[Shen et~al.(2020)Shen, Zou, Li, Zheng, and Zhang]{Paper_Shen_RP32}
Mu~Shen, Bing Zou, Xinhang Li, Yubo Zheng, and Lin Zhang.
\newblock Tensor-based eeg network formation and feature extraction for
  cross-session driving drowsiness detection*.
\newblock \emph{2020 42nd Annual International Conference of the IEEE
  Engineering in Medicine \& Biology Society (EMBC)}, pages 252--255, 2020.

\bibitem[Hang et~al.(2019)Hang, Feng, Du, Liang, Chen, Wang, and
  Liu]{Paper_Hang_RP2}
Wenlong Hang, Wei Feng, Ruoyu Du, Shuang Liang, Yan Chen, Qiong Wang, and
  Xuejun Liu.
\newblock Cross-subject eeg signal recognition using deep domain adaptation
  network.
\newblock \emph{IEEE Access}, 7:\penalty0 128273--128282, 2019.

\bibitem[Meng et~al.(2021)Meng, Hu, Gao, Kong, and Luo]{Paper_Meng_RP20}
Ming Meng, Jiahao Hu, Yunyuan Gao, Wanzeng Kong, and Zhizeng Luo.
\newblock A deep subdomain associate adaptation network for cross-session and
  cross-subject eeg emotion recognition.
\newblock \emph{SSRN Electronic Journal}, 2021.

\bibitem[He et~al.(2016)He, Zhang, Ren, and Sun]{He2016DeepRL}
Kaiming He, X.~Zhang, Shaoqing Ren, and Jian Sun.
\newblock Deep residual learning for image recognition.
\newblock \emph{2016 IEEE Conference on Computer Vision and Pattern Recognition
  (CVPR)}, pages 770--778, 2016.

\bibitem[Kuang et~al.(2021)Kuang, Shu, Hua, Wu, Zhang, Xu, Liu, and
  Jiang]{Paper_Kuang_81}
Feng Kuang, Lin Shu, Haoqiang Hua, Shibin Wu, Lulu Zhang, Xiangmin Xu, Yunhe
  Liu, and Man Jiang.
\newblock Cross-subject and cross-device wearable eeg emotion recognition using
  frontal eeg under virtual reality scenes.
\newblock \emph{2021 IEEE International Conference on Bioinformatics and
  Biomedicine (BIBM)}, pages 3630--3637, 2021.

\bibitem[Ning et~al.(2021)Ning, Chen, and Zhang]{Paper_Ning_77}
Run Ning, C.~L.~Philip Chen, and Tong Zhang.
\newblock Cross-subject eeg emotion recognition using domain adaptive few-shot
  learning networks.
\newblock \emph{2021 IEEE International Conference on Bioinformatics and
  Biomedicine (BIBM)}, pages 1468--1472, 2021.

\bibitem[Chen et~al.(2021{\natexlab{b}})Chen, Chen, and Wang]{Paper_Chen_90}
Junfu Chen, Yang Chen, and Bi~Wang.
\newblock Cross-subject domain adaptation for classifying working memory load
  with multi-frame eeg images.
\newblock \emph{ArXiv}, abs/2106.06769, 2021{\natexlab{b}}.

\bibitem[Pan et~al.(2010)Pan, Tsang, Kwok, and Yang]{TCA_2010}
Sinno~Jialin Pan, Ivor~W Tsang, James~T Kwok, and Qiang Yang.
\newblock Domain adaptation via transfer component analysis.
\newblock \emph{IEEE transactions on neural networks}, 22\penalty0
  (2):\penalty0 199--210, 2010.

\bibitem[Liu et~al.(2019)Liu, Lan, Cui, Sourina, and
  M{\"u}ller-Wittig]{Paper_Liu_RP9}
Yisi Liu, Zirui Lan, Jian Cui, Olga Sourina, and Wolfgang M{\"u}ller-Wittig.
\newblock Eeg-based cross-subject mental fatigue recognition.
\newblock \emph{2019 International Conference on Cyberworlds (CW)}, pages
  247--252, 2019.

\bibitem[He et~al.(2022{\natexlab{b}})He, Zhuang, Bao, Zeng, and
  Yan]{Paper_He_70}
Z.~He, Ning Zhuang, Guangcheng Bao, Ying Zeng, and Bin Yan.
\newblock Cross-day eeg-based emotion recognition using transfer component
  analysis.
\newblock \emph{Electronics}, 2022{\natexlab{b}}.

\bibitem[Zhou et~al.(2022{\natexlab{b}})Zhou, Xu, Niu, Wang, Wen, Wu, and
  Zhang]{Paper_Zhou_65}
Yueying Zhou, Ziming Xu, Yifan Niu, Pengpai Wang, Xuyun Wen, Xia Wu, and
  Daoqiang Zhang.
\newblock Cross-task cognitive workload recognition based on eeg and domain
  adaptation.
\newblock \emph{IEEE Transactions on Neural Systems and Rehabilitation
  Engineering}, 30:\penalty0 50--60, 2022{\natexlab{b}}.

\bibitem[Long et~al.(2013)Long, Wang, Ding, Sun, and Yu]{long2013transfer}
Mingsheng Long, Jianmin Wang, Guiguang Ding, Jiaguang Sun, and Philip~S Yu.
\newblock Transfer feature learning with joint distribution adaptation.
\newblock In \emph{Proceedings of the IEEE international conference on computer
  vision}, pages 2200--2207, 2013.

\bibitem[Wang et~al.(2018{\natexlab{b}})Wang, Chen, Hu, Peng, and
  Philip]{wang2018stratified}
Jindong Wang, Yiqiang Chen, Lisha Hu, Xiaohui Peng, and S~Yu Philip.
\newblock Stratified transfer learning for cross-domain activity recognition.
\newblock In \emph{2018 IEEE international conference on pervasive computing
  and communications (PerCom)}, pages 1--10. IEEE, 2018{\natexlab{b}}.

\bibitem[Long et~al.(2014)Long, Wang, Ding, Sun, and Yu]{long2014transfer}
Mingsheng Long, Jianmin Wang, Guiguang Ding, Jiaguang Sun, and Philip~S Yu.
\newblock Transfer joint matching for unsupervised domain adaptation.
\newblock In \emph{Proceedings of the IEEE conference on computer vision and
  pattern recognition}, pages 1410--1417, 2014.

\bibitem[Wang et~al.(2021{\natexlab{d}})Wang, Peng, and Kong]{Paper_Wang_50}
Wenjuan Wang, Yong Peng, and Wanzeng Kong.
\newblock Eeg-based emotion recognition via joint domain adaptation and
  semi-supervised rvfl network.
\newblock \emph{Advances in Intelligent Automation and Soft Computing},
  2021{\natexlab{d}}.

\bibitem[Zhang et~al.(2020{\natexlab{b}})Zhang, Wan, and
  Zhang]{Paper_Zhang_fMRI_21}
Junyi Zhang, Peng Wan, and Daoqiang Zhang.
\newblock Transport-based joint distribution alignment for multi-site autism
  spectrum disorder diagnosis using resting-state fmri.
\newblock In \emph{MICCAI}, 2020{\natexlab{b}}.

\bibitem[Comon(1994)]{comon1994independent}
Pierre Comon.
\newblock Independent component analysis, a new concept?
\newblock \emph{Signal processing}, 36\penalty0 (3):\penalty0 287--314, 1994.

\bibitem[Qu et~al.(2022)Qu, Zhang, and wei Pang]{Paper_Qu_RP29}
Hongquan Qu, Mengyu Zhang, and Li~wei Pang.
\newblock Mental workload classification method based on eeg cross-session
  subspace alignment.
\newblock \emph{Mathematics}, 2022.

\bibitem[Zhang and Wu(2020)]{Paper_Zhang_56}
Wen Zhang and Dongrui Wu.
\newblock Manifold embedded knowledge transfer for brain-computer interfaces.
\newblock \emph{IEEE Transactions on Neural Systems and Rehabilitation
  Engineering}, 28:\penalty0 1117--1127, 2020.

\bibitem[Jiang et~al.(2022{\natexlab{b}})Jiang, Zhang, and
  Zheng]{Paper_Jiang_RP33}
Qin Jiang, Yi~Zhang, and Kai Zheng.
\newblock Motor imagery classification via kernel-based domain adaptation on an
  spd manifold.
\newblock \emph{Brain Sciences}, 12, 2022{\natexlab{b}}.

\bibitem[Chen et~al.(2022)Chen, Wu, and Jiang]{chen2022multi}
Haoran Chen, Zuxuan Wu, and Yu-Gang Jiang.
\newblock Multi-prompt alignment for multi-source unsupervised domain
  adaptation.
\newblock \emph{arXiv preprint arXiv:2209.15210}, 2022.

\bibitem[Ge et~al.(2022)Ge, Huang, Xie, Lai, Song, Li, and Huang]{ge2022domain}
Chunjiang Ge, Rui Huang, Mixue Xie, Zihang Lai, Shiji Song, Shuang Li, and Gao
  Huang.
\newblock Domain adaptation via prompt learning.
\newblock \emph{arXiv preprint arXiv:2202.06687}, 2022.

\bibitem[Chambon et~al.(2018)Chambon, Galtier, and Gramfort]{Paper_Chambon_41}
Stanislas Chambon, Mathieu~N. Galtier, and Alexandre Gramfort.
\newblock Domain adaptation with optimal transport improves eeg sleep stage
  classifiers.
\newblock \emph{2018 International Workshop on Pattern Recognition in
  Neuroimaging (PRNI)}, pages 1--4, 2018.

\bibitem[Lee et~al.(2021{\natexlab{c}})Lee, Kang, Jeon, and
  Suk]{Paper_Lee_fMRI_22}
Jaein Lee, Eunsong Kang, Eunjin Jeon, and Heung-Il Suk.
\newblock Meta-modulation network for domain generalization in multi-site fmri
  classification.
\newblock In \emph{MICCAI}, 2021{\natexlab{c}}.

\bibitem[Wang et~al.(2020)Wang, Zhang, Huang, Yap, Shen, and
  Liu]{Paper_Wang_fMRI_20}
Mingliang Wang, Daoqiang Zhang, Jiashuang Huang, Pew-Thian Yap, Dinggang Shen,
  and Mingxia Liu.
\newblock Identifying autism spectrum disorder with multi-site fmri via
  low-rank domain adaptation.
\newblock \emph{IEEE Transactions on Medical Imaging}, 39:\penalty0 644--655,
  2020.

\bibitem[Lee et~al.(2022{\natexlab{b}})Lee, Hwang, Lee, Shin, Jeon, and
  Byun]{Paper_Lee_6}
Pilhyeon Lee, Sunhee Hwang, Jewook Lee, Minjung Shin, Seogkyu Jeon, and Hyeran
  Byun.
\newblock Inter-subject contrastive learning for subject adaptive eeg-based
  visual recognition.
\newblock \emph{2022 10th International Winter Conference on Brain-Computer
  Interface (BCI)}, pages 1--6, 2022{\natexlab{b}}.

\bibitem[Zhao et~al.(2021{\natexlab{c}})Zhao, Yan, and Lu]{Paper_Zhao_10}
Li-Ming Zhao, Xue Yan, and Bao-Liang Lu.
\newblock Plug-and-play domain adaptation for cross-subject eeg-based emotion
  recognition.
\newblock In \emph{AAAI}, 2021{\natexlab{c}}.

\bibitem[Li et~al.(2021{\natexlab{c}})Li, Wang, and Sourina]{Paper_Li_14}
Ruilin Li, Lipo Wang, and Olga Sourina.
\newblock Subject matching for cross-subject eeg-based recognition of driver
  states related to situation awareness.
\newblock \emph{Methods}, 2021{\natexlab{c}}.

\bibitem[Lin et~al.(2021{\natexlab{a}})Lin, Zhu, Ren, Hu, and
  Zhang]{Paper_Lin_RP19}
Guang Lin, Li~Zhu, Bin Ren, Yiteng Hu, and Jianhai Zhang.
\newblock Multi-branch network for cross-subject eeg-based emotion recognition.
\newblock In \emph{ACML}, 2021{\natexlab{a}}.

\bibitem[Tao and Dan(2021)]{Paper_Tao_15}
Jianwen Tao and Yufang Dan.
\newblock Multi-source co-adaptation for eeg-based emotion recognition by
  mining correlation information.
\newblock \emph{Frontiers in Neuroscience}, 15, 05 2021.
\newblock \doi{10.3389/fnins.2021.677106}.

\bibitem[Shen et~al.(2021{\natexlab{a}})Shen, Zou, Li, Zheng, Li, and
  Zhang]{Paper_Shen_79}
Mu~Shen, Bing Zou, Xinhang Li, Yubo Zheng, Lei Li, and Lin Zhang.
\newblock Multi-source signal alignment and efficient multi-dimensional feature
  classification in the application of eeg-based subject-independent drowsiness
  detection.
\newblock \emph{Biomed. Signal Process. Control.}, 70:\penalty0 103023,
  2021{\natexlab{a}}.

\bibitem[Xia et~al.(2022)Xia, Deng, Duch, and Wu]{Paper_Xia_73}
Kun Xia, Lingfei Deng, Wlodzislaw Duch, and Dongrui Wu.
\newblock Privacy-preserving domain adaptation for motor imagery-based
  brain-computer interfaces.
\newblock \emph{IEEE transactions on bio-medical engineering}, PP, 2022.

\bibitem[Albuquerque et~al.(2019{\natexlab{a}})Albuquerque, Monteiro, Rosanne,
  Tiwari, Gagnon, and Falk]{Paper_Albuquerque_40}
Isabela Albuquerque, Jo{\~a}o Monteiro, Olivier Rosanne, Abhishek Tiwari,
  Jean-François Gagnon, and Tiago~H. Falk.
\newblock Cross-subject statistical shift estimation for generalized
  electroencephalography-based mental workload assessment.
\newblock \emph{2019 IEEE International Conference on Systems, Man and
  Cybernetics (SMC)}, pages 3647--3653, 2019{\natexlab{a}}.

\bibitem[Liu et~al.(2021)Liu, Shen, Song, and Zhang]{Paper_Liu_RP46}
Jin Liu, Xinke Shen, Sen Song, and Dan Zhang.
\newblock Domain adaptation for cross-subject emotion recognition by subject
  clustering.
\newblock \emph{2021 10th International IEEE/EMBS Conference on Neural
  Engineering (NER)}, pages 904--908, 2021.

\bibitem[Shi et~al.(2021)Shi, wei Xin, and cai Zhang]{Paper_Shi_fMRI_17}
Chunlei Shi, Xian wei Xin, and Jia cai Zhang.
\newblock Domain adaptation using a three-way decision improves the
  identification of autism patients from multisite fmri data.
\newblock \emph{Brain Sciences}, 11, 2021.

\bibitem[Zhang et~al.(2020{\natexlab{c}})Zhang, Deng, Jia, and
  Zhang]{label_propagation}
Yabin Zhang, Bin Deng, Kui Jia, and Lei Zhang.
\newblock Label propagation with augmented anchors: A simple semi-supervised
  learning baseline for unsupervised domain adaptation.
\newblock In \emph{European Conference on Computer Vision}, pages 781--797.
  Springer, 2020{\natexlab{c}}.

\bibitem[Han et~al.(2022)Han, Gong, Feng, Zhang, Sun, and
  Zhang]{Paper_Han_fMRI_12}
Tianyi Han, Xiaoli Gong, Fan Feng, Jin Zhang, Zhe Sun, and Yu~Zhang.
\newblock Privacy preserving mutli-source domain adaptaion for medical data.
\newblock \emph{IEEE journal of biomedical and health informatics}, PP, 2022.

\bibitem[Jim{\'e}nez-Guarneros and
  G{\'o}mez-Gil(2021)]{Paper_JimnezGuarneros_76}
Magdiel Jim{\'e}nez-Guarneros and Pilar G{\'o}mez-Gil.
\newblock Standardization-refinement domain adaptation method for cross-subject
  eeg-based classification in imagined speech recognition.
\newblock \emph{Pattern Recognit. Lett.}, 141:\penalty0 54--60, 2021.

\bibitem[Bethge et~al.(2022{\natexlab{a}})Bethge, Hallgarten, {\"O}zdenizci,
  Mikut, Schmidt, and Gro{\ss}e-Puppendahl]{Paper_Bethge_67}
David Bethge, Philipp Hallgarten, Ozan {\"O}zdenizci, Ralf Mikut, Albrecht
  Schmidt, and Tobias~Alexander Gro{\ss}e-Puppendahl.
\newblock Exploiting multiple eeg data domains with adversarial learning.
\newblock \emph{ArXiv}, abs/2204.07777, 2022{\natexlab{a}}.

\bibitem[{\"O}zdenizci et~al.(2020){\"O}zdenizci, Wang, Koike-Akino, and
  Erdoğmuş]{Paper_zdenizci_RP16}
Ozan {\"O}zdenizci, Ye~Wang, Toshiaki Koike-Akino, and Deniz Erdoğmuş.
\newblock Learning invariant representations from eeg via adversarial
  inference.
\newblock \emph{IEEE Access}, 8:\penalty0 27074--27085, 2020.

\bibitem[Jia et~al.(2021)Jia, Lin, Wang, Ning, He, Zhou, Zhou, and wei
  H.~Lehman]{Paper_Jia_12}
Ziyu Jia, Youfang Lin, Jing Wang, Xiaojun Ning, Yuanlai He, Ronghao Zhou, Yuhan
  Zhou, and Li~wei H.~Lehman.
\newblock Multi-view spatial-temporal graph convolutional networks with domain
  generalization for sleep stage classification.
\newblock \emph{IEEE transactions on neural systems and rehabilitation
  engineering : a publication of the IEEE Engineering in Medicine and Biology
  Society}, 29:\penalty0 1977 -- 1986, 2021.

\bibitem[Hagad et~al.(2021)Hagad, Kimura, Fukui, and Numao]{Paper_Hagad_17}
Juan Hagad, Tsukasa Kimura, Ken-ichi Fukui, and Masayuki Numao.
\newblock Learning subject-generalized topographical eeg embeddings using deep
  variational autoencoders and domain-adversarial regularization.
\newblock \emph{Sensors}, 21:\penalty0 1792, 03 2021.
\newblock \doi{10.3390/s21051792}.

\bibitem[Higgins et~al.(2016)Higgins, Matthey, Pal, Burgess, Glorot, Botvinick,
  Mohamed, and Lerchner]{higgins2016beta}
Irina Higgins, Loic Matthey, Arka Pal, Christopher Burgess, Xavier Glorot,
  Matthew Botvinick, Shakir Mohamed, and Alexander Lerchner.
\newblock beta-vae: Learning basic visual concepts with a constrained
  variational framework.
\newblock 2016.

\bibitem[Albuquerque et~al.(2019{\natexlab{b}})Albuquerque, Monteiro, Bayazi,
  Falk, and Mitliagkas]{Paper_Albuquerque_33}
Isabela Albuquerque, Jo{\~a}o Monteiro, Mohammad Javad~Darvishi Bayazi,
  Tiago~H. Falk, and Ioannis Mitliagkas.
\newblock Generalizing to unseen domains via distribution matching.
\newblock \emph{arXiv: Learning}, 2019{\natexlab{b}}.

\bibitem[Han et~al.(2021)Han, Gu, and Lo]{Paper_Han_16}
Jinpei Han, Xiao Gu, and Benny P.~L. Lo.
\newblock Semi-supervised contrastive learning for generalizable motor imagery
  eeg classification.
\newblock \emph{2021 IEEE 17th International Conference on Wearable and
  Implantable Body Sensor Networks (BSN)}, pages 1--4, 2021.

\bibitem[Ma et~al.(2019)Ma, Li, Zheng, and Lu]{Paper_Ma_35}
Bo-Qun Ma, He~Li, Wei-Long Zheng, and Bao-Liang Lu.
\newblock Reducing the subject variability of eeg signals with adversarial
  domain generalization.
\newblock In \emph{ICONIP}, 2019.

\bibitem[Ming et~al.(2019)Ming, Ding, Pelusi, Wu, kai Wang, Prasad, and
  Lin]{Paper_Ming_RP22}
Yurui Ming, Weiping Ding, Danilo Pelusi, Dongrui Wu, Yu~kai Wang, Mukesh
  Prasad, and Chin-Teng Lin.
\newblock Subject adaptation network for eeg data analysis.
\newblock \emph{Appl. Soft Comput.}, 84, 2019.

\bibitem[Cui et~al.(2019)Cui, Xu, and Wu]{Paper_Cui_34}
Yuqi Cui, Yifan Xu, and Dongrui Wu.
\newblock Eeg-based driver drowsiness estimation using feature weighted
  episodic training.
\newblock \emph{IEEE Transactions on Neural Systems and Rehabilitation
  Engineering}, 27:\penalty0 2263--2273, 2019.

\bibitem[Ayodele et~al.(2020)Ayodele, Ikezogwo, Komolafe, and
  Ogunbona]{Paper_Ayodele_60}
Kayode~Peter Ayodele, Wisdom~O. Ikezogwo, Morenikeji~A. Komolafe, and Philip
  Ogunbona.
\newblock Supervised domain generalization for integration of disparate scalp
  eeg datasets for automatic epileptic seizure detection.
\newblock \emph{Computers in biology and medicine}, 120:\penalty0 103757, 2020.

\bibitem[Bethge et~al.(2022{\natexlab{b}})Bethge, Hallgarten,
  Grosse-Puppendahl, Kari, Mikut, Schmidt, and Ozdenizci]{Paper_Bethge_5}
David Bethge, Philipp Hallgarten, Tobias Grosse-Puppendahl, Mohamed Kari, Ralf
  Mikut, Albrecht Schmidt, and Ozan Ozdenizci.
\newblock Domain-invariant representation learning from eeg with private
  encoders, 01 2022{\natexlab{b}}.

\bibitem[Musellim et~al.(2022)Musellim, Han, Jeong, and Lee]{Paper_Musellim_7}
Serkan Musellim, Dong-Kyun Han, Ji-Hoon Jeong, and Seong-Whan Lee.
\newblock Prototype-based domain generalization framework for
  subject-independent brain-computer interfaces.
\newblock \emph{2022 44th Annual International Conference of the IEEE
  Engineering in Medicine \& Biology Society (EMBC)}, pages 711--714, 2022.

\bibitem[Zhang et~al.(2019{\natexlab{b}})Zhang, Yao, Chen, and
  Monaghan]{Paper_Zhang_31}
Dalin Zhang, Lina Yao, Kaixuan Chen, and Jessica J.~M. Monaghan.
\newblock A convolutional recurrent attention model for subject-independent eeg
  signal analysis.
\newblock \emph{IEEE Signal Processing Letters}, 26:\penalty0 715--719,
  2019{\natexlab{b}}.

\bibitem[Yousefnezhad et~al.(2020)Yousefnezhad, Selvitella, Zhang, Greenshaw,
  and Greiner]{Paper_Yousefnezhad_fMRI_35}
Muhammad Yousefnezhad, Alessandro~Maria Selvitella, Daoqiang Zhang,
  Andrew~James Greenshaw, and Russell Greiner.
\newblock Shared space transfer learning for analyzing multi-site fmri data.
\newblock \emph{ArXiv}, abs/2010.15594, 2020.

\bibitem[Li et~al.(2020{\natexlab{c}})Li, Liu, Chen, and
  Zhang]{Paper_Li_fMRI_36}
Weida Li, Mingxia Liu, Fang Chen, and Daoqiang Zhang.
\newblock Graph-based decoding model for functional alignment of unaligned fmri
  data.
\newblock In \emph{AAAI}, 2020{\natexlab{c}}.

\bibitem[Zeng et~al.(2018)Zeng, Wang, pan Hu, Yang, Pu, Shen, Chen, Liu, Yin,
  rong Tan, Wang, and Hu]{Paper_Zeng_fMRI_29}
Lingli Zeng, Huaning Wang, Pan pan Hu, Bo~Yang, Weidan Pu, Hui Shen, Xingui
  Chen, Zhening Liu, Hong Yin, Qing rong Tan, Kai Wang, and Dewen Hu.
\newblock Multi-site diagnostic classification of schizophrenia using
  discriminant deep learning with functional connectivity mri.
\newblock \emph{EBioMedicine}, 30:\penalty0 74 -- 85, 2018.

\bibitem[chun Huang et~al.(2022)chun Huang, Busch, Wallenstein, Gerasimiuk,
  Benz, Lajoie, Wolf, Turk-Browne, and Krishnaswamy]{Paper_Huang_fMRI_27}
Je~chun Huang, Erica~L. Busch, Tom Wallenstein, Michal Gerasimiuk, Andrew Benz,
  Guillaume Lajoie, Guy Wolf, Nicholas~B. Turk-Browne, and Smita Krishnaswamy.
\newblock Learning shared neural manifolds from multi-subject fmri data.
\newblock \emph{ArXiv}, abs/2201.00622, 2022.

\bibitem[Zhang et~al.(2021{\natexlab{b}})Zhang, Wang, and
  Ma]{Paper_Zhang_fMRI_18}
Li~Zhang, Jia-Rui Wang, and Yue Ma.
\newblock Graph convolutional networks via low-rank subspace for multi-site
  rs-fmri asd diagnosis.
\newblock \emph{2021 14th International Congress on Image and Signal
  Processing, BioMedical Engineering and Informatics (CISP-BMEI)}, pages 1--6,
  2021{\natexlab{b}}.

\bibitem[Wang et~al.(2017)Wang, Xue, Li, Luo, Huang, Cui, and
  Huang]{Paper_Wang_42}
Lina Wang, Weining Xue, Y.~Li, Mei-Lin Luo, Jie Huang, Weigang Cui, and Chao
  Huang.
\newblock Automatic epileptic seizure detection in eeg signals using
  multi-domain feature extraction and nonlinear analysis.
\newblock \emph{Entropy}, 19:\penalty0 222, 2017.

\bibitem[Li et~al.(2022{\natexlab{b}})Li, Gao, and Suganthan]{Paper_Li_66}
Ruilin Li, Ruobin Gao, and Ponnuthurai~Nagaratnam Suganthan.
\newblock A decomposition-based hybrid ensemble cnn framework for improving
  cross-subject eeg decoding performance.
\newblock \emph{ArXiv}, abs/2203.09477, 2022{\natexlab{b}}.

\bibitem[Zhu et~al.(2021{\natexlab{b}})Zhu, Li, Lu, and Li]{Paper_Zhu_RP28}
Yuanlu Zhu, Ying Li, Jinling Lu, and Pengcheng Li.
\newblock Eegnet with ensemble learning to improve the cross-session
  classification of ssvep based bci from ear-eeg.
\newblock \emph{IEEE Access}, 9:\penalty0 15295--15303, 2021{\natexlab{b}}.

\bibitem[Lawhern et~al.(2018{\natexlab{b}})Lawhern, Solon, Waytowich, Gordon,
  Hung, and Lance]{eegnet_2018}
Vernon~J Lawhern, Amelia~J Solon, Nicholas~R Waytowich, Stephen~M Gordon,
  Chou~P Hung, and Brent~J Lance.
\newblock {EEGNet}: a compact convolutional neural network for {EEG}-based
  brain–computer interfaces.
\newblock \emph{Journal of Neural Engineering}, 15\penalty0 (5):\penalty0
  056013, October 2018{\natexlab{b}}.
\newblock ISSN 1741-2560, 1741-2552.
\newblock \doi{10.1088/1741-2552/aace8c}.
\newblock URL
  \url{https://iopscience.iop.org/article/10.1088/1741-2552/aace8c}.

\bibitem[Roots et~al.(2020)Roots, Muhammad, and Muhammad]{Paper_Roots_RP6}
Karel Roots, Yar Muhammad, and Naveed Muhammad.
\newblock Fusion convolutional neural network for cross-subject eeg motor
  imagery classification.
\newblock \emph{Comput.}, 9:\penalty0 72, 2020.

\bibitem[Luo et~al.(2022)Luo, Wei, and liang Lu]{Paper_Luo_22}
Yun Luo, Gengchen Wei, and Bao liang Lu.
\newblock {PDAML}: A pseudo domain adaptation paradigm for subject-independent
  {EEG}-based emotion recognition, 2022.
\newblock URL \url{https://openreview.net/forum?id=TscS0R8QzfG}.

\bibitem[Finn et~al.(2017)Finn, Abbeel, and Levine]{MAML_2017}
Chelsea Finn, Pieter Abbeel, and Sergey Levine.
\newblock Model-agnostic meta-learning for fast adaptation of deep networks,
  2017.
\newblock URL \url{https://arxiv.org/abs/1703.03400}.

\bibitem[Lemkhenter and Favaro(2022)]{Paper_Lemkhenter_RP31}
Abdelhak Lemkhenter and Paolo Favaro.
\newblock Towards sleep scoring generalization through self-supervised
  meta-learning.
\newblock \emph{2022 44th Annual International Conference of the IEEE
  Engineering in Medicine \& Biology Society (EMBC)}, pages 2961--2966, 2022.

\bibitem[Duan et~al.(2020)Duan, Shaikh, Chauhan, Chu, Srihari, Pathak, and
  Srihari]{Paper_Duan_RP10}
Tiehang Duan, Mohammad~Abuzar Shaikh, Mihir Chauhan, Jun Chu, Rohini~K.
  Srihari, Archita Pathak, and Sargur~N. Srihari.
\newblock Meta learn on constrained transfer learning for low resource cross
  subject eeg classification.
\newblock \emph{IEEE Access}, 8:\penalty0 224791--224802, 2020.

\bibitem[Tian et~al.(2021)Tian, Chen, and Ganguli]{Tian2021UnderstandingSL}
Yuandong Tian, Xinlei Chen, and Surya Ganguli.
\newblock Understanding self-supervised learning dynamics without contrastive
  pairs.
\newblock \emph{ArXiv}, abs/2102.06810, 2021.

\bibitem[Shen et~al.(2021{\natexlab{b}})Shen, Liu, Hu, Zhang, and
  Song]{Paper_Shen_19}
Xinke Shen, Xianggen Liu, Xin Hu, Dan Zhang, and Sen Song.
\newblock Contrastive learning of subject-invariant eeg representations for
  cross-subject emotion recognition, 09 2021{\natexlab{b}}.

\bibitem[Cheng et~al.(2020)Cheng, Goh, Dogrusoz, Tuzel, and
  Azemi]{Paper_Cheng_RP8}
Joseph~Y. Cheng, Hanlin Goh, Kaan Dogrusoz, Oncel Tuzel, and Erdrin Azemi.
\newblock Subject-aware contrastive learning for biosignals.
\newblock \emph{ArXiv}, abs/2007.04871, 2020.

\bibitem[Wagh et~al.(2021)Wagh, Wei, Rawal, Berry, Barnard, Brinkmann, Worrell,
  Jones, and Varatharajah]{Paper_Wagh_RP45}
Neeraj Wagh, Jionghao Wei, Samarth Rawal, Brent~M. Berry, Leland~R. Barnard,
  Benjamin~H. Brinkmann, Gregory~A. Worrell, David~T. Jones, and Yogatheesan
  Varatharajah.
\newblock Domain-guided self-supervision of eeg data improves downstream
  classification performance and generalizability.
\newblock In \emph{ML4H@NeurIPS}, 2021.

\bibitem[Banluesombatkul et~al.(2021)Banluesombatkul, Ouppaphan, Leelaarporn,
  Lakhan, Chaitusaney, Jaimchariyatam, Chuangsuwanich, Chen, Phan,
  Dilokthanakul, and Wilaiprasitporn]{Banluesombatkul2021MetaSleepLearnerAP}
Nannapas Banluesombatkul, Pichayoot Ouppaphan, Pitshaporn Leelaarporn,
  Payongkit Lakhan, Busarakum Chaitusaney, Nattapong Jaimchariyatam, Ekapol
  Chuangsuwanich, Wei Chen, Huy~P Phan, Nat Dilokthanakul, and Theerawit
  Wilaiprasitporn.
\newblock Metasleeplearner: A pilot study on fast adaptation of
  bio-signals-based sleep stage classifier to new individual subject using
  meta-learning.
\newblock \emph{IEEE Journal of Biomedical and Health Informatics},
  25:\penalty0 1949--1963, 2021.

\bibitem[Song et~al.(2021{\natexlab{b}})Song, Yang, Jia, and
  Xie]{Paper_Song_21}
Yonghao Song, Lie Yang, Xueyu Jia, and Longhan Xie.
\newblock Common spatial generative adversarial networks based eeg data
  augmentation for cross-subject brain-computer interface.
\newblock \emph{ArXiv}, abs/2102.04456, 2021{\natexlab{b}}.

\bibitem[Pan et~al.(2021)Pan, Liu, Shi, Wong, and Chan]{Paper_Pan_fMRI_16}
Li~Pan, Jundong Liu, Mingqin Shi, Chi~Wah Wong, and Kei Hang~Katie Chan.
\newblock Identifying autism spectrum disorder based on individual-aware
  down-sampling and multi-modal learning.
\newblock \emph{ArXiv}, abs/2109.09129, 2021.

\bibitem[Subah et~al.(2021)Subah, Deb, Dhar, and Koshiba]{Paper_Subah_fMRI_14}
Faria~Zarin Subah, Kaushik Deb, Pranab~Kumar Dhar, and Takeshi Koshiba.
\newblock A deep learning approach to predict autism spectrum disorder using
  multisite resting-state fmri.
\newblock \emph{Applied Sciences}, 11:\penalty0 3636, 2021.

\bibitem[Bhaumik et~al.(2018)Bhaumik, Pradhan, Das, and
  Bhaumik]{Paper_Bhaumik_fMRI_23}
Runa Bhaumik, Ashish Pradhan, Soptik Das, and Dulal~K. Bhaumik.
\newblock Predicting autism spectrum disorder using domain-adaptive cross-site
  evaluation.
\newblock \emph{Neuroinformatics}, 16:\penalty0 197--205, 2018.

\bibitem[Yang et~al.(2019)Yang, Zhao, Jiang, Gao, and Liu]{Paper_Yang_RP4}
Fu~Yang, Xingcong Zhao, Wenge Jiang, Pengfei Gao, and Guangyuan Liu.
\newblock Multi-method fusion of cross-subject emotion recognition based on
  high-dimensional eeg features.
\newblock \emph{Frontiers in Computational Neuroscience}, 13, 2019.

\bibitem[Fdez et~al.(2020)Fdez, Guttenberg, Witkowski, and
  Pasquali]{Paper_Fdez_RP5}
Javier Fdez, Nicholas Guttenberg, Olaf Witkowski, and Antoine Pasquali.
\newblock Cross-subject eeg-based emotion recognition through neural networks
  with stratified normalization.
\newblock \emph{Frontiers in Neuroscience}, 15, 2020.

\bibitem[Dissanayake et~al.(2021)Dissanayake, Fernando, Denman, Sridharan, and
  Fookes]{Paper_Dissanayake_62}
Theekshana Dissanayake, Tharindu Fernando, Simon Denman, Sridha Sridharan, and
  Clinton Fookes.
\newblock Deep learning for patient-independent epileptic seizure prediction
  using scalp eeg signals.
\newblock \emph{IEEE Sensors Journal}, 21:\penalty0 9377--9388, 2021.

\bibitem[Shoeb(2009)]{Dataset_CHBMIT}
Ali~H. Shoeb.
\newblock Application of machine learning to epileptic seizure onset detection
  and treatment.
\newblock 2009.

\bibitem[Cui et~al.(2022)Cui, Lan, Sourina, and Muller-Wittig]{Paper_Cui_RP13}
Jian Cui, Zirui Lan, Olga Sourina, and Wolfgang Muller-Wittig.
\newblock Eeg-based cross-subject driver drowsiness recognition with an
  interpretable convolutional neural network.
\newblock \emph{IEEE transactions on neural networks and learning systems}, PP,
  2022.

\bibitem[Zhang et~al.(2020{\natexlab{d}})Zhang, Li, Li, Peng, Kang, Jiang, Li,
  Zhu, Yao, Biswal, and Xu]{Paper_Zhang_fMRI_24}
Tao Zhang, Cunbo Li, Peiyang Li, Yueheng Peng, Xiaodong Kang, Chenyang Jiang,
  Fali Li, Xuyang Zhu, Dezhong Yao, Bharat~B. Biswal, and Peng Xu.
\newblock Separated channel attention convolutional neural network
  (sc-cnn-attention) to identify adhd in multi-site rs-fmri dataset.
\newblock \emph{Entropy}, 22, 2020{\natexlab{d}}.

\bibitem[Jiang et~al.(2021{\natexlab{b}})Jiang, Wang, Shi, Wu, Hu, Chen, Hu,
  Wang, and Qiu]{Paper_Jiang_fMRI_3}
Zhoufan Jiang, Yanming Wang, Chenwei Shi, Yueyang Wu, Rongjie Hu, Shishuo Chen,
  Sheng Hu, Xiaoxiao Wang, and Bensheng Qiu.
\newblock Attention module improves both performance and interpretability of 4d
  fmri decoding neural network.
\newblock \emph{ArXiv}, abs/2110.00920, 2021{\natexlab{b}}.

\bibitem[Jana et~al.(2022)Jana, Sabath, and Agrawal]{Paper_Jana_RP14}
Gopal~Chandra Jana, Anshuman Sabath, and Anupam Agrawal.
\newblock Capsule neural networks on spatio-temporal eeg frames for
  cross-subject emotion recognition.
\newblock \emph{Biomed. Signal Process. Control.}, 72:\penalty0 103361, 2022.

\bibitem[Sabour et~al.(2017{\natexlab{b}})Sabour, Frosst, and
  Hinton]{Sabour2017DynamicRB}
Sara Sabour, Nicholas Frosst, and Geoffrey~E. Hinton.
\newblock Dynamic routing between capsules.
\newblock \emph{ArXiv}, abs/1710.09829, 2017{\natexlab{b}}.

\bibitem[Li et~al.(2021{\natexlab{d}})Li, Li, Pan, and Wang]{Paper_Li_RP24}
Jingcong Li, Shuqi Li, Jiahui Pan, and Fei Wang.
\newblock Cross-subject eeg emotion recognition with self-organized graph
  neural network.
\newblock \emph{Frontiers in Neuroscience}, 15, 2021{\natexlab{d}}.

\bibitem[Cao et~al.(2021)Cao, Yang, Qin, Zhu, Chen, Wang, and
  Liu]{Paper_Cao_fMRI_13}
Menglin Cao, Ming Yang, Chi Qin, Xiaofei Zhu, Yanni Chen, Jue Wang, and Tian
  Liu.
\newblock Using deepgcn to identify the autism spectrum disorder from
  multi-site resting-state data.
\newblock \emph{Biomed. Signal Process. Control.}, 70:\penalty0 103015, 2021.

\bibitem[Rong et~al.(2020)Rong, Huang, Xu, and Huang]{Rong2020DropEdgeTD}
Yu~Rong, Wenbing Huang, Tingyang Xu, and Junzhou Huang.
\newblock Dropedge: Towards deep graph convolutional networks on node
  classification.
\newblock In \emph{ICLR}, 2020.

\bibitem[Joshi et~al.(2022)Joshi, Ghongade, Joshi, and
  Kulkarni]{Paper_Joshi_RP36}
Vaishali~M. Joshi, Rajesh~B. Ghongade, Aditi~M Joshi, and Rushikesh Kulkarni.
\newblock Deep bilstm neural network model for emotion detection using
  cross-dataset approach.
\newblock \emph{Biomed. Signal Process. Control.}, 73:\penalty0 103407, 2022.

\bibitem[Joshi and Ghongade(2022)]{Joshi2022IDEAID}
Vaishali~M. Joshi and Rajesh~B. Ghongade.
\newblock Idea: Intellect database for emotion analysis using eeg signal.
\newblock \emph{J. King Saud Univ. Comput. Inf. Sci.}, 34:\penalty0 4433--4447,
  2022.

\bibitem[Lee et~al.(2021{\natexlab{d}})Lee, Han, Kim, Jeong, and
  Lee]{Paper_Lee_84}
Dae-Hyeok Lee, Dong-Kyun Han, Sung-Jin Kim, Ji-Hoon Jeong, and Seong-Whan Lee.
\newblock Subject-independent brain-computer interface for decoding high-level
  visual imagery tasks.
\newblock \emph{2021 IEEE International Conference on Systems, Man, and
  Cybernetics (SMC)}, pages 3396--3401, 2021{\natexlab{d}}.

\bibitem[Schirrmeister et~al.(2017{\natexlab{a}})Schirrmeister, Springenberg,
  Fiederer, Glasstetter, Eggensperger, Tangermann, Hutter, Burgard, and
  Ball]{Schirrmeister2017DeepLW}
Robin~Tibor Schirrmeister, Jost~Tobias Springenberg, Lukas Dominique~Josef
  Fiederer, Martin Glasstetter, Katharina Eggensperger, Michael Tangermann,
  Frank Hutter, Wolfram Burgard, and Tonio Ball.
\newblock Deep learning with convolutional neural networks for eeg decoding and
  visualization.
\newblock \emph{Human Brain Mapping}, 38:\penalty0 5391 -- 5420,
  2017{\natexlab{a}}.

\bibitem[Dar et~al.(2022)Dar, Akram, Rajamanickam, Khawaja, and
  Murugappan]{Paper_Dar_RP35}
Muhammad~Najam Dar, Muhammad~Usman Akram, Yuvaraj Rajamanickam, Sajid~Gul
  Khawaja, and Murugappan Murugappan.
\newblock Eeg-based emotion charting for parkinson's disease patients using
  convolutional recurrent neural networks and cross dataset learning.
\newblock \emph{Computers in biology and medicine}, 144:\penalty0 105327, 2022.

\bibitem[Lin et~al.(2021{\natexlab{b}})Lin, Jie, Dong, Ding, Bian, and
  Liu]{Paper_Lin_fMRI_33}
Kai-Yi Lin, Biao Jie, Peng Dong, Xintao Ding, Weixin Bian, and Mingxia Liu.
\newblock Extracting sequential features from dynamic connectivity network with
  rs-fmri data for ad classification.
\newblock In \emph{MLMI@MICCAI}, 2021{\natexlab{b}}.

\bibitem[Harrison et~al.(2020)Harrison, Bijsterboch, Segerdahl, Fitzgibbon,
  Farahibozorg, Duff, Smith, and Woolrich]{Paper_Harrison_fMRI_4}
Samuel~J. Harrison, Janine~D. Bijsterboch, Andrew~R. Segerdahl, Sean~P.
  Fitzgibbon, Seyedeh-Rezvan Farahibozorg, Eugene~P. Duff, Stephen~M. Smith,
  and Mark~W. Woolrich.
\newblock Modelling subject variability in the spatial and temporal
  characteristics of functional modes.
\newblock \emph{Neuroimage}, 222, 2020.

\bibitem[Wang et~al.(2022{\natexlab{c}})Wang, Yao, Ma, and
  Liu]{Paper_Wang_fMRI_9}
Nan Wang, Dongren Yao, Lizhuang Ma, and Mingxia Liu.
\newblock Multi-site clustering and nested feature extraction for identifying
  autism spectrum disorder with resting-state fmri.
\newblock \emph{Medical image analysis}, 75:\penalty0 102279,
  2022{\natexlab{c}}.

\bibitem[Koelstra et~al.(2012)Koelstra, M{\"u}hl, Soleymani, Lee, Yazdani,
  Ebrahimi, Pun, Nijholt, and Patras]{Dataset_DEAP}
Sander Koelstra, Christian M{\"u}hl, M.~Soleymani, Jong-Seok Lee, Ashkan
  Yazdani, Touradj Ebrahimi, Thierry Pun, Anton Nijholt, and I.~Patras.
\newblock Deap: A database for emotion analysis ;using physiological signals.
\newblock \emph{IEEE Transactions on Affective Computing}, 3:\penalty0 18--31,
  2012.

\bibitem[Katsigiannis and Ramzan(2018)]{Dataset_DREAMER}
Stamos Katsigiannis and Naeem Ramzan.
\newblock Dreamer: A database for emotion recognition through eeg and ecg
  signals from wireless low-cost off-the-shelf devices.
\newblock \emph{IEEE Journal of Biomedical and Health Informatics}, 22\penalty0
  (1):\penalty0 98--107, 2018.
\newblock \doi{10.1109/JBHI.2017.2688239}.

\bibitem[Zheng and Lu(2015)]{Dataset_SEED}
Wei-Long Zheng and Bao-Liang Lu.
\newblock Investigating critical frequency bands and channels for eeg-based
  emotion recognition with deep neural networks.
\newblock \emph{IEEE Transactions on Autonomous Mental Development}, 7\penalty0
  (3):\penalty0 162--175, 2015.
\newblock \doi{10.1109/TAMD.2015.2431497}.

\bibitem[Zheng et~al.(2018)Zheng, Liu, Lu, Lu, and Cichocki]{Dataset_SEEDIV}
W.~Zheng, W.~Liu, Y.~Lu, B.~Lu, and A.~Cichocki.
\newblock Emotionmeter: A multimodal framework for recognizing human emotions.
\newblock \emph{IEEE Transactions on Cybernetics}, pages 1--13, 2018.
\newblock ISSN 2168-2267.
\newblock \doi{10.1109/TCYB.2018.2797176}.

\bibitem[Hu et~al.(2022)Hu, Wang, and Zhang]{Dataset_THU-EP}
Xin Hu, Fei Wang, and Dan Zhang.
\newblock Similar brains blend emotion in similar ways: Neural representations
  of individual difference in emotion profiles.
\newblock \emph{NeuroImage}, 247:\penalty0 118819, 01 2022.

\bibitem[Soleymani et~al.(2012)Soleymani, Lichtenauer, Pun, and
  Pantic]{Dataset_MAHNOB}
Mohammad Soleymani, Jeroen Lichtenauer, Thierry Pun, and Maja Pantic.
\newblock A multimodal database for affect recognition and implicit tagging.
\newblock \emph{IEEE Transactions on Affective Computing}, 3\penalty0
  (1):\penalty0 42--55, 2012.
\newblock \doi{10.1109/T-AFFC.2011.25}.

\bibitem[Brunner et~al.(2008)Brunner, Leeb, M{\"u}ller-Putz, Schl{\"o}gl, and
  Pfurtscheller]{Dataset_BCI_2a}
Clemens Brunner, Robert Leeb, Gernot M{\"u}ller-Putz, Alois Schl{\"o}gl, and
  Gert Pfurtscheller.
\newblock Bci competition 2008--graz data set a.
\newblock \emph{Institute for Knowledge Discovery (Laboratory of Brain-Computer
  Interfaces), Graz University of Technology}, 16:\penalty0 1--6, 2008.

\bibitem[Leeb and Brunner(2008)]{Dataset_BCI_2b}
Robert Leeb and Clemens Brunner.
\newblock Bci competition 2008 graz data set b.
\newblock 2008.

\bibitem[Lee et~al.(2019)Lee, Kwon, Kim, Kim, Lee, Williamson, Fazli, and
  Lee]{Dataset_KU}
Min-Ho Lee, O-Yeon Kwon, Yong-Jeong Kim, Hong-Kyung Kim, Young-Eun Lee, John
  Williamson, Siamac Fazli, and Seong-Whan Lee.
\newblock {EEG dataset and OpenBMI toolbox for three BCI paradigms: an
  investigation into BCI illiteracy}.
\newblock \emph{GigaScience}, 8\penalty0 (5), 01 2019.
\newblock ISSN 2047-217X.
\newblock \doi{10.1093/gigascience/giz002}.
\newblock URL \url{https://doi.org/10.1093/gigascience/giz002}.
\newblock giz002.

\bibitem[Cho et~al.(2017)Cho, Ahn, Ahn, Kwon, and Jun]{Dataset_GIST}
Hohyun Cho, Minkyu Ahn, Sangtae Ahn, Moonyoung Kwon, and Sung~Chan Jun.
\newblock {EEG datasets for motor imagery brain–computer interface}.
\newblock \emph{GigaScience}, 6\penalty0 (7), 05 2017.
\newblock ISSN 2047-217X.
\newblock \doi{10.1093/gigascience/gix034}.
\newblock URL \url{https://doi.org/10.1093/gigascience/gix034}.
\newblock gix034.

\bibitem[O’Reilly et~al.(2014)O’Reilly, Gosselin, Carrier, and
  Nielsen]{Dataset_MASS}
Christian O’Reilly, Nadia Gosselin, Julie Carrier, and Tore Nielsen.
\newblock Montreal archive of sleep studies: an open‐access resource for
  instrument benchmarking and exploratory research.
\newblock \emph{Journal of Sleep Research}, 23, 2014.

\bibitem[Khalighi et~al.(2015)Khalighi, Sousa, Santos, and
  Nunes]{Dataset_ISRUC}
Sirvan Khalighi, Teresa Sousa, José Santos, and Urbano Nunes.
\newblock Isruc-sleep: A comprehensive public dataset for sleep researchers.
\newblock \emph{Computer Methods and Programs in Biomedicine}, 124, 11 2015.
\newblock \doi{10.1016/j.cmpb.2015.10.013}.

\bibitem[Cao et~al.(2019)Cao, Chuang, King, and Lin]{Dataset_Taiwan}
Zehong Cao, Chun-Hsiang Chuang, Jung-Kai King, and Chin-Teng Lin.
\newblock Multi-channel eeg recordings during a sustained-attention driving
  task.
\newblock \emph{Scientific Data}, 6, 2019.

\bibitem[Zheng and Lu(2017)]{Dataset_SEED_VIG}
Wei-Long Zheng and Bao-Liang Lu.
\newblock A multimodal approach to estimating vigilance using {EEG} and
  forehead {EOG}.
\newblock \emph{Journal of Neural Engineering}, 14\penalty0 (2):\penalty0
  026017, feb 2017.
\newblock \doi{10.1088/1741-2552/aa5a98}.
\newblock URL \url{https://doi.org/10.1088/1741-2552/aa5a98}.

\bibitem[Rezaei et~al.(2017)Rezaei, Mohammadi, and Khazaie]{Dataset_Kermanshah}
Mohammad Rezaei, Hiwa Mohammadi, and Habibolah Khazaie.
\newblock Eeg/eog/emg data from a cross sectional study on psychophysiological
  insomnia and normal sleep subjects.
\newblock \emph{Data in Brief}, 15, 09 2017.
\newblock \doi{10.1016/j.dib.2017.09.033}.

\bibitem[Albuquerque et~al.(2020)Albuquerque, Tiwari, Parent, Cassani, Gagnon,
  Lafond, Tremblay, and Falk]{Dataset_WAUC}
Isabela Albuquerque, Abhishek Tiwari, Mark Parent, Raymundo Cassani,
  Jean-François Gagnon, Daniel Lafond, Sébastien Tremblay, and Tiago~H. Falk.
\newblock Wauc: A multi-modal database for mental workload assessment under
  physical activity.
\newblock \emph{Frontiers in Neuroscience}, 14, 2020.
\newblock ISSN 1662-453X.
\newblock \doi{10.3389/fnins.2020.549524}.
\newblock URL
  \url{https://www.frontiersin.org/articles/10.3389/fnins.2020.549524}.

\bibitem[Lehnertz(2008)]{Dataset_Bonn}
Klaus Lehnertz.
\newblock Epilepsy and nonlinear dynamics.
\newblock \emph{Journal of biological physics}, 34:\penalty0 253--66, 09 2008.
\newblock \doi{10.1007/s10867-008-9090-3}.

\bibitem[Song et~al.(2019)Song, Zheng, Lu, Zong, Zhang, and Cui]{Dataset_MPED}
Tengfei Song, Wenming Zheng, Cheng Lu, Yuan Zong, Xilei Zhang, and Zhen Cui.
\newblock Mped: A multi-modal physiological emotion database for discrete
  emotion recognition.
\newblock \emph{IEEE Access}, 7:\penalty0 12177--12191, 2019.

\bibitem[Shah et~al.(2018)Shah, von Weltin, Lopez, McHugh, Veloso,
  Golmohammadi, Obeid, and Picone]{Dataset_TUSZ}
Vinit Shah, Eva von Weltin, Silvia Lopez, James McHugh, Lily Veloso, Meysam
  Golmohammadi, Iyad Obeid, and Joseph Picone.
\newblock The temple university hospital seizure detection corpus.
\newblock \emph{Frontiers in Neuroinformatics}, 12, 01 2018.
\newblock \doi{10.3389/fninf.2018.00083}.

\bibitem[Schirrmeister et~al.(2017{\natexlab{b}})Schirrmeister, Springenberg,
  Fiederer, Glasstetter, Eggensperger, Tangermann, Hutter, Burgard, and
  Ball]{Dataset_HGD}
Robin~Tibor Schirrmeister, Jost~Tobias Springenberg, Lukas Dominique~Josef
  Fiederer, Martin Glasstetter, Katharina Eggensperger, Michael Tangermann,
  Frank Hutter, Wolfram Burgard, and Tonio Ball.
\newblock Deep learning with convolutional neural networks for eeg decoding and
  visualization.
\newblock \emph{Human Brain Mapping}, 38\penalty0 (11):\penalty0 5391--5420,
  2017{\natexlab{b}}.
\newblock \doi{https://doi.org/10.1002/hbm.23730}.
\newblock URL \url{https://onlinelibrary.wiley.com/doi/abs/10.1002/hbm.23730}.

\bibitem[Torres-García et~al.(2016)Torres-García, Reyes-Garcia,
  Villaseñor-Pineda, and Garcia-Aguilar]{Dataset_Spanish}
Alejandro Torres-García, Carlos~Alberto Reyes-Garcia, Luis Villaseñor-Pineda,
  and Gregorio Garcia-Aguilar.
\newblock Implementing a fuzzy inference system in a multi-objective eeg
  channel selection model for imagined speech classification.
\newblock \emph{Expert Systems with Applications}, 59:\penalty0 1--12, 10 2016.
\newblock \doi{10.1016/j.eswa.2016.04.011}.

\bibitem[Nguyen et~al.(2017)Nguyen, Karavas, and Artemiadis]{Dataset_Long}
Chuong Nguyen, Georgios Karavas, and Panagiotis Artemiadis.
\newblock Inferring imagined speech using eeg signals: a new approach using
  riemannian manifold features.
\newblock \emph{Journal of Neural Engineering}, 15, 07 2017.
\newblock \doi{10.1088/1741-2552/aa8235}.

\bibitem[Kemp et~al.(2000)Kemp, Zwinderman, Tuk, Kamphuisen, and
  Oberye]{Dataset_Sleep-EDF}
B.~Kemp, A.H. Zwinderman, B.~Tuk, H.A.C. Kamphuisen, and J.J.L. Oberye.
\newblock Analysis of a sleep-dependent neuronal feedback loop: the slow-wave
  microcontinuity of the eeg.
\newblock \emph{IEEE Transactions on Biomedical Engineering}, 47\penalty0
  (9):\penalty0 1185--1194, 2000.
\newblock \doi{10.1109/10.867928}.

\bibitem[Quan et~al.(1998)Quan, Howard, Iber, Kiley, Nieto, O'Connor, Rapoport,
  Redline, Robbins, Samet, and Wahl]{Dataset_SHHS}
Stuart Quan, Barbara Howard, Conrad Iber, James Kiley, F.~Nieto, George
  O'Connor, David Rapoport, Susan Redline, John Robbins, Jonathan Samet, and
  ‡Patricia Wahl.
\newblock The sleep heart health study: Design, rationale, and methods.
\newblock \emph{Sleep}, 20:\penalty0 1077--85, 01 1998.
\newblock \doi{10.1093/sleep/20.12.1077}.

\bibitem[Schelter et~al.(2007)Schelter, Winterhalder, Maiwald, Brandt, Schad,
  Timmer, and Schulze-Bonhage]{Dataset_Freiburg}
Björn Schelter, Matthias Winterhalder, Thomas Maiwald, Armin Brandt, Ariane
  Schad, Jens Timmer, and Andreas Schulze-Bonhage.
\newblock Do false predictions of seizures depend on the state of vigilance? a
  report from two seizure-prediction methods and proposed remedies.
\newblock \emph{Epilepsia}, 47:\penalty0 2058--70, 01 2007.
\newblock \doi{10.1111/j.1528-1167.2006.00848.x}.

\bibitem[Gu et~al.(2022)Gu, Cai, Gao, Jiang, Ning, and Qian]{Paper_Gu_3}
Xiaoqing Gu, Weiwei Cai, Ming Gao, Yizhang Jiang, Xin Ning, and Pengjiang Qian.
\newblock Multi-source domain transfer discriminative dictionary learning
  modeling for electroencephalogram-based emotion recognition.
\newblock \emph{IEEE Transactions on Computational Social Systems}, pages 1--9,
  2022.
\newblock \doi{10.1109/TCSS.2022.3153660}.

\bibitem[Li et~al.(2022{\natexlab{c}})Li, Hua, qiang Xu, Shu, Xu, Kuang, and
  Wu]{Paper_Li_72}
Jinyu Li, Haoqiang Hua, Zhi qiang Xu, Lin Shu, Xiangmin Xu, Feng Kuang, and
  Shibin Wu.
\newblock Cross-subject eeg emotion recognition combined with connectivity
  features and meta-transfer learning.
\newblock \emph{Computers in biology and medicine}, 145:\penalty0 105519,
  2022{\natexlab{c}}.

\bibitem[di~Martino et~al.(2014)di~Martino, Yan, Li, Denio, Castellanos,
  Alaerts, Anderson, Assaf, Bookheimer, Dapretto, Deen, Delmonte, Dinstein,
  Ertl-Wagner, Fair, Gallagher, Kennedy, Keown, Keysers, Lainhart, Lord, Luna,
  Menon, Minshew, Monk, Mueller, M{\"u}ller, Nebel, Nigg, O'Hearn, Pelphrey,
  Peltier, Rudie, Sunaert, Thioux, Tyszka, Uddin, Verhoeven, Wenderoth,
  Wiggins, Mostofsky, and Milham]{Dataset_ABIDEI}
Adriana di~Martino, Chaogan Yan, Qingyang Li, Erin~B. Denio, Francisco~Xavier
  Castellanos, Kaat Alaerts, Jeffrey~S. Anderson, Michal Assaf, Susan~Y.
  Bookheimer, Mirella Dapretto, Ben Deen, Sonja Delmonte, Ilan Dinstein,
  Birgit~B. Ertl-Wagner, Damien~A. Fair, Louise Gallagher, Daniel~P. Kennedy,
  Christopher~Lee Keown, Christian Keysers, Janet~E. Lainhart, Catherine Lord,
  Beatriz Luna, V.~Menon, Nancy~J. Minshew, Christopher~S. Monk, Sophia
  Mueller, Ralph-Axel M{\"u}ller, Mary~Beth Nebel, Joel~T. Nigg, Kirsten
  O'Hearn, Kevin~A. Pelphrey, Scott~J. Peltier, Jeffrey~D. Rudie, Stefan
  Sunaert, Marc Thioux, Julian~Michael Tyszka, Lucina~Q. Uddin, Judith~S.
  Verhoeven, Nicole Wenderoth, Jillian~Lee Wiggins, Stewart~H. Mostofsky, and
  Michael~Peter Milham.
\newblock The autism brain imaging data exchange: Towards large-scale
  evaluation of the intrinsic brain architecture in autism.
\newblock \emph{Molecular psychiatry}, 19:\penalty0 659 -- 667, 2014.

\bibitem[di~Martino et~al.(2017)di~Martino, O’Connor, Chen, Alaerts,
  Anderson, Assaf, Balsters, Baxter, Beggiato, Bernaerts, Blanken, Bookheimer,
  Braden, Byrge, Castellanos, Dapretto, Delorme, Fair, Fishman, Fitzgerald,
  Gallagher, Keehn, Kennedy, Lainhart, Luna, Mostofsky, M{\"u}ller, Nebel,
  Nigg, O'Hearn, Solomon, Toro, Vaidya, Wenderoth, White, Craddock, Lord,
  Leventhal, and Milham]{Dataset_ABIDEII}
Adriana di~Martino, David O’Connor, Bosi Chen, Kaat Alaerts, Jeffrey~S.
  Anderson, Michal Assaf, Joshua~Henk Balsters, Leslie~C. Baxter, Anita
  Beggiato, Sylvie Bernaerts, Laura M.~E. Blanken, Susan~Y. Bookheimer,
  B.~Blair Braden, Lisa Byrge, Francisco~Xavier Castellanos, Mirella Dapretto,
  Richard Delorme, Damien~A. Fair, Inna Fishman, Jacqueline Fitzgerald, Louise
  Gallagher, R.~Joanne~Jao Keehn, Daniel~P. Kennedy, Janet~E. Lainhart, Beatriz
  Luna, Stewart~H. Mostofsky, Ralph-Axel M{\"u}ller, Mary~Beth Nebel, Joel~T.
  Nigg, Kirsten O'Hearn, Marjorie Solomon, Roberto Toro, Chandan~J. Vaidya,
  Nicole Wenderoth, Tonya White, Richard~Cameron Craddock, Catherine Lord,
  Bennett~L. Leventhal, and Michael~Peter Milham.
\newblock Enhancing studies of the connectome in autism using the autism brain
  imaging data exchange ii.
\newblock \emph{Scientific Data}, 4, 2017.

\bibitem[{Van Essen} et~al.(2013){Van Essen}, Smith, Barch, Behrens, Yacoub,
  and Ugurbil]{Dataset_HCP}
David~C. {Van Essen}, Stephen~M. Smith, Deanna~M. Barch, Timothy~E.J. Behrens,
  Essa Yacoub, and Kamil Ugurbil.
\newblock The wu-minn human connectome project: An overview.
\newblock \emph{NeuroImage}, 80:\penalty0 62--79, 2013.
\newblock ISSN 1053-8119.
\newblock \doi{https://doi.org/10.1016/j.neuroimage.2013.05.041}.
\newblock URL
  \url{https://www.sciencedirect.com/science/article/pii/S1053811913005351}.
\newblock Mapping the Connectome.

\bibitem[Schonberg et~al.(2012)Schonberg, Fox, Mumford, Congdon, Trepel, and
  Poldrack]{Dataset_OpenfMRIBalloon}
Tom Schonberg, Craig~R. Fox, Jeanette~A. Mumford, Eliza Congdon, Christopher
  Trepel, and Russell~A. Poldrack.
\newblock Decreasing ventromedial prefrontal cortex activity during sequential
  risk-taking: An fmri investigation of the balloon analog risk task.
\newblock \emph{Frontiers in Neuroscience}, 6, 2012.

\bibitem[Brown et~al.(2012)Brown, Sidhu, Greiner, Asgarian, Bastani,
  Silverstone, Greenshaw, and Dursun]{Dataset_ADHD}
Matthew R.~G. Brown, Gagan Preet~Singh Sidhu, Russell Greiner, Nasimeh
  Asgarian, Meysam Bastani, Peter~H. Silverstone, Andrew~James Greenshaw, and
  Serdar~M. Dursun.
\newblock Adhd-200 global competition: diagnosing adhd using personal
  characteristic data can outperform resting state fmri measurements.
\newblock \emph{Frontiers in Systems Neuroscience}, 6, 2012.

\bibitem[Petersen et~al.(2010)Petersen, Aisen, Beckett, Donohue, Gamst, Harvey,
  Jack, Jagust, Shaw, Toga, Trojanowski, and Weiner]{Dataset_ADNII}
Ronald~C. Petersen, Paul~S. Aisen, Laurel~A Beckett, Michael~C. Donohue,
  Anthony~Collins Gamst, Danielle~J. Harvey, Clifford~R. Jack, William~J.
  Jagust, Leslie~M. Shaw, Arthur~W. Toga, John~Q. Trojanowski, and Michael~W
  Weiner.
\newblock Alzheimer's disease neuroimaging initiative (adni).
\newblock \emph{Neurology}, 74:\penalty0 201 -- 209, 2010.

\bibitem[Beckett et~al.(2015)Beckett, Donohue, Wang, Aisen, Harvey, and
  Saito]{Dataset_ADNIII}
Laurel~A Beckett, Michael~C. Donohue, Cathy Wang, Paul~S. Aisen, Danielle~J.
  Harvey, and Naomi~H. Saito.
\newblock The alzheimer's disease neuroimaging initiative phase 2: Increasing
  the length, breadth, and depth of our understanding.
\newblock \emph{Alzheimer's \& Dementia}, 11:\penalty0 823--831, 2015.

\bibitem[Chang et~al.(2019)Chang, Pyles, Marcus, Gupta, Tarr, and
  Aminoff]{Dataset_Bold5000}
Nadine Chang, John~A. Pyles, Austin Marcus, Abhinav~Kumar Gupta, Michael~J.
  Tarr, and Elissa~M. Aminoff.
\newblock Bold5000, a public fmri dataset while viewing 5000 visual images.
\newblock \emph{Scientific Data}, 6, 2019.

\bibitem[Chen et~al.(2017)Chen, Leong, Honey, Yong, Norman, and
  Hasson]{Dataset_Sherlock}
Janice Chen, Yuan~Chang Leong, Christopher~John Honey, Chung~Hen Yong,
  Kenneth~A. Norman, and Uri Hasson.
\newblock Shared memories reveal shared structure in neural activity across
  individuals.
\newblock \emph{Nature neuroscience}, 20:\penalty0 115 -- 125, 2017.

\bibitem[Keator et~al.(2016)Keator, Erp, Turner, Glover, Mueller, Liu,
  Voyvodic, Rasmussen, Calhoun, Lee, Toga, McEwen, Ford, Mathalon, Diaz,
  O'Leary, Bockholt, Gadde, Preda, Wible, Stern, and
  et~al.]{Dataset_AuditoryOddball}
David~B. Keator, T.A.G.M. Erp, Jessica~A. Turner, Gary~H. Glover, Bryon~A.
  Mueller, Thomas~T. Liu, James Voyvodic, Jerod~M Rasmussen, Vince~D. Calhoun,
  Hyo-Jong Lee, Arthur~W. Toga, Sarah~C. McEwen, Judith~M. Ford, Daniel~H.
  Mathalon, Michele~T. Diaz, Daniel~S. O'Leary, H.~Jeremy Bockholt, Syam Gadde,
  Adrian Preda, Cynthia~G. Wible, Hal~S. Stern, and et~al.
\newblock The function biomedical informatics research network data repository.
\newblock \emph{NeuroImage}, 124:\penalty0 1074--1079, 2016.

\bibitem[Pinho et~al.(2018)Pinho, Amadon, Ruest, Fabre, Dohmatob, Denghien,
  Ginisty, Becuwe-Desmidt, Roger, Laurier, Joly-Testault, M{\'e}diouni-Cloarec,
  Doubl{\'e}, Martins, Pinel, Eger, Varoquaux, Pallier, Dehaene, Hertz-Pannier,
  and Thirion]{Dataset_IBC}
Ana~Lu{\'i}sa Pinho, Alexis Amadon, Torsten Ruest, Murielle Fabre, Elvis
  Dohmatob, Isabelle Denghien, Chantal Ginisty, S{\'e}verine Becuwe-Desmidt,
  S{\'e}verine Roger, Laurence Laurier, V{\'e}ronique Joly-Testault, Ga{\"e}lle
  M{\'e}diouni-Cloarec, Christine Doubl{\'e}, Bernadette Martins, Philippe
  Pinel, Evelyn Eger, Ga{\"e}l Varoquaux, Christophe Pallier, Stanislas
  Dehaene, Lucie Hertz-Pannier, and Bertrand Thirion.
\newblock Individual brain charting, a high-resolution fmri dataset for
  cognitive mapping.
\newblock \emph{Scientific Data}, 5, 2018.

\bibitem[Wang et~al.(2022{\natexlab{d}})Wang, Yao, Ma, and
  Liu]{Paper_wang_multi-site_2022}
Nan Wang, Dongren Yao, Lizhuang Ma, and Mingxia Liu.
\newblock Multi-site clustering and nested feature extraction for identifying
  autism spectrum disorder with resting-state {fMRI}.
\newblock \emph{Medical Image Analysis}, 75:\penalty0 102279, January
  2022{\natexlab{d}}.
\newblock ISSN 13618415.
\newblock \doi{10.1016/j.media.2021.102279}.
\newblock URL
  \url{https://linkinghub.elsevier.com/retrieve/pii/S1361841521003248}.

\end{thebibliography}

\end{document}